\documentclass[times,astrobib,amssymb,usenatbib]{mn2e}
\usepackage{graphicx}
\usepackage{txfonts}
\usepackage{natbib}
\usepackage{supertabular}
\usepackage{longtable}
\usepackage{url}
\usepackage{dcolumn}
\usepackage{placeins}
\usepackage{multirow} 
\usepackage{rotating}
\begin{document}
\newcommand{\sqcm}{cm$^{-2}$}  
\newcommand{\lya}{Ly$\alpha$}
\newcommand{\lyb}{Ly$\beta$}
\newcommand{\lyg}{Ly$\gamma$}
\newcommand{\heo}{\mbox{He\,{\sc i}}}
\newcommand{\he}{\mbox{He\,{\sc ii}}}
\newcommand{\hi}{\mbox{H\,{\sc i}}}
\newcommand{\hw}{\mbox{H\,{\sc ii}}}
\newcommand{\os}{\mbox{O\,{\sc vi}}}
\newcommand{\ose}{\mbox{O\,{\sc vii}}}
\newcommand{\cf}{\mbox{C\,{\sc iv}}}
\newcommand{\cto}{\mbox{C\,{\sc ii}}}
\newcommand{\ct}{\mbox{C\,{\sc iii}}}
\newcommand{\sit}{\mbox{Si\,{\sc iii}}}
\newcommand{\sif}{\mbox{Si\,{\sc iv}}}
\newcommand{\sito}{\mbox{Si\,{\sc ii1260}}}
\newcommand{\sia}{\mbox{Si\,{\sc ii1526}}}
\newcommand{\nf}{\mbox{N\,{\sc v}}}
\newcommand{\nt}{\mbox{N\,{\sc iii}}}
\newcommand{\neo}{\mbox{Ne\,{\sc viii}}}
\newcommand{\mgt}{\mbox{Mg\,{\sc ii}}}
\newcommand{\zabs}{$z_{\rm abs}$}
\newcommand{\zmin}{$z_{\rm min}$}
\newcommand{\zmax}{$z_{\rm max}$}
\newcommand{\zqso}{$z_{\rm qso}$}
\newcommand{\subHe}{_{\it HeII}}
\newcommand{\subH}{_{\it HI}}
\newcommand{\subHLy}{_{\it H Ly}}
\newcommand{\degree}{\ensuremath{^\circ}}
\newcommand{\lapp}{\mbox{\raisebox{-0.3em}{$\stackrel{\textstyle <}{\sim}$}}}
\newcommand{\gapp}{\mbox{\raisebox{-0.3em}{$\stackrel{\textstyle >}{\sim}$}}}
\newcommand{\be}{\begin{equation}}
\newcommand{\en}{\end{equation}}
\newcommand{\di}{\displaystyle}
\def\tworule{\noalign{\medskip\hrule\smallskip\hrule\medskip}} 
\def\onerule{\noalign{\medskip\hrule\medskip}} 
\def\bl{\par\vskip 12pt\noindent}
\def\bll{\par\vskip 24pt\noindent}
\def\blll{\par\vskip 36pt\noindent}
\def\rot{\mathop{\rm rot}\nolimits}
\def\alf{$\alpha$}
\def\refff{\leftskip20pt\parindent-20pt\parskip4pt}
\def\zabs{$z_{\rm abs}$}
\def\zem{$z_{\rm em}$}
\def\mgii{Mg\,{\sc ii}~}
\def\feiia{Fe\,{\sc ii}$\lambda$2600}
\def\mgia{Mg\,{\sc i}$\lambda$2852}
\def\mgiia{Mg\,{\sc ii}$\lambda$2796}
\def\mgiib{Mg\,{\sc ii}$\lambda$2803}
\def\mgiiab{Mg\,{\sc ii}$\lambda\lambda$2796,2803}
\def\wobs{$w_{\rm obs}$}
\def\kms{km~s$^{-1}$}
\def\chisq{$\chi^{2}$}
\def\dtype{$\delta_{\rm type}$}

\title[High redshift \os\ absorbers]{A high resolution study of intergalactic 
\os\ absorbers at $z \sim$~2.3}
\author[S. Muzahid et al.]{S. Muzahid$^{1}$\thanks{E-mail: sowgat@iucaa.ernet.in},  
R. Srianand$^{1}$, J. Bergeron$^{2}$  and  P. Petitjean$^{2}$ \\  
$^{1}$ Inter-University Centre for Astronomy and Astrophysics, Post Bag 4, 
Ganeshkhind, Pune 411\,007, India \\
$^{2}$ Universit\'e Paris 6, UMR 7095, Institut d'Astrophysique de Paris-CNRS, 
98bis Boulevard Arago, 75014 Paris, France \\
}
\date{Accepted. Received; in original form }
\maketitle
\label{firstpage}
\begin {abstract} 
 
We present a detailed study of the largest sample of intervening \os\ 
systems in the redshift range 1.9 $\le z \le$ 3.1 detected in high  
resolution ($R$~$\sim$~45,000) spectra of 18 bright QSOs observed with 
VLT/UVES. Based on Voigt profile and apparent optical depth analysis 
we find that (i) the Doppler parameters of the \os\ absorption are usually   
broader than those of \cf\ (ii) the column density distribution 
of \os\ is steeper than that of \cf\ (iii) line spread ($\delta v$) 
of the \os\ and \cf\ are strongly correlated (at $5.3\sigma$ level) 
with $\delta v(\os)$ being systematically larger than 
$\delta v(\cf)$ and (iv) $\delta v(\os)$ and $\delta v(\cf)$ are also 
correlated (at $>5\sigma$ level) with their respective column densities 
and with $N$(\hi) (3 and 4.5$\sigma$ respectively). 
The median column densities of \hi, \os, and \cf\ are found to be 
higher when low ions are present. $N$(\cf) and $N$(\hi) 
are strongly correlated (at $4.3\sigma$ level).
However, no significant correlation is found between $N$(\os) 
and $N$(\hi). These findings favor the idea that \cf\ and \os\ absorption 
originate from different phases of a correlated structure and systems 
with large velocity spread are probably associated with overdense regions. 
The velocity offset between optical depth weighted redshifts of \cf\ 
and \os\ absorption is found to be in the 
range $0\le~|\Delta v (\os-\cf)|~\le48$ \kms\ with a median value of 
8~\kms. 

We do not find any evidence for the ratios $N$(\os)/$N$(\hi), 
$N$(\os)/$N$(\cf) and $N$(\cf)/$N$(\hi) to evolve with $z$ over the 
redshift range considered here. But a lack of systems 
with high $N(\os)/N(\hi)$ ratio (i.e., $\ge -$0.5 dex) for $z>$ 2.5 
is noticeable. Similar trend is also seen for the $N(\cf)/N(\hi)$ ratio. 
We compare the properties of \os\ systems in our sample with that of 
low redshift ($z < 0.5$) samples from the literature and find that  
(i) the \os\ components at low-$z$ are systematically wider than at
high-$z$ with an enhanced non-thermal contribution to their $b$-parameter, 
(ii) the slope of the column density distribution functions for high and 
low-$z$ are consistent,  
(iii) the range in gas temperature estimated from a subsample of well aligned 
absorbers is similar at both high and low-$z$, and   
(iv) $\Omega_{\os}$~=~$(1.0\pm0.2)\times10^{-7}$ for $N(\os)>10^{13.7}$  
cm$^{-2}$, estimated in our high-$z$ sample, 
is very similar to low-$z$ estimations. 
\end {abstract}


\section{Introduction} 
The study of low density intergalactic medium (IGM) is extremely important 
because it forms the primary reservoir of baryons throughout the cosmic ages. 
These baryons get accumulated into galaxies in the process of structure formation. 
The heavy elements produced in galaxies got transported to the IGM by means of 
outflows driven by supernovae or tidal interactions. Thus the IGM enrichment 
history provides useful constraints on the star formation history and 
contribution of various 
feedback mechanisms at different epochs. The tenuous IGM is detectable in the 
form of Ly$\alpha$ and heavy element absorption lines in the QSO spectra. 
Hence the observation of Ly$\alpha$ and metal lines are crucial to understand 
the interaction between galaxies and the surrounding IGM. 

The observations of \cf\ and \os\ absorption in QSO spectra have established 
the presence of heavy elements in the IGM unequivocally \citep[]{Cowie95a,
Cowie95b,Songaila96,Bergeron02,Simcoe02,Carswell02}.   
The early observations of heavy elements in the high redshift universe were 
focused on \cf\ absorption, since it has 
strongest lines falling in the region free from Ly$\alpha$ contamination. 
Several high redshift surveys have ascertained that the cosmic density of 
\cf\ absorbers has not evolved substantially from redshift $z = 5$ 
to $z = 1.5$ \citep[see][]{Songaila01,Songaila05,Boksenberg03,Pettini03,Schaye03}.  
Due to lack of sufficient QSO sight lines, a clear picture 
is yet to emerge regarding the evolution of \cf\ at $z > 6$. However the 
limited sample of \cf\ absorbers available until now indeed indicates that a 
good fraction of metals 
may already be present in the IGM even at these high redshifts \citep{Songaila06,
Simcoe06a,Ryan-Weber06,Becker06,Becker09}. The enrichment level found by these 
studies is consistent with [C/H] $\sim -2.8$.  

Given the low metallicity, the direct detection of metals in the 
underdense regions [with overdensity, 
$\delta (\equiv \rm {n_{\rm H}/\bar n_{\rm H}}) 
\ll 10$], which occupy most of the volume of the universe at any given epoch, 
is beyond the reach of the present day large telescopes. 
Statistical methods like pixel analysis are used instead 
\citep{Ellison00,Schaye03,Aracil04,Aguirre05,Aguirre08,Scannapieco06,Pieri06}.  
They show that metals must be present even in the underdense regions. 
However the fractional volume occupied by the metals is still unknown. 

Even in regions where metal absorption is detected directly, it is unclear  
what are the main physical processes that maintain the ionization state  
of the gas. In general, it is believed that photoionization by 
the meta-galactic UV background keeps the gas ionized.  
On the other hand, the winds that seed the IGM with metals and the accretion 
shocks in the evolving density fields may also provide sufficient 
mechanical feedback to collisionally ionize the gas. 
Therefore, it is crucial to simultaneously 
study different species covering a wide range of ionization states to get a better 
understanding of the metal enrichment and the different ionizing mechanisms at play. 

Under photoionization by UV background, the \os\ absorption is generally 
produced from regions of low-density having high ionization parameter. 
In addition, the high cosmic abundance of oxygen makes \os\ a good 
tracer of metal-enrichment in the low-density IGM. In fact photoionization 
seems to be a viable process for most of the high redshift \os\ absorbers 
\citep[]{Bergeron02,Carswell02,Bergeron05}. 
On the other hand, hydrodynamical simulations 
\citep{Cen99,Dave99,Dave01,Fang01,Kang05,
Smith11,Cen11} suggest that a considerable amount 
of baryons could reside in the warm-hot phase of the intergalactic 
medium (called WHIM with $T \approx 10^{5}-10^{7}$~K) and this fraction
evolves with redshift. Highly ionized 
species of oxygen such as O~{\sc vi}, O~{\sc vii} and O~{\sc viii} can be 
useful probes of the WHIM. In the optical regime, \os\ is the best 
species to probe relatively cooler phase (i.e., $T \sim 3\times10^{5}$ K) of 
the WHIM because, the ionization fraction of \os\ has its maximum around 
this temperature in case of collisional ionization. 
Indeed, it has been 
suggested that a large fraction of the \os\ absorption associated 
with the IGM \citep[][]{Simcoe02,Simcoe06b} and high-$z$ damped Ly$\alpha$ 
systems \citep[][]{Fox07b} may originate from collisionally ionized gas. 
Thus the origin of the ionization of the intergalactic \os\ 
absorbers is a matter of debate.  

Numerous extragalactic \os\ systems have been detected at $z < 0.5$ with the 
{\it Far Ultraviolet Spectroscopic Explorer} ($FUSE$) and 
{\it Hubble Space Telescope} ($HST$) \citep[e.g.,][]{Tripp00,Stocke06,
Danforth06,Tripp08,Thom08,Wakker09,Lehner09,Tumlinson11}. 
However these studies could not establish convincingly whether \os\ absorbers 
are predominantly tracing the WHIM gas or not. They seem to arise 
in either high metallicity photoionized \citep[e.g.,][]{Oppenheimer09} or low 
metallicity collisionally ionized \citep[e.g.,][]{Smith11} gas loosely 
associated with galaxies. An alternative tool to detect the WHIM, which is 
independent of the metal enrichment, is to search for 
thermally-broadened Ly$\alpha$ absorbers (BLAs)  
\citep[]{Sembach04,Richter04,Richter06,Danforth10,Savage10,Savage11}. 
However, until now the WHIM detection through BLAs has also been ambiguous. 
The few detections of \neo\ in the FUV regime 
\citep[]{Savage05a,Narayanan09,Narayanan11} and O~{\sc vii} in the soft 
X-ray regime \citep[e.g.,][]{Nicastro05a,Nicastro05b,Buote09,Fang10,
Zappacosta10} possibly hint the existence of the WHIM. 

At high redshift ($z>2$) previous studies \citep[]{Simcoe02,Simcoe04,
Carswell02,Bergeron02,Bergeron05,Schaye07,Frank10a,Frank10b} have already 
provided important insights into the properties of \os\ absorbers. 
Here we present a detailed analysis of \os\ absorbers using 
a sample which is twice as large as the previous sample of high-$z$ 
intervening \os\ systems studied at high resolution. 

This paper is organized as follows. In section \ref{obs} we describe 
the observations and the data reduction procedure for our  
data sample. In section \ref{data} we describe the line  
identification strategy and present the \os\ and \cf\ sample and various 
physically motivated subsamples. In section \ref{dist} we analyze the 
distributions of \os\ Voigt profile parameters and compared them with those 
of \cf. In this section we also compare the properties of \os\ absorption 
at high and low redshift. In section \ref{kinemat} we discuss the 
line kinematics of \os\ and \cf\ absorption. 
In section \ref{model} we present analysis based on total column 
densities with results of photoionization model as guidelines. In section 
\ref{discuss} we summarize our results. 

Throughout this paper we use the following cosmological parameters 
for a flat universe : $\Omega_{\rm m}$ = 0.3, $\Omega_{\Lambda}$ = 0.7, 
$\Omega_{b}h^{2}$ = 0.02 and $H_0$ = 71 \kms Mpc$^{-1}$. The solar 
relative abundances are taken to be default values used in 
CLOUDY v(07.02), i.e., log~(C/H)$_{\odot}$ = $-$3.61 and 
log~(O/H)$_{\odot}$ = $-$3.31. 
%

\begin{center}
\begin{table*} 
\vskip -0.2in
\scriptsize 
{\bf  
\caption{Details of \os\ systems}
\begin{flushleft}
\begin{tabular}{lcccccccccccccc} 
\hline
QSO & \zem\ & \zmin\ & \zmax\ & $z_{\rm sys}$ & log~$N(\hi)$ & log~$N(\os)$ & log~$N(\cf)$ & Class\footnotemark[1] & Case\footnotemark[2] & 
\dtype\ \footnotemark[3] & $\delta v(\os)$ & $\delta v(\cf)$ & $|\Delta v(\os - \cf)|$ & Low Ions\footnotemark[6] \\  
    &      &       &       &       &       &        &        &        &       &     & (\kms)   & (\kms) & (\kms) & \\ 
\hline \hline
\underline{HE~1341--1020} &2.135&1.983&2.083& 1.9982& 13.77$\pm$0.04& 13.71$\pm$0.04& 12.22$\pm$0.05& $dd$&   A&    0&  24.3&  19.7&  4.6 & No \\
              &     &     &     & 2.0414& 15.70$\pm$0.07& 14.01$\pm$0.12& 13.25$\pm$0.09& $bd$&   B& $+$1&  88.6&  36.9&  ... & Yes \\ 
	      &     &     &     & 2.0850& 15.10$\pm$0.02& 13.79$\pm$0.08& 12.98$\pm$0.06& $bb$&   B& $-$1&  23.6&  91.8&  ... & No \\ 
\underline{Q~0122--380}   &2.190&1.989&2.137& 2.0349& 15.54$\pm$0.06& 13.70$\pm$0.09& 13.10$\pm$0.04& $dd$&   B& $-$1&  59.8&  77.7&  ... & No \\ 
	      &     &     &     & 2.0626& 12.49$\pm$0.04& 13.47$\pm$0.05& 12.94$\pm$0.01& $bd$&   A&    0&  45.1&  19.3&  2.0 & No \\ 
\underline{PKS~1448--232} &2.220&2.008&2.166& 2.1099& 13.86$\pm$0.02& 14.48$\pm$0.07& 13.16$\pm$0.03& $dd$&   A&    0&  56.0&  35.4&  3.3 & No \\ 
              &     &     &     & 2.1660& 15.52$\pm$0.36& 14.29$\pm$0.13& 13.48$\pm$0.18& $dd$& A/B&    0& 272.8& 166.4& 13.4 & Yes \\ 
\underline{PKS~0237--23}  &2.222&1.954&2.168& 1.9878\footnotemark[4]& 13.64$\pm$0.01& 13.41$\pm$0.06& $\le$~11.97& $dd$&   A&    0&  41.3&   0.0&  ... & No \\ 
              &     &     &     & 2.0108\footnotemark[4]& 14.51$\pm$0.06& 13.15$\pm$0.05& $\le$~11.93& $dd$&   A&    0&  26.5&   0.0&  ... & No \\ 
	      &     &     &     & 2.0412& 16.24$\pm$0.04& 13.16$\pm$0.07& 12.61$\pm$0.06& $bd$&   B& $-$1 &  31.0&  36.8&  ... & No \\ 
	      &     &     &     & 2.0422& $\le$~13.48& 14.30$\pm$0.07& 13.66$\pm$0.05& $bd$&   A&    0&  47.7&  40.7&  7.2 & No \\  
\underline{HE~0001--2340} &2.263&1.982&2.209& 2.0323& 16.07$\pm$0.16& 13.89$\pm$0.06& 12.85$\pm$0.13& $bd$&   B&    0&  50.3&  64.2& 15.3 & No \\  
              &     &     &     & 2.1617& 15.61$\pm$0.22& 13.82$\pm$0.06& 12.76$\pm$0.06& $dd$&   B&    0&  80.4&  80.7& 19.2 & No \\   
\underline{Q~0109--3518}  &2.404&1.980&2.347& 2.0226& 14.99$\pm$0.05& 14.24$\pm$0.14& 12.52$\pm$0.03& $bd$&   B&   +1&  60.1&  41.0&  ... & Yes \\  
              &     &     &     & 2.1415& 15.01$\pm$0.53& 13.52$\pm$0.06& 11.70$\pm$0.08& $dd$&   B&    0&  60.1&  16.6&  5.3 & No \\  
\underline{HE~2217--2818} &2.414&1.959&2.357& 2.0160& 15.03$\pm$0.02& 13.77$\pm$0.13& 12.80$\pm$0.05& $bb$&   B& 0  &  86.8&  86.0&  0.1 & No \\  
	      &     &     &     & 2.0748& 14.12$\pm$0.01& 14.32$\pm$0.01& 12.85$\pm$0.01& $dd$& A/B& $-$1&  37.8&  92.0&  ... & No  \\ 
	      &     &     &     & 2.1808& 16.06$\pm$0.04& 14.19$\pm$0.23& 13.53$\pm$0.07& $dd$& A/B&    0& 178.2& 115.1& 46.4 & Yes \\ 
\underline{Q~0329--385}   &2.435&2.018&2.378& 2.0764& 13.70$\pm$0.08& 13.26$\pm$0.05& 13.21$\pm$0.01& $bd$&   A&    0&  21.3&  19.2&  4.2 & No \\ 
              &     &     &     & 2.1470\footnotemark[4]& 14.73$\pm$0.21& 13.95$\pm$0.18& $\le$~12.89& $dd$&   A&    0& 120.1&   0.0&  ... & No \\ 
              &     &     &     & 2.2489\footnotemark[4]& 13.51$\pm$0.30& 14.25$\pm$0.49& $\le$~12.82& $dd$&   A&    0& 107.4&   0.0&  ... & No \\ 
	      &     &     &     & 2.2510& 15.82$\pm$0.55& 14.86$\pm$0.16& 14.41$\pm$0.08& $dd$&   B&    0& 127.4&  84.4& 10.4 & Yes \\ 
	      &     &     &     & 2.3139& 14.23$\pm$0.10& 13.30$\pm$0.10& 12.35$\pm$0.12& $dd$&   B&    0&  48.2&  43.1& 12.9 & No \\ 
	      &     &     &     & 2.3520& 13.05$\pm$0.01& 14.09$\pm$0.09& 13.50$\pm$0.08& $dd$&   A&    0&  41.2&  27.3&  1.0 & Yes \\ 
	      &     &     &     & 2.3639& 14.85$\pm$0.09& 13.73$\pm$0.03& 12.40$\pm$0.10& $dd$&   A&    0&  64.8&  57.8&  1.2 & No \\ 
	      &     &     &     & 2.3738& 15.26$\pm$0.01& 14.26$\pm$0.08& 12.66$\pm$0.06& $dd$&   A&    0& 153.0&  96.6&  0.8 & No \\ 
HE~1158--1843 &2.449&1.980&2.391& 2.2354& 14.74$\pm$0.02& 13.71$\pm$0.17& 12.99$\pm$0.12& $dd$&   A&    0&  47.1&  35.4&  2.1 & No  \\ 
              &     &     &     & 2.2660& 15.88$\pm$0.05& 13.68$\pm$0.05& 13.68$\pm$0.04& $bd$& A/B& $+1$& 142.4&  29.7&  ... & Yes \\ 
\underline{HE~1347--2457} &2.611&1.985&2.551& 2.1162& 15.16$\pm$0.07& 14.57$\pm$0.07& 13.43$\pm$0.07& $bd$& A/B& $+1$&  91.2&  63.2&  ... & Yes \\
	      &     &     &     & 2.2349& 15.06$\pm$0.02& 14.43$\pm$0.47& 13.24$\pm$0.11& $dd$&   A&    0& 195.9& 187.3& 10.6 & No \\ 
	      &     &     &     & 2.3289& 16.45$\pm$0.04& 14.63$\pm$0.04& 14.26$\pm$0.04& $bb$&   B& $-$1&  59.1&  99.5&  ... & Yes \\ 
	      &     &     &     & 2.3327\footnotemark[4]& 13.89$\pm$0.02& 13.26$\pm$0.03& $\le$~12.02& $dd$&   A&    0&  45.9&   0.0&  ... & No \\  
	      &     &     &     & 2.3422\footnotemark[4]& 14.82$\pm$0.30& 13.45$\pm$0.09& $\le$~12.30& $dd$&   A&    0&  56.7&   0.0&  ... & No\\  
	      &     &     &     & 2.3700& 14.96$\pm$0.10& 14.08$\pm$0.05& 12.95$\pm$0.08& $bd$&   A&    0&  77.8&  62.7& 16.9 & Yes \\  
	      &     &     &     & 2.4455& 15.14$\pm$0.47& 13.69$\pm$0.26& 12.62$\pm$0.17& $bd$&   B&    0&  67.7&  64.7&  0.8 & No \\  
\underline{Q~0453--423}  &2.658&2.001&2.597& 2.1694& 13.75$\pm$0.03& 14.27$\pm$0.05& 12.65$\pm$0.05& $bb$&   A& $+1$ & 119.2&  80.5&  ... & No \\ 
	      &     &     &     & 2.2765& 16.29$\pm$0.06& 14.87$\pm$0.32& 14.78$\pm$0.23& $bb$&   B& $-$1& 179.6& 149.8&  ... & Yes \\ 
	      &     &     &     & 2.3978& 15.14$\pm$0.09& 14.97$\pm$0.36& 14.47$\pm$0.18& $bb$&   A&    0& 186.0& 156.2& 48.5 & Yes \\ 
	      &     &     &     & 2.4435& 15.37$\pm$0.05& 14.44$\pm$0.24& 13.76$\pm$0.35& $bb$& A/B&    0& 181.4& 142.8&  6.9 & Yes \\ 
	      &     &     &     & 2.5028& 15.77$\pm$0.02& 13.99$\pm$0.03& 13.71$\pm$0.34& $bb$&   B& $-$1&  74.7&  78.3&  ... & Yes \\ 
	      &     &     &     & 2.5214& 15.31$\pm$0.04& 13.58$\pm$0.19& 12.58$\pm$0.15& $bb$&   B& $-$1&  51.6&  32.5&  ... & No \\ 
	      &     &     &     & 2.5371\footnotemark[4]& 14.67$\pm$0.08& 13.42$\pm$0.07& $\le$~12.29& $dd$&   A&    0&  45.2&   0.0&  ... & No \\ 
\underline{PKS~0329--255} &2.703&2.080&2.641& 2.2044& 15.44$\pm$0.22& 14.61$\pm$0.24& 13.32$\pm$0.21& $bd$&   B& $+$1& 176.9& 144.7&  ... & No \\
              &     &     &     & 2.3284& 16.20$\pm$0.12& 13.96$\pm$0.30& 13.49$\pm$0.70& $bb$&   B& $-$1 &  93.9& 186.8&  ... & Yes \\ 
              &     &     &     & 2.3896\footnotemark[4]& 14.22$\pm$0.12& 13.37$\pm$0.02& $\le$~12.00& $bb$&   A& $+$1&   0.0&   0.0&  ... & No \\ 
              &     &     &     & 2.4253& 15.58$\pm$0.05& 13.86$\pm$0.12& 13.25$\pm$0.19& $dd$&   B&    0&  65.8&  50.0&  9.7 & Yes \\ 
              &     &     &     & 2.5687& 14.78$\pm$0.04& 13.58$\pm$0.44& 12.52$\pm$0.26& $dd$&   A&    0&  42.8&  48.0&  6.8 & No \\ 
Q~0002--422   &2.767&2.064&2.704& 2.1767\footnotemark[4]& 13.62$\pm$0.01& 13.72$\pm$0.04& $\le$~12.02& $bd$&   A&    0&  52.6&   0.0&  ... & No \\ 
	      &     &     &     & 2.2200\footnotemark[4]& 14.95$\pm$0.02& 13.80$\pm$0.32& $\le$~12.09& $dd$& A/B&    0&  51.8&   0.0&  ... & No  \\ 
	      &     &     &     & 2.2621& 15.47$\pm$0.04& 13.67$\pm$0.11& 12.43$\pm$0.03& $bd$&   B&    0&  44.5&  38.5& 12.0 & No \\ 
              &     &     &     & 2.5395& 14.55$\pm$0.03& 13.79$\pm$0.01& 12.48$\pm$0.02& $dd$&   A&    0&  61.5&  29.1&  7.2 & No \\ 
              &     &     &     & 2.6075& 14.72$\pm$0.03& 14.10$\pm$0.05& 12.14$\pm$0.05& $bd$&   A&    0&  70.4&  34.9&  7.7 & No \\ 
\underline{HE~0151--4326} &2.789&2.043&2.726& 2.0885& 15.16$\pm$0.06& 13.75$\pm$0.10& 12.81$\pm$0.29& $dd$&   B&    0&  35.3&  33.8&  2.7 & No \\  
              &     &     &     & 2.1699& 15.25$\pm$0.02& 13.86$\pm$0.07& 13.04$\pm$0.01& $bd$&   B&    0&  43.5&  27.5&  9.7 & No \\   
	      &     &     &     & 2.2010& 15.39$\pm$0.03& 14.23$\pm$0.13& 13.12$\pm$0.19& $bd$&   B&    0& 131.6& 146.9&  1.1 & Yes \\ 
	      &     &     &     & 2.3598\footnotemark[4]& 14.03$\pm$0.03& 13.45$\pm$0.10& $\le$~12.24& $bd$&   A&    0&  47.6&   0.0&  ... & No \\  
	      &     &     &     & 2.4681& 13.25$\pm$0.22& 14.38$\pm$0.54& 13.43$\pm$0.06& $dd$&   A&    0&  73.3&  65.9& 15.0 & Yes \\ 
	      &     &     &     & 2.4927& 14.77$\pm$0.01& 14.16$\pm$0.27& 13.51$\pm$0.03& $bd$&   A&    0& 120.6&  42.6& 18.1 & No \\  
	      &     &     &     & 2.5053& 15.01$\pm$0.02& 14.19$\pm$0.08& 12.47$\pm$0.16& $bb$&   A& $-$1& 118.1& 117.5&  ... & No \\ 
	      &     &     &     & 2.5235& 14.81$\pm$0.56& 13.94$\pm$0.23& 12.30$\pm$0.15& $bb$&   A& $-$1&   0.0&   0.0&  ... & No \\ 
HE~2347--4342 &2.871&2.301&2.806& 2.3475& 16.17$\pm$0.11& 13.74$\pm$0.11& 13.49$\pm$0.01& $bd$&   B&    0&  34.7&  25.6&  9.7 & Yes \\ 
	      &     &     &     & 2.4382& 14.84$\pm$0.02& 14.38$\pm$0.11& 12.98$\pm$0.15& $bd$&   B&    0& 126.8&  74.7& 11.0 & No \\  
	      &     &     &     & 2.6346& 14.84$\pm$0.38& 14.31$\pm$0.28& 13.08$\pm$0.14& $bb$&   A&    0& 126.0&  64.4& 26.5 & No \\ 
	      &     &     &     & 2.6498& 15.12$\pm$0.06& 14.07$\pm$0.08& 12.66$\pm$0.06& $bd$& A/B&    0& 114.8&  79.8&  2.1 & No \\  
	      &     &     &     & 2.7121& 14.85$\pm$0.07& 13.96$\pm$0.06& 12.27$\pm$0.11& $bd$&   B& $-$1&  86.8&  46.2&  ... & No \\ 
	      &     &     &     & 2.7356& 16.50$\pm$0.28& 14.41$\pm$0.11& 14.08$\pm$0.10\footnotemark[5]& $bb$&   B& $-$1&  58.2&   0.0&  ... & Yes \\ 
	      &     &     &     & 2.7456& 14.42$\pm$0.04& 13.47$\pm$0.04& .....\footnotemark[7] & $bd$&   A&    0&  71.0&   0.0&  ... & No \\ 
HE~0940--1050 &3.084&2.458&3.016& 2.5167& 15.37$\pm$0.03& 13.89$\pm$0.01& 12.84$\pm$0.07& $bb$&   B&    0&  62.2&  74.8& 16.6 & No \\ 
              &     &     &     & 2.6433& 15.56$\pm$0.15& 13.94$\pm$0.07& 13.16$\pm$0.07& $bb$&   B& $-$1&  90.9&  89.5&  ... & No \\  
	      &     &     &     & 2.6580& 15.41$\pm$0.02& 13.79$\pm$0.01& 13.40$\pm$0.05& $dd$&   B&    0&  68.5&  32.8&  5.1 & Yes \\  
	      &     &     &     & 2.8265& 16.72$\pm$0.37& 14.50$\pm$0.06& 14.17$\pm$0.16& $bb$&   B& $-$1& 136.6& 377.0&  ... & Yes \\  
	      &     &     &     & 2.8345& 16.66$\pm$0.06& 14.40$\pm$0.01& 14.27$\pm$0.24& $bb$&   B& $-$1&  83.2&  91.4&  ... & Yes \\  
	      &     &     &     & 2.9377& 14.77$\pm$0.03& 13.86$\pm$0.12& 12.94$\pm$0.05& $bb$&   A& $-$1& 101.3&  70.2&  ... & Yes \\  
	      &     &     &     & 2.9401& 14.67$\pm$0.01& 14.16$\pm$0.06& 12.64$\pm$0.09& $bb$& A/B& $+$1&  92.2&  71.9&  ... & Yes \\  
Q~0420--388   &3.117&2.625&3.048& 2.6615& 15.44$\pm$0.03& 14.58$\pm$0.03& 14.37$\pm$0.18& $bb$&   B& $-$1& 229.2&  89.6&  ... & Yes \\ 
              &     &     &     & 2.8100& 15.29$\pm$0.03& 13.48$\pm$0.05& 12.53$\pm$0.04& $bd$&   B&    0&  46.3&  17.3&  5.1 & No \\ 
PKS~2126--158 &3.280&2.508&3.209& 2.9073& 16.16$\pm$0.04& 13.92$\pm$0.17& 13.63$\pm$0.23& $bd$&   B&    0&  68.7&  82.5&  7.6 & Yes \\ 
\hline 
\hline
\end{tabular} 
\end{flushleft}
Table Notes -- 
Underlined lines of sight are common to \citet{Bergeron05}. 
\footnotemark[1] {Class is based on the \os\ profiles. The systems with both the 
doublets are unblended (partially blended) are marked by ``$dd$" (``$bb$"). In  ``$bd$" 
systems one of the doublets is blended.}  
\footnotemark[2] {Case is based on the presence of unsaturated Lyman-series line. ``Case A" systems 
have at least one of the available Lyman-series line unsaturated. In ``Case B", systems all 
the available Lyman-series are saturated. ``Case A/B" are the cases where some parts of \hi\ 
absorption are unsaturated.} 
\footnotemark[3] {\dtype\ is based on the robust measurement of \os\ line spread. \dtype\ of $+1$ and 
$-1$ indicate upper and lower limits on line spread whereas \dtype\ of 0 represents robust measurement.} 
\footnotemark[4] {``\os\ only" system.} 
\footnotemark[5] {\cf\ column density is taken from \citet{Agafonova07}.} 
\footnotemark[6] {Indicating the presence of low ions.} 
\footnotemark[7] {\cf\ falls in the spectral gap.}
\label{allsystems} 
}
\end{table*} 
\end{center}


\section{Observations} 
\label{obs}

The spectra used in this study were obtained with the Ultra-Violet and 
Visible Echelle Spectrograph (UVES) \citep{Dekker00} mounted on the 
ESO Kueyen 8.2 m telescope at the Paranal observatory in the course of 
the ESO-VLT large programme ``The Cosmic Evolution of the IGM" 
\citep{Bergeron04}. This large programme provided a homogeneous 
set of 18 QSO sight lines with QSO emission redshifts ranging from 2.1 to 3.3.  
The raw data were reduced using the 
UVES pipeline \citep{Ballester00} which is available as a dedicated context 
of the MIDAS data reduction software. The main function of the pipeline is 
to perform a precise inter-order background subtraction for science frames 
and master flat fields, and to apply an optimal extraction to retrieve the 
object signal, rejecting cosmic ray impacts and performing sky 
subtraction at the same time. The reduction is checked step-by-step. 
Wavelengths are corrected to vacuum-heliocentric values using standard 
conversion  equations \citep{Edlen66,Stumpff80}. 
Combination of individual exposures is performed by adjusting the flux in 
each individual exposures to the same level and inverse variance weighting the 
flux in each pixel. Great care was taken in computing the error spectrum 
while combining the individual exposures. Our final error in each pixel is
 the quadratic sum of the weighted mean of errors in the different spectra 
and the scatter in the individual flux measurements. Errors in individual 
pixels obtained by this method are consistent with the rms dispersion 
around the best fitted continuum in regions free of absorption lines. 
The final combined spectrum covers the wavelength range of 3000 to
 $10,000$~{\AA} with occasional narrow gaps in the red. 
During the observations, the $2\times2$ binning mode was used yielding a 
binned pixel size of 2.0 -- 2.4 \kms.  A typical signal-to-noise ratio 
(S/N) $\sim$~30 -- 40 and 60 -- 70 per pixel was achieved at 3300 and 5500~{\AA} 
respectively. The typical final spectral resolution is 
$R \sim 45,000$ (FWHM $\sim$ 6.6 \kms) 
over the entire wavelength range. This spectral resolution allows us to 
resolve lines with $b$-parameter as narrow as $\sim$~4~\kms.  
The unabsorbed QSO continuum is then fitted 
using low order polynomials extrapolated from wavelength ranges devoid of 
strong absorption lines. The detailed description of 
data calibration is presented in \citet{Aracil04} and \citet{Chand04}. 
%

\begin{table}
\begin{center}
\scriptsize
\caption{ Systems with upper limits on $N(\os)$}
\begin{tabular}{l c c c c p{.5cm} p{.3cm}} 
\hline 
    &        & \multicolumn{3}{c}{log (column density)} & $\delta v(\cf)$ & Low \\ \cline{3-5} 
QSO & $z_{\rm sys}$  & $N(\hi)$ & $N(\cf)$ &  $\le$ $N(\os)$ & (\kms) &  Ions \\ 
\hline \hline 
PKS~1448--232  & 1.9516 & 14.94 $\pm$ 0.05 & 12.89 $\pm$ 0.01&  13.90$^{a}$ &  23.9 & No \\
PKS~1448--232  & 1.9781 & 15.07 $\pm$ 0.07 & 12.64 $\pm$ 0.03&  13.77$^{a}$ &  35.6 & No \\
HE~0001--2340  & 2.1634 & 14.71 $\pm$ 0.04 & 11.92 $\pm$ 0.06&  12.81$^{a}$ &  20.2 & No \\
Q~0109--3518   & 2.0463 & 16.09 $\pm$ 0.41 & 14.07 $\pm$ 0.10&  14.38       & 172.3 & Yes \\
HE~2217--2818  & 2.0374 & 15.46 $\pm$ 0.07 & 12.16 $\pm$ 0.06&  12.91$^{a}$ &  23.3 & No \\
HE~2217--2818  & 2.1553 & 14.13 $\pm$ 0.02 & 12.48 $\pm$ 0.02&  12.94       &  23.9 & No \\
HE~1158--1843  & 2.0348 & 15.43 $\pm$ 0.29 & 12.61 $\pm$ 0.03&  13.28$^{a}$ &  38.8 & No \\
HE~1158--1843  & 2.0407 & 15.52 $\pm$ 0.07 & 12.51 $\pm$ 0.03&  13.17$^{a}$ &  29.1 & No \\
HE~1347--2457  & 1.9750 & 14.76 $\pm$ 0.02 & 12.64 $\pm$ 0.04&  13.60$^{a}$ &  45.7 & No \\
Q~0453--423    & 2.4163 & 15.00 $\pm$ 0.01 & 12.56 $\pm$ 0.02&  13.81       &  28.5 & No \\
PKS~0329--255  & 2.1611 & 15.95 $\pm$ 0.09 & 12.41 $\pm$ 0.05&  13.23       &  29.4 & Yes \\
PKS~0329--255  & 2.2953 & 14.82 $\pm$ 0.03 & 11.96 $\pm$ 0.09&  12.57$^{a}$ &  19.1 & No \\
PKS~0329--255  & 2.4208 & 14.86 $\pm$ 0.02 & 12.53 $\pm$ 0.16&  14.31       &  41.7 & Yes \\
PKS~0329--255  & 2.5868 & 15.14 $\pm$ 0.02 & 12.29 $\pm$ 0.04&  12.76$^{a}$ &  15.9 & Yes \\
Q~0002--422    & 2.3647 & 12.28 $\pm$ 0.02 & 12.14 $\pm$ 0.03&  12.56$^{a}$ &  15.3 & No \\
HE~0151--4326  & 2.4013 & 15.13 $\pm$ 0.01 & 12.56 $\pm$ 0.08&  13.97       &  60.5 & Yes \\
HE~0151--4326  & 2.4158 & 13.35 $\pm$ 0.01 & 13.03 $\pm$ 0.01&  14.19       &  23.4 & Yes \\
HE~0151--4326  & 2.4196 & 13.04 $\pm$ 0.02 & 12.75 $\pm$ 0.01&  12.57$^{a}$ &  21.7 & Yes \\
HE~0151--4326  & 2.5199 & 15.40 $\pm$ 0.02 & 12.28 $\pm$ 0.03&  12.56$^{a}$ &  25.9 & Yes \\
HE~2347--4342  & 2.3132 & 15.89 $\pm$ 0.71 & 13.73 $\pm$ 0.07&  14.82       & 194.8 & Yes \\
HE~2347--4342  & 2.3317 & 15.58 $\pm$ 0.06 & 12.38 $\pm$ 0.04&  14.31       &  35.9 & Yes \\
HE~0940--1050  & 2.3307 & 16.32 $\pm$ 0.43 & 14.85 $\pm$ 0.23&  15.19       & 197.2 & Yes \\
HE~0940--1050  & 2.4090 & 15.94 $\pm$ 0.14 & 13.56 $\pm$ 0.16&  14.18       &  35.2 & Yes \\
HE~0940--1050  & 2.6136 & 15.32 $\pm$ 0.02 & 12.56 $\pm$ 0.02&  13.58       &  31.6 & No \\
HE~0940--1050  & 2.6679 & 15.60 $\pm$ 0.06 & 13.78 $\pm$ 0.09&  14.11       &  68.6 & Yes \\
Q~0420--388    & 2.8235 & 15.52 $\pm$ 0.03 & 13.73 $\pm$ 0.10&  14.17       & 132.4 & Yes \\
Q~0420--388    & 2.8496 & 14.25 $\pm$ 0.04 & 12.70 $\pm$ 0.13&  14.06       &  64.9 & No \\
Q~0420--388    & 2.9519 & 15.35 $\pm$ 0.01 & 12.72 $\pm$ 0.10&  13.93       &  58.2 & Yes \\
PKS~2126--158  & 2.3889 & 13.92 $\pm$ 0.02 & 13.05 $\pm$ 0.01&  13.66       &  19.7 & No \\
PKS~2126--158  & 2.5537 & 14.05 $\pm$ 0.04 & 13.10 $\pm$ 0.05&  14.55       &  28.2 & No \\
PKS~2126--158  & 2.6790 & 14.00 $\pm$ 0.03 & 14.24 $\pm$ 0.03&  14.24       &  33.9 & Yes \\
PKS~2126--158  & 2.8194 & 15.61 $\pm$ 0.02 & 13.50 $\pm$ 0.08&  14.60       & 170.5 & No  \\
PKS~2126--158  & 2.9634 & 15.85 $\pm$ 0.03 & 13.27 $\pm$ 0.09&  14.08       &  91.3 & Yes \\ 
\hline 
\hline
\end{tabular} 
\label{upp_lim}  
Table Note -- $^{a}$Unabsorbed continuum is seen at least in one the \os\ doublets. 
\end{center}
\end{table}  

%

\section{Data sample and observables} 
\label{data}
In this section we describe our line identification strategy, 
absorption line measurement techniques and various physically 
motivated subsamples we use for statistical studies. 

For the analysis presented here we concentrate on the intervening
\os\ and \cf\ systems defined as those with apparent ejection 
velocity larger than 5000 \kms\ relative to the QSO emission redshift   
\citep[using \zem\ given in][]{Rollinde05,Scannapieco06}. 
The detailed discussions of the associated systems towards QSOs in
our sample can be found in \citet{Fox08}. 

Following \citet{Scannapieco06} we define a system by grouping together 
all the components whose separation from their nearest neighbor is less 
than a linking length, $\rm v_{\rm link}~=~100$~\kms. These authors have 
shown that the absorber's properties and in particular the velocity 
clustering do not change much for velocities smaller than this.
Note that the same convention was also adopted by \citet[]{Songaila05}. 

We have searched for \os\ following two different approaches.  

\par\noindent
(1) First we identify metal line  doublets (e.g. \cf\ and/or \sif) 
redshifted beyond the wavelength of the QSO \lya\ emission and we look 
for \os\ doublets around this redshift (i.e. within $\sim 100$ \kms\ to 
the \cf\ or \sif\ redshift). The presence of \os\ is confirmed after checking 
the consistency in the optical depths of the two \os\ lines. 
There are 104 \cf\ systems \citep[see also][]{Dodorico10} along the 
lines of sight we study over the redshift range where \os\ is detectable 
with S/N $> 10$. 
In addition, there is a system (\zabs\ = 2.7356 towards HE~2347$-$4342)  
that is identified by the presence of \sif\ doublets where the corresponding 
\cf\ doublets fall in the narrow wavelength range not covered by our UVES 
spectrum. The presence of \cf\ in this system has been confirmed 
by \citet{Agafonova07}. The total \cf\ column density in this 
system is taken from their measurement. 
Out of these 105 \cf\ systems, 72 show detectable \os\ absorption. 
Details of the systems are given in Table~\ref{allsystems}. 
For the other 33 \cf\ systems, only upper limits on \os\ column density 
can be obtained. These systems are listed in Table~\ref{upp_lim}. 
Whenever possible we use rms error in the unabsorbed continuum 
at the expected position of the \os\ doublets to estimate 3$\sigma$ 
limit on column density using the same number of components and 
$b$-values as seen in \cf. In rest of the cases where there is strong 
\lya\ absorption we use the $z$ and $b$-values of \cf\ and 
generated the minimum \os\ profile that explains the observed 
spectrum. This allows us to get only a conservative upper limit 
on $N(\os)$.

\begin{table} 
\begin{center}
\caption{ List of Lyman limit systems with \os\ detections}
\begin{flushleft}
\begin{tabular}{lcccc} 
\hline
QSO           & $z_{\rm sys}$ & log~$N(\hi)$ & log~$N(\os)$ & log~$N(\cf)$ \\ 
\hline \hline 
PKS~0329--255 & 2.1584&17.53&12.99&12.86\\ 
Q~0002--422   & 2.1683&18.46&13.64&14.38\\ 
              & 2.3023&17.75&15.21&15.47\\  
	      & 2.4633&18.49&14.63&14.79\\ 
PKS~2126--158 & 2.6377&19.23&14.85&....$^{a}$\\ 
              & 2.7692&18.62&14.80&14.71\\  
\hline 
\hline
\end{tabular} 
\end{flushleft}
\label{tab_lls} 
Table Note -- $^{a}$ \cf\ falls in the spectral gap 
\end{center}
\end{table} 
\begin{figure} 
\centerline{
\vbox{
\centerline{\hbox{ 
\includegraphics[height= 8.4cm,width= 8.4cm,angle= 0]{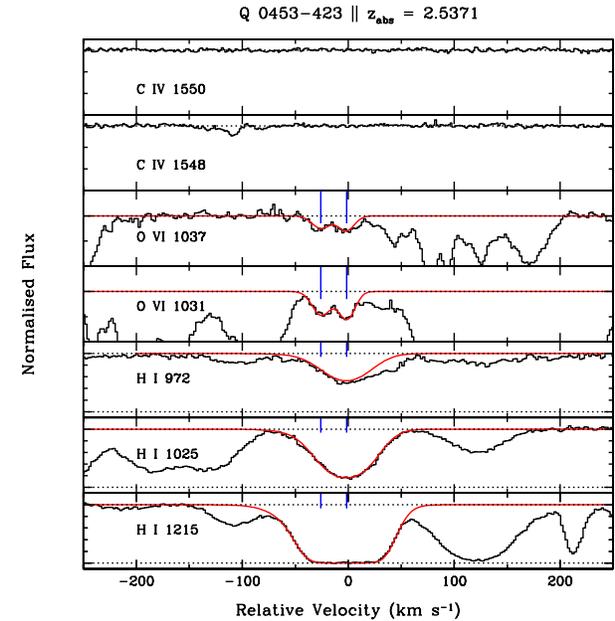}
}}
}}
\caption{An example of \os\ system without detectable associated \cf\ 
	and other heavy elements absorption. The best fitted Voigt 
	profiles of \os\ and \hi\ are over-plotted on the observed data.  
	The vertical tick marks show 
	the positions of the individual components. The absorption redshift 
	that defines the zero velocity and the name of the background QSO 
	are indicated at the top. We found 11 such systems in our sample.} 
\label{ploto6only} 
\end{figure} 
%
%
\begin{figure} 
\centerline{
\vbox{
\centerline{\hbox{ 
\includegraphics[height=8.4cm,width= 9.0cm,angle= 0]{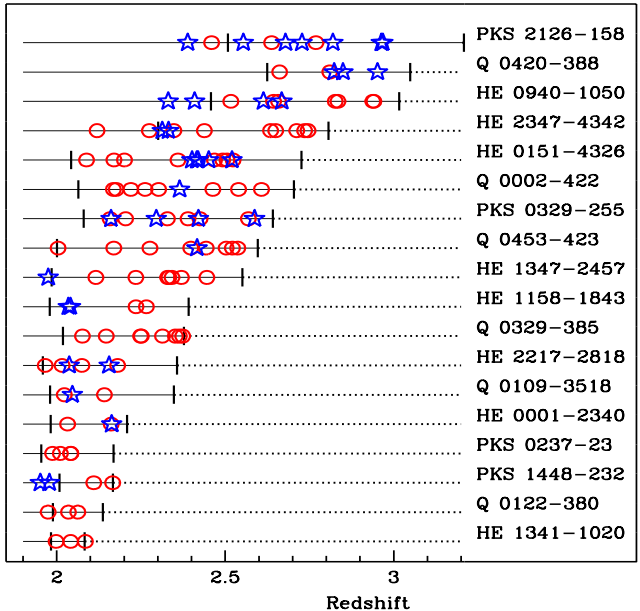} 
}}
}}
\caption{Schematic diagram showing the positions of the absorbers in 
         redshift space along the lines of sight of the QSOs in our sample. 
	 The (red) circles indicate the redshifts of \os\ systems whereas 
	 (blue) stars indicate the redshift of \cf\ systems where we  
	 have upper limits on $N(\os)$. The left and right vertical 
	 lines are the minimum and maximum redshifts along each line of sight.
	 The horizontal dotted lines are just to guide the reader's eye.} 
\label{cluster} 
\end{figure} 
%

\par\noindent   
(2) Secondly, we directly search for the \os\ doublets in the 
\lya\ forest. In fact, \os\ absorption lines without detectable associated 
\cf\ and/or \sif\ (and H~{\sc i}) can  trace the highly ionized gas that 
originates from the WHIM with 
characteristic temperature $\sim 10^{5}-10^{7}$K predicted in some 
simulations \citep[e.g.,][]{Cen99,Cen06,Dave99,Dave01}. For each of these  
identified coincidences we checked the consistency of the shape and 
optical depth ratios of the \os\ doublets. We then checked for the 
presence of associated \lya\ (and possibly higher Lyman series lines) 
at the redshift of the chosen \os\ doublets. While the presence of 
associated \lya\ absorption confirms the \os\ identification, it need 
not be detectable in case \os\ comes from collisionally ionized gas.
Therefore, we do not impose the detection of \lya\ absorption as a necessary 
criterion to confirm the \os\ doublets.
We found 12 systems from the presence of the \os\ doublet only (an example 
is shown in Fig.~\ref{ploto6only}). In all these cases associated 
\hi\ is detected. 
For one of these systems (\zabs\ = 2.7456 towards HE~2347--4342) 
the wavelength range corresponding to \cf\ absorption fall in a spectral 
gap, so that we could not probe the presence of \cf\ in this system.
For the other 11 systems we do not detect any other metal.
Apart for the \zabs~=~2.1767 and 2.3598 systems towards Q~0002--422 and 
HE~0151--4326, respectively, all other systems show absorption from at 
least one of the higher Lyman series lines in addition to \lya\  
(i.e. at least \lyb). Thus we are confident that these identifications of 
\os\ are secure. However, as we look for the presence of both lines in the 
doublet it is possible that our method has missed some of the \os\ only 
absorbers where absorption from one of the transitions (or both) is 
contaminated by \lya\ absorption.

We have found six Lyman-limit systems (LLS) with detectable \os\ 
listed in Table~\ref{tab_lls}. There are 3 more LLS (\zabs~=~2.7278 \& 2.9676 
towards PKS~2126--158 and \zabs~=~2.4512 towards HE~0151--4326) where only 
upper limits on \os\ column densities could be obtained. Note that 
all the LLS are excluded in the analysis presented in this paper. This is 
because $N(\hi)$ measurement in these systems are uncertain and they may 
not trace IGM gas which is of prime interest in this study. 

The redshifts of all the intervening \os\ absorbers (both detections and 
upper limits) are summarized in Fig.~\ref{cluster}. 
For each line of sight (mentioned in the extreme right) the vertical tick 
mark on the left indicates the redshift above which the S/N per pixel of the 
corresponding spectrum is~$> 10$. This defines the minimum redshift 
(\zmin) for each line of sight as given in column \#3 of Table~\ref{allsystems} 
\& \ref{tab_lowsnr}. 
There are 7 systems listed in Table~\ref{tab_lowsnr} for which the 
\os\ absorption falls in the spectral region where S/N~$\le$~10.  
To get robust Voigt profile parameters we restrict ourselves to 
systems detected when the spectrum has S/N~$> 10$. 
The vertical tick mark on the right indicates the redshift at which the 
velocity difference from the QSO emission redshift (\zem) is 5000~\kms. 
This defines the maximum redshift (\zmax) for each line of sight as given in 
column \#4 of Table~\ref{allsystems} \& \ref{tab_lowsnr}. 
This cut is applied to remove the \os\ systems associated with the QSO or 
QSO neighborhood. 
Note that the two systems at 
\zabs\ = 2.0850 towards HE~1341$-$1020 and \zabs\ = 2.1660 towards 
PKS~1448$-$232 falling $\sim$~5000 ~\kms\ from the respective emission 
redshifts have been included in the sample. 

The maximum redshift path covered by the observations with 
S/N~$>$~10 is $\Delta z = 7.62$ or $\Delta X = 24.85$. However, 
this should be treated as upper limits as line blanketing by  
Ly$\alpha$ lines reduce the available redshift path length.  
%
\begin{table*}
\begin{center}
\scriptsize 
\caption{Details of additional \os\ systems detected in the spectral range where continuum S/N $\le 10$}
\begin{flushleft}
\begin{tabular}{lcccccccccccccc} 
\hline 
QSO & \zem\ & \zmin\ & \zmax\ & $z_{\rm sys}$ & log~$N(\hi)$ & log~$N(\os)$ & log~$N(\cf)$ & Class & Case & 
\dtype\ & $\delta v(\os)$ & $\delta v(\cf)$ & $|\Delta v(\os - \cf)|$ & Low Ions \\  
    &    &    &     &     &     &    &      &     &     &    & (\kms)   & (\kms) & (\kms) &  \\ 
\hline \hline
Q~0122--380   & 2.190 & 1.995 & 2.137 & 1.9746&15.43$\pm$ 0.16&14.72$\pm$ 0.07&15.30$\pm$0.20& $bd$ &B  & $-$1 &  90.4&281.5& .... & Yes \\ 
HE~2217--2818 & 2.414 & 1.963 & 2.357 & 1.9658&15.82$\pm$ 0.57&14.95$\pm$ 0.13&14.42$\pm$0.04& $bd$ &A/B&   0  & 328.3&171.1&42.66 & Yes \\ 
Q~0453--423   & 2.658 & 2.014 & 2.597 & 2.0041&15.10$\pm$ 0.09&14.26$\pm$ 0.21&13.16$\pm$0.04& $bd$ &B  &   +1 & 166.8&121.6& .... & No \\
HE~2347--4342 & 2.871 & 2.311 & 2.807 & 2.1198&14.26$\pm$ 0.04&14.42$\pm$ 0.43&13.42$\pm$0.05& $bb$ &A  &   +1 &  34.9& 33.5& .... & No \\ 
              &       &       &       & 2.2750&13.97$\pm$ 0.13&14.59$\pm$ 0.77&13.86$\pm$0.20& $bb$ &A  & $-$1 & 117.6&204.0& .... & No \\ 
PKS~2126--158 & 3.280 & 2.520 & 3.208 & 2.4596&14.25$\pm$ 0.05&14.30$\pm$ 0.10&13.46$\pm$0.02& $bd$ &B  &    0 &  48.4& 41.8& 1.55 & No \\ 
              &       &       &       & 2.4855&15.11$\pm$ 0.16&14.47$\pm$ 0.11&12.61$\pm$0.06& $bb$ &B  &    0 &  80.1& 62.2&16.18 & No \\ 
\hline 
\hline
\end{tabular} 
\end{flushleft}
\label{tab_lowsnr}
\end{center}
\end{table*} 


\subsection{Absorption line measurement techniques} 

\subsubsection{Voigt profile fitting}

We use standard Voigt profile fitting and apparent optical depth techniques  
to derive absorption line parameters. The Voigt profile fit provides 
best fitted values of the column density ($N$), velocity dispersion ($b$) and 
redshift ($z$) for each component. The absorption lines originating from 
individual species (\hi, \cf\ and \os) are fitted using all the detected 
transitions with minimum number of components required to get 
the reduced $\chi^2$ close to 1. 
Whenever possible, we have tied the \os\ and \cf\ components in redshift. 
However, most of the systems are best fitted by components with different 
sets of parameters for \os\ and \cf.  
We use all the available Lyman series lines to extract $N$(H~{\sc i}). 
In this case also whenever possible component structure from metal lines were 
used to constrain the redshifts of individual H~{\sc i} components.

We also have independent Voigt profile decompositions for 51 \os\ systems 
along 12 lines of sight (indicated by underlined QSO names in column \#1 of 
Table~\ref{allsystems}) performed by \citet{Bergeron05} using  
VPFIT \footnote{See http://www.ast.cam.ac.uk/$\sim$rfc/vpfit.html} 
\citep{Webb87,Rauch92}. 
In this case fits to H~{\sc i}, \cf\  and \os\ absorption were performed
independently without constraining the redshifts of any components.
As the \os\ absorption lines fall in the \lya\ forest, some amount of 
subjectivity (wavelength range used, placement of components and number 
of components etc.,) is involved in the Voigt profile decomposition. 
However, the availability of decompositions derived using two independent 
procedures allow us to investigate the statistical influence of the 
fitting procedures.

In total we find 239 individual Voigt profile components for \os\ 
(with  $12.75\le {\rm log}~N(\os) [{\rm cm^{-2}}] \le 14.49$) and 318 components 
for \cf\ (with $11.58\le {\rm log}~N(\os) [{\rm cm^{-2}}] \le 14.76$). 
Total column densities of \hi, \os\ and \cf\ are listed in 
Tables~\ref{allsystems} and \ref{tab_lowsnr} in columns \#6, \#7 and \#8, 
respectively.
These are obtained by summing the column densities in individual Voigt profile 
components in a given system.

\begin{figure} 
\centerline{
\vbox{
\centerline{\hbox{ 
\includegraphics[width= 8.4cm,angle= 0]{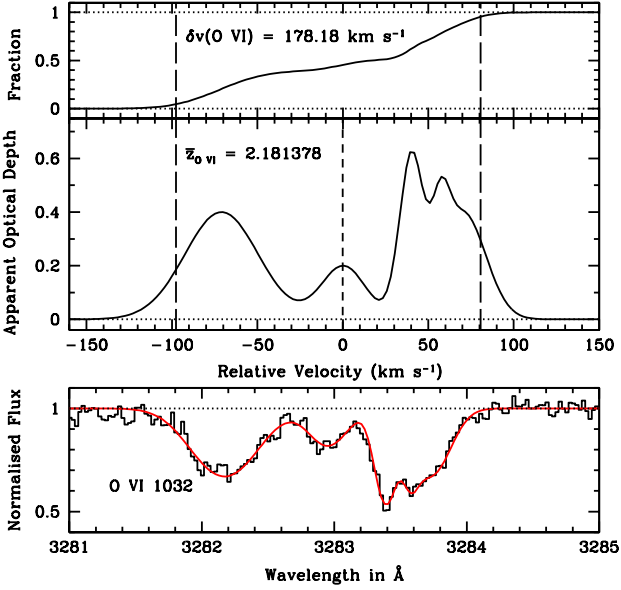}  
}}
}}
\caption{$Bottom :$ The \os\ $\lambda 1032$ absorption profile of the system at 
\zabs~=~2.1808 towards HE~2217$-$2818. The best fitted Voigt profile is 
over-plotted. $Middle :$ The apparent optical depth profile [$\tau_a(v)$] 
estimated from the best fitted Voigt profile is shown. The zero velocity 
corresponds to the optical depth weighted redshift $\bar z_{\os}=2.181378$. 
$Top :$ Integration of the apparent optical depth.   
The line profile velocity width, $\delta v$, has been calculated as 
[$v(95\%)$ $-$ $v(5\%)$] where $v(95\%)$ and $v(5\%)$ are the velocities 
corresponding to 95 and 5\% percentiles (vertical long dashed lines) 
of the apparent optical depth distribution.}
\label{delv_demo} 
\end{figure} 

\subsubsection{Apparent optical depth (AOD)}
\label{vsa}

For each of these systems we have calculated the velocity width for \os\ and \cf\ 
(when detected) absorption using AOD technique. 
Following \citet{Ledoux06a}, the absorption velocity 
width, $\delta v$, has been calculated as [$v(95\%)$ $-$ $v(5\%)$], 
where $v(95\%)$ and $v(9\%)$ are the velocities corresponding to the 95 and 5 
percent percentiles of the apparent optical depth distribution. Note this
definition is slightly different from that used by \citet{Songaila06},
who defined the velocity spread as the velocity range over which the optical 
depth is larger than some fraction of the peak optical depth. As pointed out 
by \citet{Songaila06}, the velocity spread measurements are very sensitive to 
the velocity range over which the absorption is studied.
\citet{Songaila06} used regions within 350~\kms~ either side of
the peak optical depth. This in turn restricts the maximum measurable 
width to 700~\kms. In order to be consistent with our definition of
a system, we consider here the velocity range covered by the Voigt profile
components with no gap larger than 100~\kms\ (i.e. $\rm v_{\rm link} \sim$~100 \kms). 

The $\delta {v} (\os)$ is sometimes difficult to measure because of blending in 
the \lya\ forest. Therefore, we introduce an index $\delta_{\rm type}$ 
(see column \#11 of Tables~\ref{allsystems} \& \ref{tab_lowsnr}) 
which takes the value 0, $+1$ and $-1$.
The systems with $\delta_{\rm type}=0$ are the ones for which
\os\ absorption is well defined based on both
the \os\ lines. Systems with $\delta_{\rm type}=+1$
and $-1$ are those for which the measured width should be considered as
an upper and lower limit,  respectively. 
There are 54 systems with $\delta_{\rm type}=0$.  

We also calculate the first moment of the apparent optical depth distribution 
which gives us the optical depth weighted redshift 
for both \cf\ and \os\ (i.e. $\bar{z}_{\cf}$ and $\bar{z}_{\os}$). 
This allows us  to calculate the velocity shift 
(called $|\Delta v(\os - \cf)|$)
between the \os\ and \cf\ absorption originating from the same system. 

Fig.~\ref{delv_demo} illustrates the way we measure $\delta v$.
Bottom panel shows the observed \os\ profile and the best fitted
Voigt profile.  In the middle panel we plot the AOD profile as a
function of velocity. The top panel shows the
integrated optical depth starting from a minimum velocity. The
left and right vertical dashed lines mark the velocities 
at which the integrated optical depth is 5\% and 95\% of the 
total optical depth respectively. The velocity difference between
these two lines gives $\delta v$. For clarity, the
zero velocity is fixed at  $\bar{z}_{\os}$ so that integrated
optical depth is 50\% at $v =0$ \kms.

If both \os\ and \cf\ absorption originate from the same gas with 
constant density and homogeneous ionization conditions  then 
the velocity offset measured between  $\bar{z}_{\os}$ and 
$\bar{z}_{\cf}$ should be zero (i.e. $|\Delta v(\os - \cf)|$~= 0).
Any mismatch of the optical depth weighted redshifts between species 
could be interpreted as (a) relative line of sight velocity between 
the two species (as expected in the
case of \os\ originating from different interfaces like evaporating 
region, cooling front or shocked gas) or (b) ionization inhomogeneity 
along the line of sight. The measured $\delta v(\os)$, $\delta v(\cf)$ 
and $|\Delta v(\os - \cf)|$ in our sample are summarized in column \#12,
\#13 and \#14 of Tables~\ref{allsystems} \& \ref{tab_lowsnr}, respectively. 
%
\begin{figure*} 
\centerline{
\vbox{
\centerline{\hbox{ 
\includegraphics[height= 6.5cm,width= 6.0cm,angle= 0]{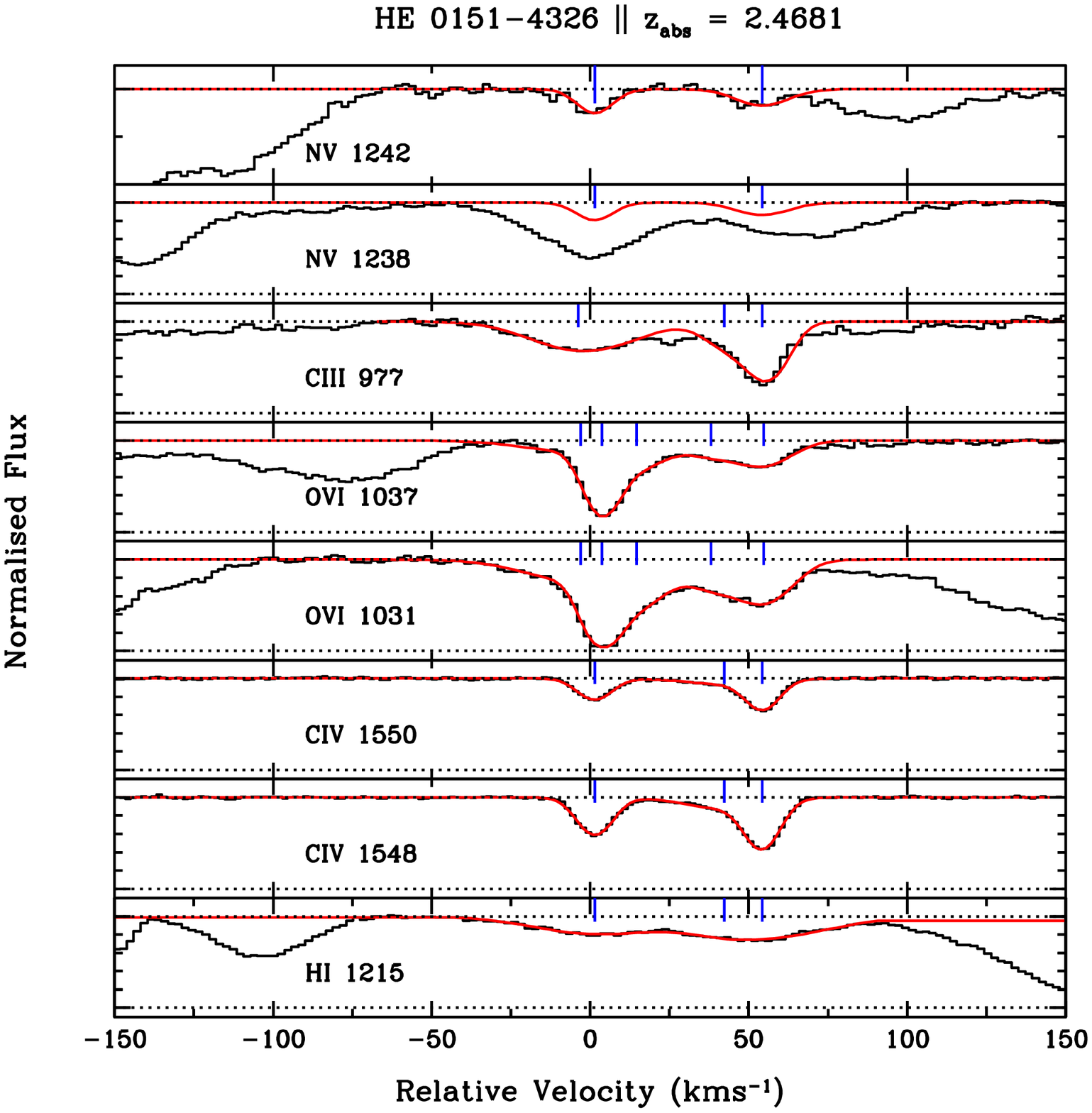} 
\includegraphics[height= 6.5cm,width= 6.0cm,angle= 0]{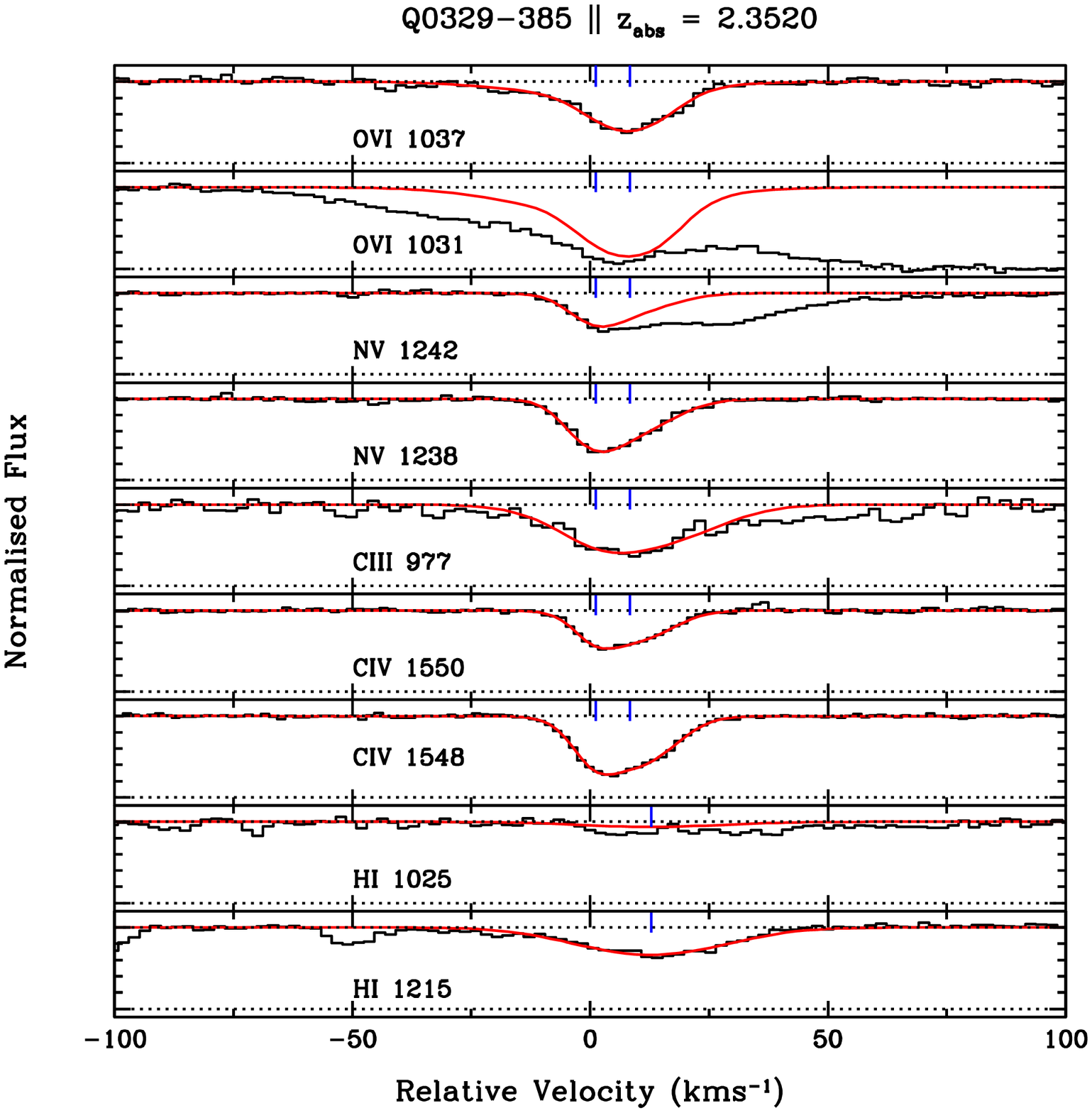} 
\includegraphics[height= 6.5cm,width= 6.0cm,angle= 0]{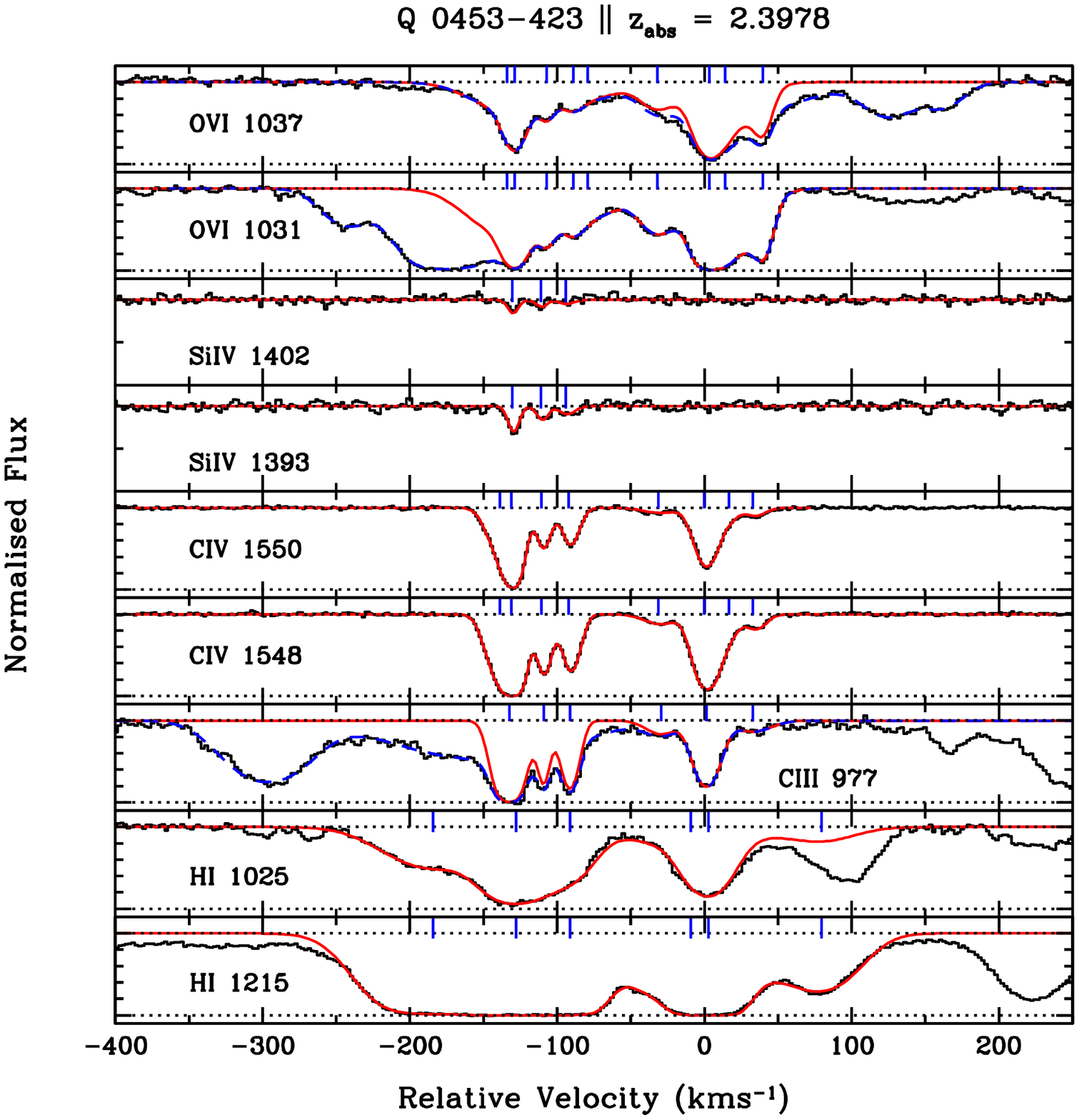} 
}}
}}
\caption{Examples of \os\ systems where both  lines in the doublet are 
unblended (left; $dd$), one of the line in the  doublet is blended 
(middle; $bd$) and both lines in the doublet are partially blended 
(right; $bb$). The smooth curves are the best fitted Voigt profiles. 
The horizontal tick marks indicate the centroids of the individual 
components. The absorption redshift that defines the zero velocity 
and the name of the background QSO are indicated at the top of each 
panel. 
} 
\label{systems} 
\end{figure*} 

\subsection{Sub-samples of \os\ systems} 
%

As \os\ originates from a whole range of physical conditions and 
because of blending, the characteristics of the \os\ absorption are
difficult to extract. Hence, we build different subsamples depending on the 
issue we want to address.

The H~{\sc i} column density and $b$-parameter are very important for 
understanding the physical state of \os\ gas \citep[see e.g.,][]{Tripp08}. 
In particular, to investigate whether the gas is photoionized, we need to 
know $N$(H~{\sc i}) in individual \os\  components. So we define ``Case-A'' 
systems as the ones where at least one of the Lyman series line is unsaturated. 
In principle these systems also enable us to derive $b$-values for most of the
\hi\ components. There are 35 systems classified as ``Case-A'' 
systems in our sample. The systems shown in the left and middle panels of 
Fig.~\ref{systems} are examples of ``Case-A'' systems. There are 10 systems 
where at least one \hi\ absorption is unsaturated for some of the components.
We call these systems at ``Case-A/B''. The right most panel in Fig.~\ref{systems} 
is an example of such a case. 
In the remaining systems most of the \hi\ absorption is saturated and Voigt 
profile measurements are uncertain. We call these systems ``Case-B''. 
We notice that several of the ``Case-B'' systems show absorption lines from 
low ionization species such as \cto, \ct, \sit/\sif\  etc. 
The information regarding the presence of low ions is also given in 
column \#15 of Tables~\ref{allsystems}, \ref{upp_lim} and \ref{tab_lowsnr}. 
In total there are 46 systems where low ions are detected. Note that the  
\ct\ information is not available in our sample for $z \lesssim 2.2$.    

We also classify the systems based on the presence of unblended \os\ 
doublets. Out of 84 systems only 29 systems have both \os\ lines unblended 
for which we are able to perform robust Voigt profile fits. 
We call this sample as ``robust sample'' as far as Voigt profile parameters 
are concerned (these are marked as ``$dd$'' in column \#9 of 
Tables~\ref{allsystems} and \ref{tab_lowsnr}). 
An example of such a system is shown in the left panel of Fig.~\ref{systems}. 
In 30 other systems one of the \os\ lines is blended (as in the middle panel 
of Fig.~\ref{systems}). Here the \os\ fit relies mainly on the unblended line. 
We denote these systems as ``$bd$'' in column \#9 of 
Tables~\ref{allsystems} and \ref{tab_lowsnr}.
In the remaining 25 systems both lines are partially blended (as in the 
right panel of Fig.~\ref{systems}). In this case (referred to as ``$bb$'' 
in column \#9 of Tables~\ref{allsystems} and \ref{tab_lowsnr}) we use part of 
the profiles of both \os\ lines to fit the Voigt profiles. In most of these cases 
the additional absorption is treated as due to intervening \lya\ absorption. 
Therefore, the Voigt profile parameters (number of components and $b$-values) 
are uncertain. 
%


\section{Analysis based on Voigt profile fit}
\label{dist}

In this section we discuss the distributions of various parameters
derived from our Voigt profile fits. 

\subsection{$b$-parameter distribution}

The $b$-parameter derived from Voigt profile fit 
provides only an upper limit to the kinetic temperature of the gas. 
Thus, we do not attempt to constrain the physical state of the \os\ 
gas using individual $b$-values, instead we draw some broad 
conclusions using the $b$-parameter distributions.  

First, we compare our \os\ $b$-parameter distribution for the 12 lines of 
sight with that of the \citet{Bergeron05}. The Kolmogorov-Smirnov (KS) test 
suggests that the two distributions are drawn from the same parent 
population with a 32\% probability for the observed deviation in the 
cumulative distributions to occur by chance.

\begin{figure} 
\centerline{
\vbox{
\centerline{\hbox{ 
\includegraphics[height= 8.6cm,width= 8.4cm,angle= 0]{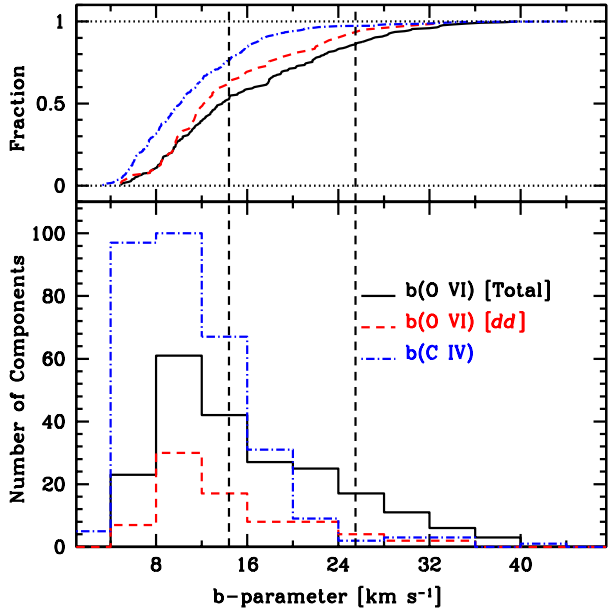} 
}}
}}
\caption{$Bottom$~: Comparison of the $b$-parameter distributions of \cf\ 
         (dot-dashed histogram) and \os\ (solid histogram). The dashed 
	 histogram is for the \os\ components originating from $dd$ systems. 
	 The vertical dashed lines correspond to $b(\os)$ = 14.4 and  
	 25.5 \kms\ respectively.     
	 $Top$~: The cumulative $b$-parameter distributions of \cf, 
	 \os\ (both total \& $dd$) components are shown following the 
	 same line style as in the bottom panel. 
	 } 
\label{b_dist1} 
\end{figure}

In the bottom panel of Fig.~\ref{b_dist1} we plot the  
$b$-parameter distributions measured in individual Voigt profile 
components for \os\ (solid histogram)  and \cf\ (dot-dashed histogram) 
in our full sample. The (red) dashed histogram gives the $b(\os)$ 
distribution in the robust components (i.e. from ``{\it dd}" subsample).  
The \os\ components arising from systems with S/N~$\le 10$ 
(listed in Table~\ref{tab_lowsnr}) are not included here.  
Median values of $b(\os)$ = 13.8 \kms\ and $b(\cf)$ = 10.1 \kms\ 
correspond to $T$~$\sim1.8\times10^{5}$ K and $7.4\times 10^4$ K 
for \os\ and \cf\ respectively in the case of pure thermal broadening.  
The temperature corresponding to median $b(\os)$ agrees well with 
$T$~$\sim 2.1\times 10^5$ K found by \citet{Simcoe02}. 
The median $b$-value of \os\ components becomes 12.5 \kms\ 
if we restrict ourselves to the robust sample.  
In the case of pure photoionization heating the expected temperature 
(i.e. $T \sim 2\times10^{4}$ K) implies  $b(\os) \sim 5$ \kms. 
The lowest $b(\os)$ measured (4.9$\pm$3.2 \kms) in our sample is 
consistent with this. 
%

%

In the case of collisional ionization, the fraction of oxygen in \os\ peaks 
around $T \sim (2-3)\times 10^5$ K, which corresponds to $b(\os) \sim 14.4$~\kms.  
This is very close to the median $b(\os)$-value of  
our full sample. This is shown by the left vertical dashed line in 
Fig.~\ref{b_dist1}. There are 52\% (and 62\% for the $dd$ subsample) 
components having $b$-parameters less than 14.4 \kms\ suggesting 
photoionization (and/or non-equilibrium collisional ionization at high 
metallicities) is the dominant process in these systems. 
On the other hand, the \os\ fraction will be less than 0.01 when 
$T$~$\ge 6\times 10^5$ K \citep[see][]{Gnat07}, or $b(\os) \gtrsim 25.5$ 
\kms\ (second vertical line in Fig.~\ref{b_dist1}). 
Therefore under the purely thermally broadened case 
one does not expect the \os\ $b$-parameter\ to be higher than this 
unless the metallicity and/or $N(\sc H)$ is very high. 
These systems will also have broad and shallow associated \lya.  
We find 14\% (and 8\% for the $dd$ subsample) of the components 
having $b$-values higher than this. 
Most of these high $b$ components in our sample are part of blends where 
the \os\ profile is decomposed into multiple Voigt profile components. 
Isolated \os\ components with $b(\os) >25$~\kms\ are very rare.  
Note that very few such isolated broad \os\ components together with broad 
albeit shallow \lya\ absorption are detected at low-$z$ 
\citep[see e.g.,][]{Savage10,Savage11}. These are interpreted as 
collisionally ionized gas with $T\sim 10^{6}$~K. The non-detection of such 
systems in our sample may be attributed to the bias introduced by 
line blanketing and blending due to the \lya\ forest absorption.   

The temperature for which the ionization fraction of \cf\ peaks under 
collisional ionization equilibrium is $\sim1.1\times10^5$ K, or  
 $b(\cf)$ = 12.40 \kms. Sixty seven percent of the \cf\ components have 
$b$-parameter less than this suggesting that a considerable fraction of \cf\ 
originates either from photoionized gas or from non-equilibrium 
cooling gas with high metallicity (i.e. $Z \sim 1.0~Z_{\odot}$). 
It is apparent from Fig.~\ref{b_dist1} that the $b$-parameter distribution
is wider in case of \os\ compared to \cf. This is evident in the
cumulative distributions plotted in the top panel.
The KS-test indicates that the two distributions are drawn 
from significantly different populations with the probability of this 
difference occurring by chance being less than 0.1\%. 
Ideally one would expect $b(\cf)\ge b(\os)$ if \cf\ and \os\ 
trace the same gaseous phase.  
Our finding is consistent with \os\ and \cf\ originating from different 
phases of the gas associated with the absorption system 
\citep[see also,][]{Simcoe02,Bergeron05}. 
%
\begin{figure} 
\centerline{
\vbox{
\centerline{\hbox{ 
\includegraphics[height= 8.6cm,width= 8.4cm,angle= 0]{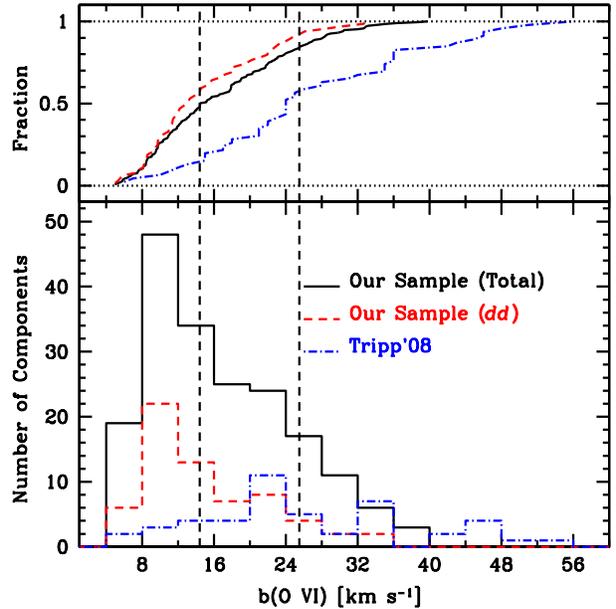} 
}}
}}
\caption{Comparison of the $b$-parameter distributions of \os\ at 
high and low redshifts. The low-$z$ sample is the STIS data sample of 
\citet{Tripp08}. Only those \os\ components, originating from systems 
with log~$N(\os) {\rm (cm^{-2})} > 13.6$ (corresponding to a rest frame 
EW of $\sim$ 50 m\AA) are considered here. The vertical dashed lines are 
as in Fig.~\ref{b_dist1}. The overall $b$-distribution of our sample is 
significantly different from that of the low-$z$ sample as can be seen 
from the cumulative distributions plotted in the top panel  
with the same line style as in the bottom panel.} 
\label{b_dist3} 
\end{figure} 

Next we compare the $b(\os)$ distribution in our sample with 
the low-$z$ sample of \citet{Tripp08} obtained using the Space Telescope 
Imaging Spectrograph (STIS) on board  {\it HST}. 
The spectral resolution of STIS data is $\sim$~7~\kms, very much 
similar to the one used here but the STIS spectra are slightly under-sampled.
Even though the S/N of STIS spectra are typically a 
factor of two lower than those of our UVES data, the contamination by 
intervening \hi\ absorption is less severe in the low-$z$ sample.
In the bottom panel of Fig.~\ref{b_dist3} we show the two \os\ 
$b$-distributions. We restrict the comparison to systems with 
log~$N$(\os) (cm$^{-2}$)~$>$~13.6 which is the completeness limit 
of the STIS sample.
The median $b$-values for the low and high-$z$ \os\ components 
are 24.0 and 14.6 \kms\ (13.2 \kms\ for the $dd$ subsample), respectively. 
It is interesting to note that almost 87\% of the low redshift $b$-parameters 
are consistent with the temperature expected from collisional ionization, 
i.e., $b(\os)>$~14.4 \kms\ 
(first dashed vertical line), only 53\% (44\% for the $dd$ subsample) of 
the high redshift components satisfy this. 
%

\begin{table*}
\caption {Results of decomposition of thermal and non-thermal contributions to the line broadening}
\begin{tabular}{cccccccccllc} 
\hline
QSO &  $z_{\rm sys}$ & $v_{\rm rel}$ & $b(\hi)$ & $b(\os)$ & $b(\cf)$ &\multicolumn{3}{c}{log~$N$ (cm$^{-2}$)}  & $b_{\rm nt}$ & log~$T$ & log~$T_{\rm max}$\footnotemark[2] \\ \cline{7-9} 
& &(\kms) & (\kms) & (\kms) & (\kms) & \hi & \os & \cf & (\kms) & ($T$ in K) & ($T_{\rm max}$ in K) \\  
\hline 
\hline
PKS1448-232   &	 2.1099&  $-$4.7 & 14.2 &  8.1 & 7.3 & 13.05 & 14.28 & 13.11 &  7.5 (6.3) & 3.94 (3.99)                & $\le$4.09 \\   
PKS~0237--23  &  1.9878\footnotemark[1]&    +0.4 & 20.6 &  9.5 & ... & 13.64 & 13.41 & ...   &  8.2       & 4.33       & $\le$4.41 \\ 
PKS~0237--23  &  2.0108\footnotemark[1]&  $-$3.8 & 31.3 &  9.4 & ... & 14.31 & 13.15 & ...   &  5.4       & 4.76       & $\le$4.77 \\ 
Q~0329--385   &  2.0764&    +4.9 & 19.7 &  7.9 & 6.2 & 13.64 & 13.26 & 13.21 &  6.4 (2.6) & 4.32 (4.36)                & $\le$4.37 \\ 
Q~0329--385   &  2.2489\footnotemark[1]& $-$50.3 & 18.6 &  8.2 & ... & 12.76 & 13.68 & ...   &  7.0       & 4.26       & $\le$4.32 \\ 
Q~0329--385   &  2.2489\footnotemark[1]& $-$19.4 & 10.1 & 10.3 & ... & 11.95 & 13.47 & ...   &  ...       & ....       & $\le$3.79 \\ 
Q~0329--385   &  2.2489\footnotemark[1]& $-$ 3.1 & 33.1 & 21.9 & ... & 13.11 & 13.76 & ...   & 20.9       & 4.60       & $\le$4.82 \\ 
Q~0329--385   &  2.2489\footnotemark[1]&   +43.4 & 24.7 & 12.1 & ... & 13.10 & 13.64 & ...   & 10.7       & 4.48       & $\le$4.57 \\ 
Q~0329--385   &	 2.3139&    +4.0 & 38.6 & 11.5 & 4.3 & 14.02 & 13.16 & 11.79 &  6.5 (...) & 4.94 (...)                 & $\le$4.96 \\ 
Q~0329--385   &	 2.3139&   +34.8 & 16.9 &  8.8 & 5.8 & 13.23 & 12.75 & 12.21 &  8.0 (3.3) & 4.13 (4.22)                & $\le$4.24 \\ 
Q~0329--385   &  2.3639& $-$46.3 & 27.8 & 15.3 & 9.2 & 14.40 & 13.37 & 11.86 & 14.1 (4.7) & 4.54 (4.66)                & $\le$4.67 \\ 
Q~0329--385   &	 2.3639&  $-$4.1 & 19.0 &  8.7 & 9.1 & 13.91 & 13.52 & 12.25 &  7.5 (7.6) & 4.27 (4.26)                & $\le$4.34 \\ 
HE~1347--2457 &  2.3327\footnotemark[1]&    +3.2 & 23.5 & 18.5 & ... & 13.78 & 13.26 & ...   & 18.1       & 4.13       & $\le$4.52 \\ 
HE~1347--2457 &  2.3422\footnotemark[1]&   +10.8 & 18.3 &  9.9 & ... & 14.40 & 12.95 & ...   &  9.1       & 4.18       & $\le$4.31 \\ 
Q~0453--423   &  2.5371\footnotemark[1]& $-$26.3 & 38.7 & 10.3 & ... & 13.48 & 13.05 & ...   &  3.6       & 4.95       & $\le$4.96 \\ 
Q~0453--423   &  2.5371\footnotemark[1]&  $-$1.6 & 28.5 & 11.7 & ... & 14.62 & 13.17 & ...   &  9.6       & 4.64       & $\le$4.69 \\ 
PKS~0329--255 &  2.5687&  $-$3.9 & 30.0 & 10.1 & 6.1 & 14.62 & 13.12 & 12.15 &  7.0 (...) & 4.71 (...)                 & $\le$4.74 \\ 
PKS~0329--255 &  2.5687&   +14.2 & 29.9 & 13.8 &20.8 & 14.25 & 13.35 & 12.27 & 12.0 (19.8)& 4.66 (4.48)                & $\le$4.73 \\ 
HE~0151--4326 &  2.5053&   +42.6 & 28.0 & 21.7 &10.5 & 14.71 & 13.52 & 12.10 & 21.2 (7.0) & 4.31 (4.65)                & $\le$4.68 \\ 
HE~2217-2818  &  2.0748&   +64.4 & 29.0 & 12.6 & ... & 14.12 & 14.32 & ...   & 10.6       & 4.64                       & $\le$4.71 \\ 
\hline 
\hline
\end{tabular} 
~~~~~~~~ Table Notes -- 
\footnotemark[1]Components from ``\os\ only" systems. ~~~~~~~ 
\footnotemark[2]Calculated from $b(\hi)$ assuming pure thermal broadening. ~~~~~~~~ \\  
Values in the parenthesis are calculated using $b(\hi)$ -- $b(\cf)$ pairs.  
\label{tab_turb} 
\end{table*}


Almost 43\% of the low redshift components are consistent with 
$b>$ 25.5 \kms\ (second dashed vertical line). In our high redshift 
sample, only 16\% (9\% for the $dd$ subsample) components show 
$b$-value greater than 25.5 \kms. 
The overall $b$-distribution of our sample is significantly different from 
that of the low-$z$ sample as can be seen from the cumulative distributions 
plotted in the top panel of the Fig.~\ref{b_dist3}. 
A two sided KS-test gives a maximum departure between the two 
distributions, D = 0.41 with a probability, P = $3.9\times10^{-6}$ that 
the two samples are drawn from the same parent population. 
The difference is even more when we compare the 
low-$z$ sample to our $dd$ subsample as can be seen from the 
cumulative distributions. 

In summary, we find that at high redshift, the $b$-parameter distributions 
of \cf\ and \os\ are significantly different with \os\ absorption being 
wider than \cf, suggesting that the two species trace different phases 
of the absorbing gas. 
The $b(\os)$ distribution at high-$z$ is very different from that at  
low-$z$ as measured by \citet{Tripp08}. 
Recently \citet{Fox11} has drawn a similar conclusion using the high-$z$  
Voigt profile fitting results of \citet{Bergeron05}. 
%

\begin{figure*} 
\centerline{
\vbox{
\centerline{\hbox{ 
\includegraphics[height= 5.4cm,width= 4.1cm,bb=40 163 590 718,clip=]{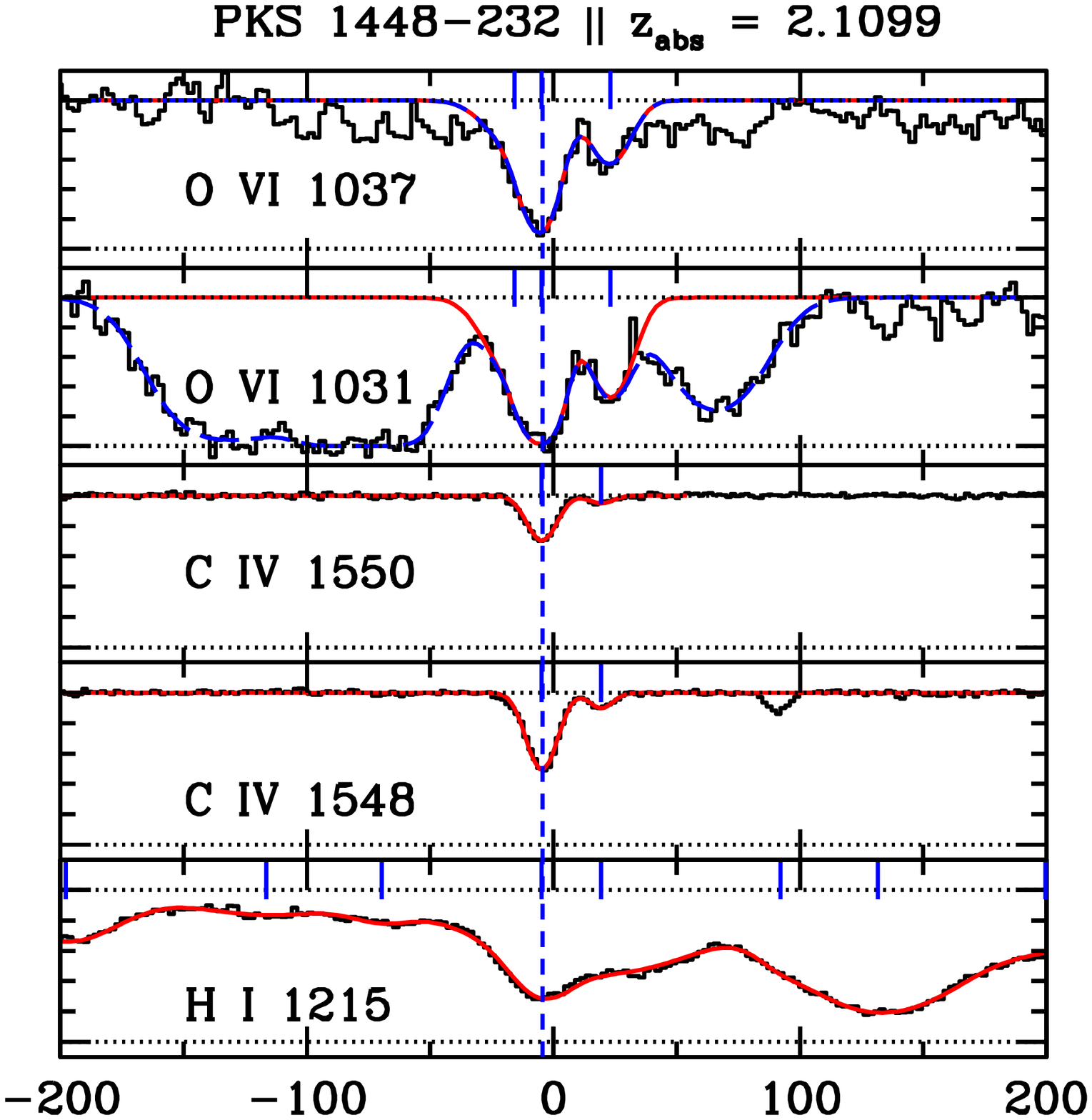} 
\includegraphics[height= 5.4cm,width= 4.1cm,bb=40 163 590 718,clip=]{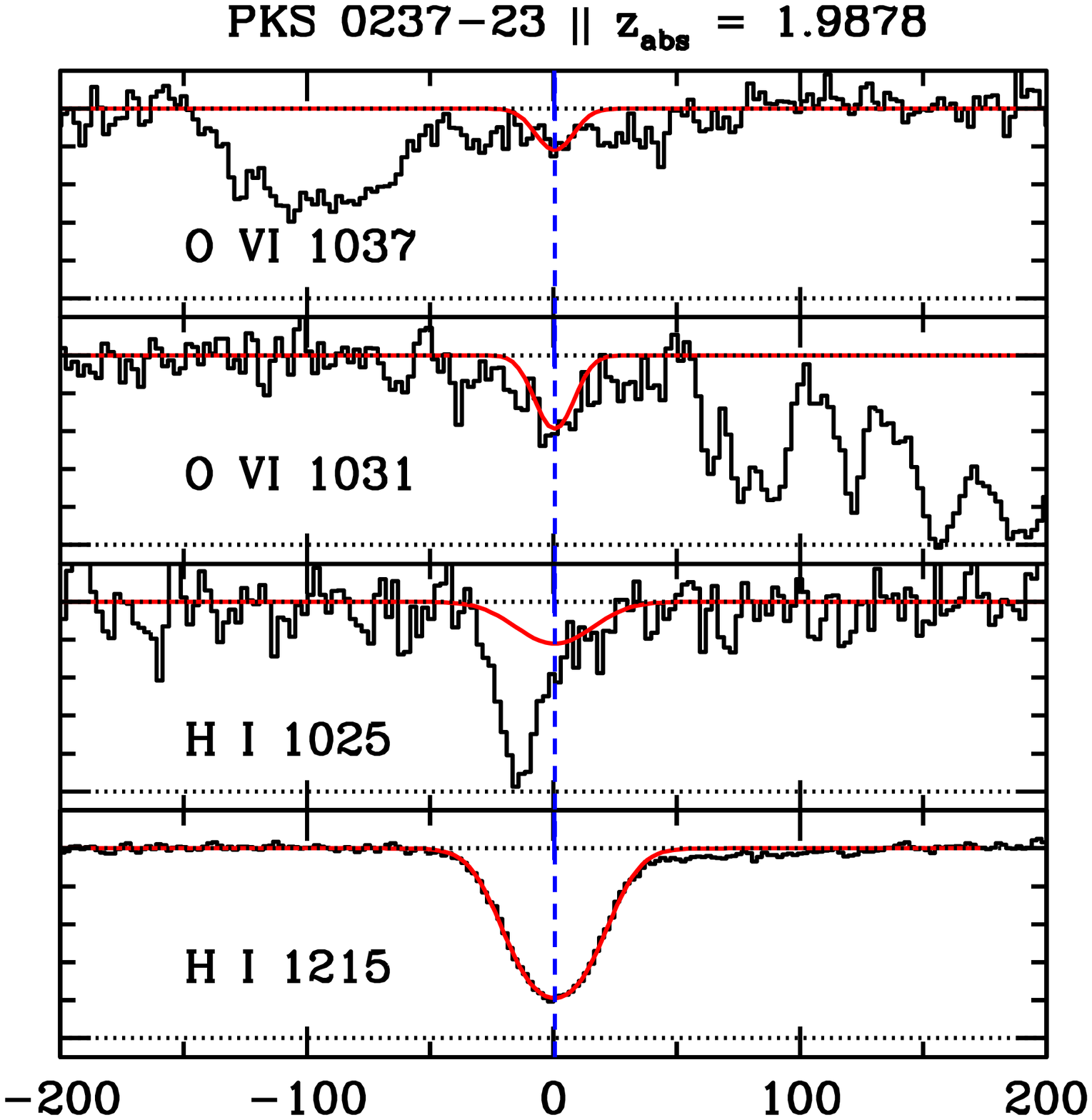} 
\includegraphics[height= 5.4cm,width= 4.1cm,bb=40 163 590 718,clip=]{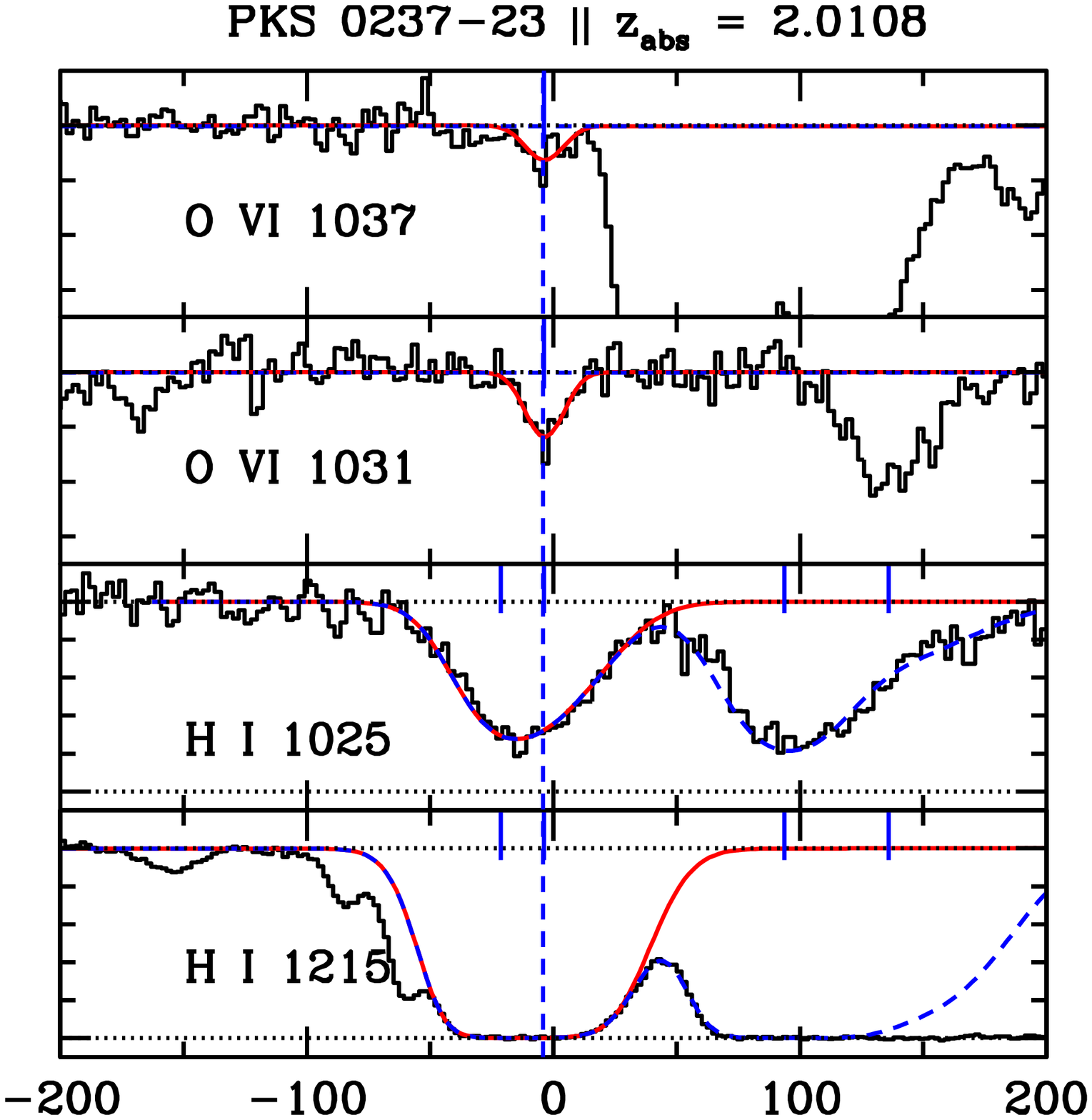} 
}}
\centerline{\hbox{ 
\includegraphics[height= 5.4cm,width= 4.1cm,bb=40 163 590 718,clip=]{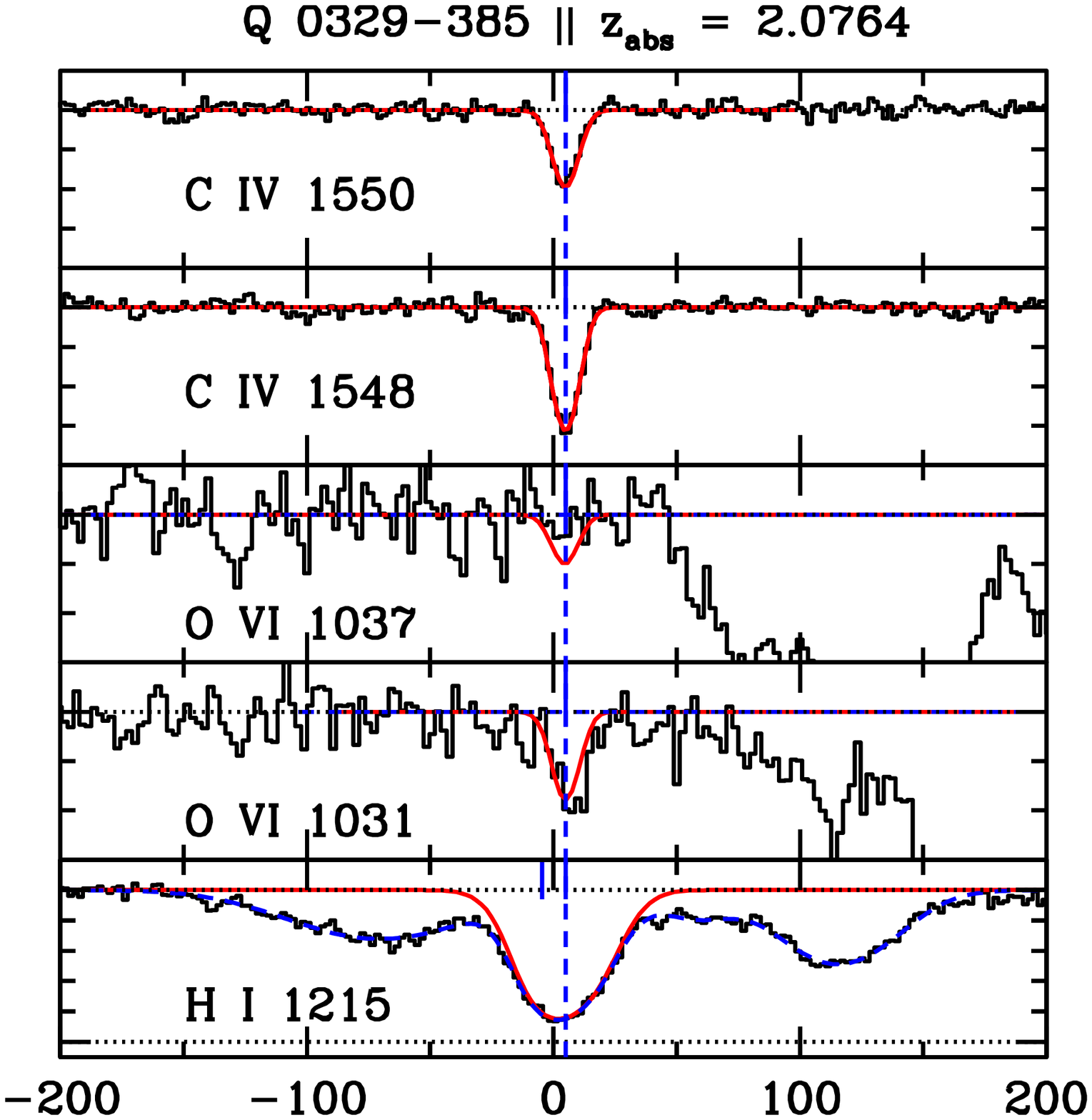} 
\includegraphics[height= 5.4cm,width= 4.1cm,bb=40 163 590 718,clip=]{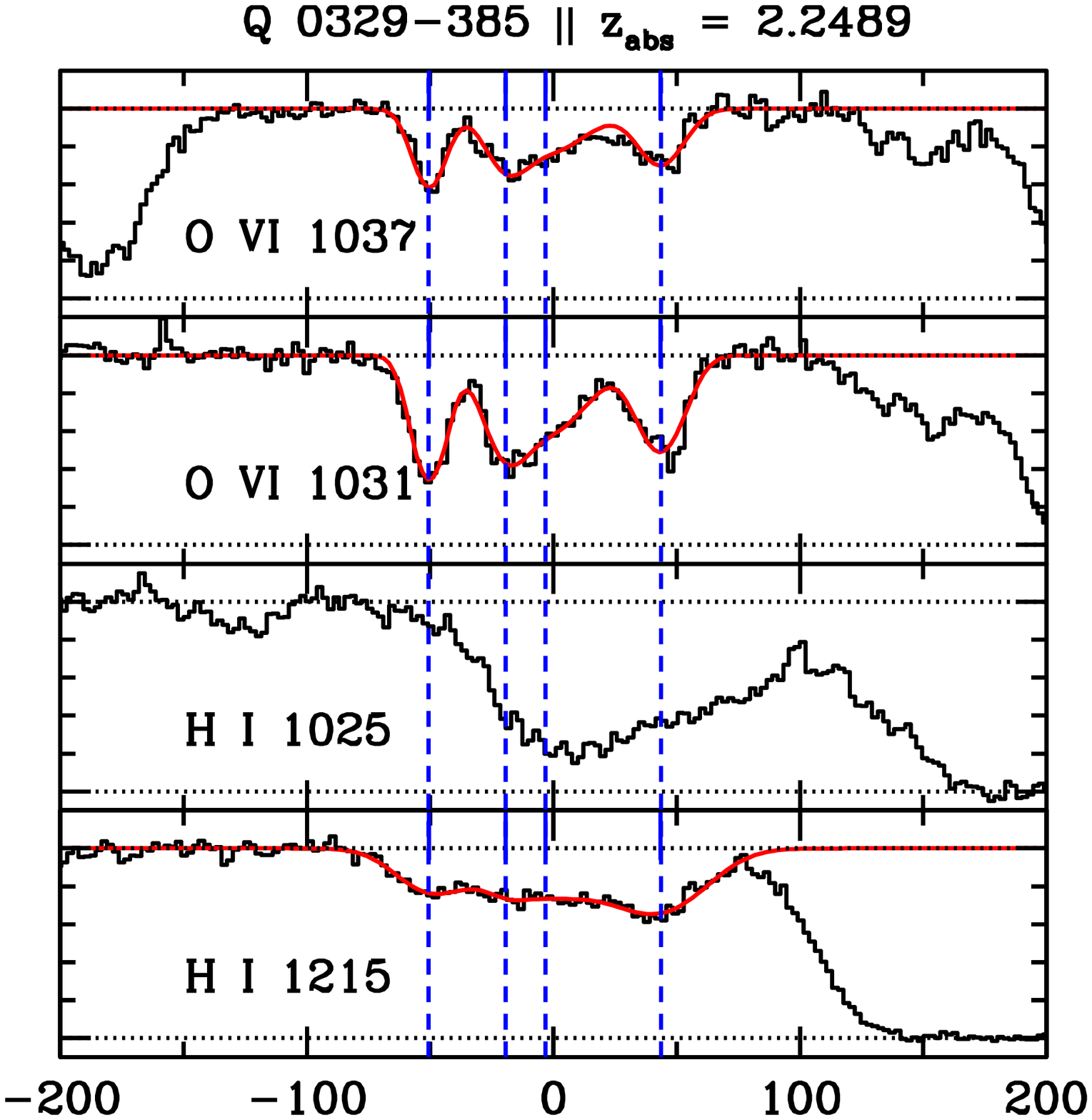} 
\includegraphics[height= 5.4cm,width= 4.1cm,bb=40 163 590 718,clip=]{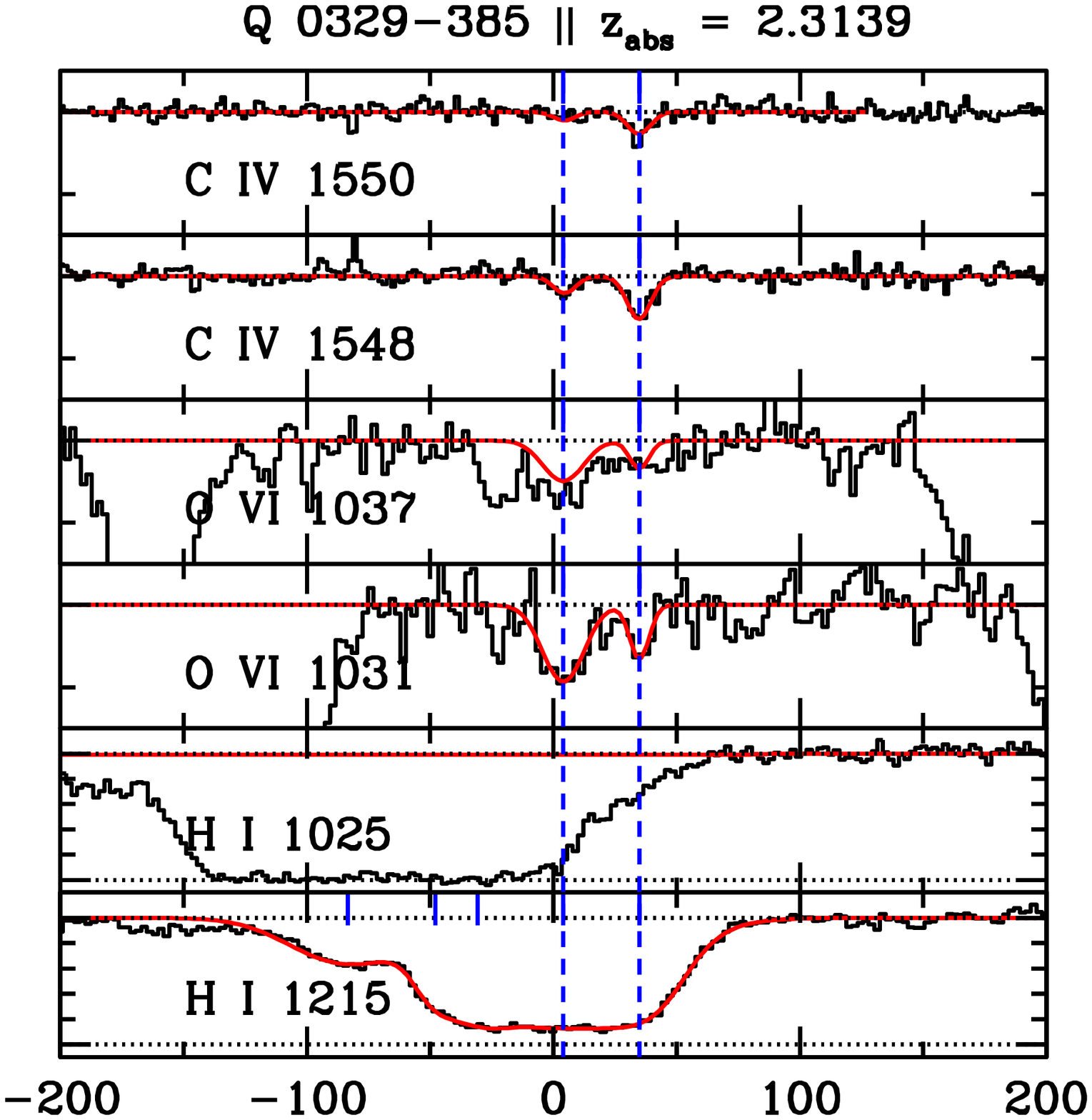} 
}}
\centerline{\hbox{ 
\includegraphics[height= 5.4cm,width= 4.1cm,bb=40 163 590 718,clip=]{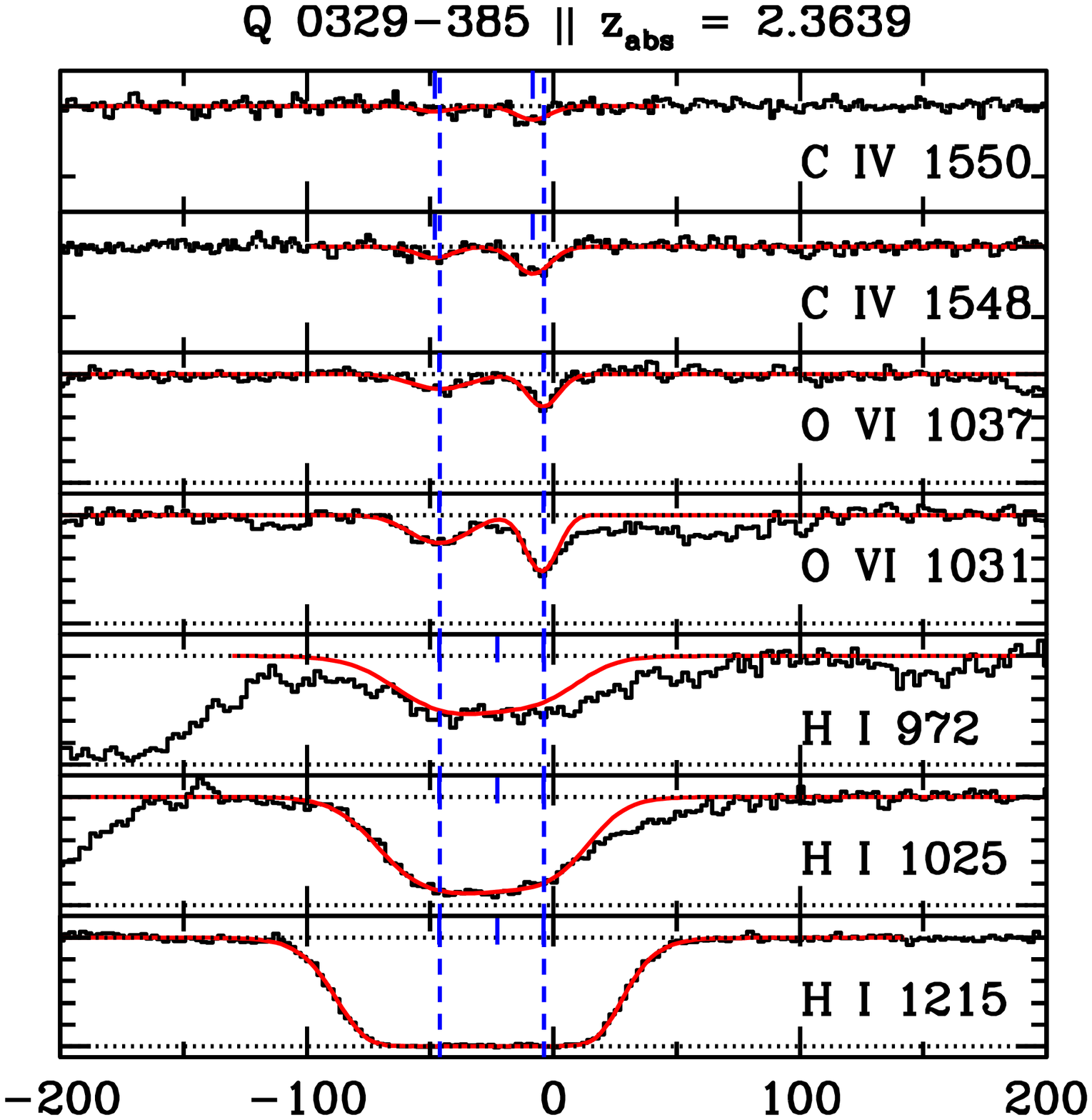} 
\includegraphics[height= 5.4cm,width= 4.1cm,bb=40 163 590 718,clip=]{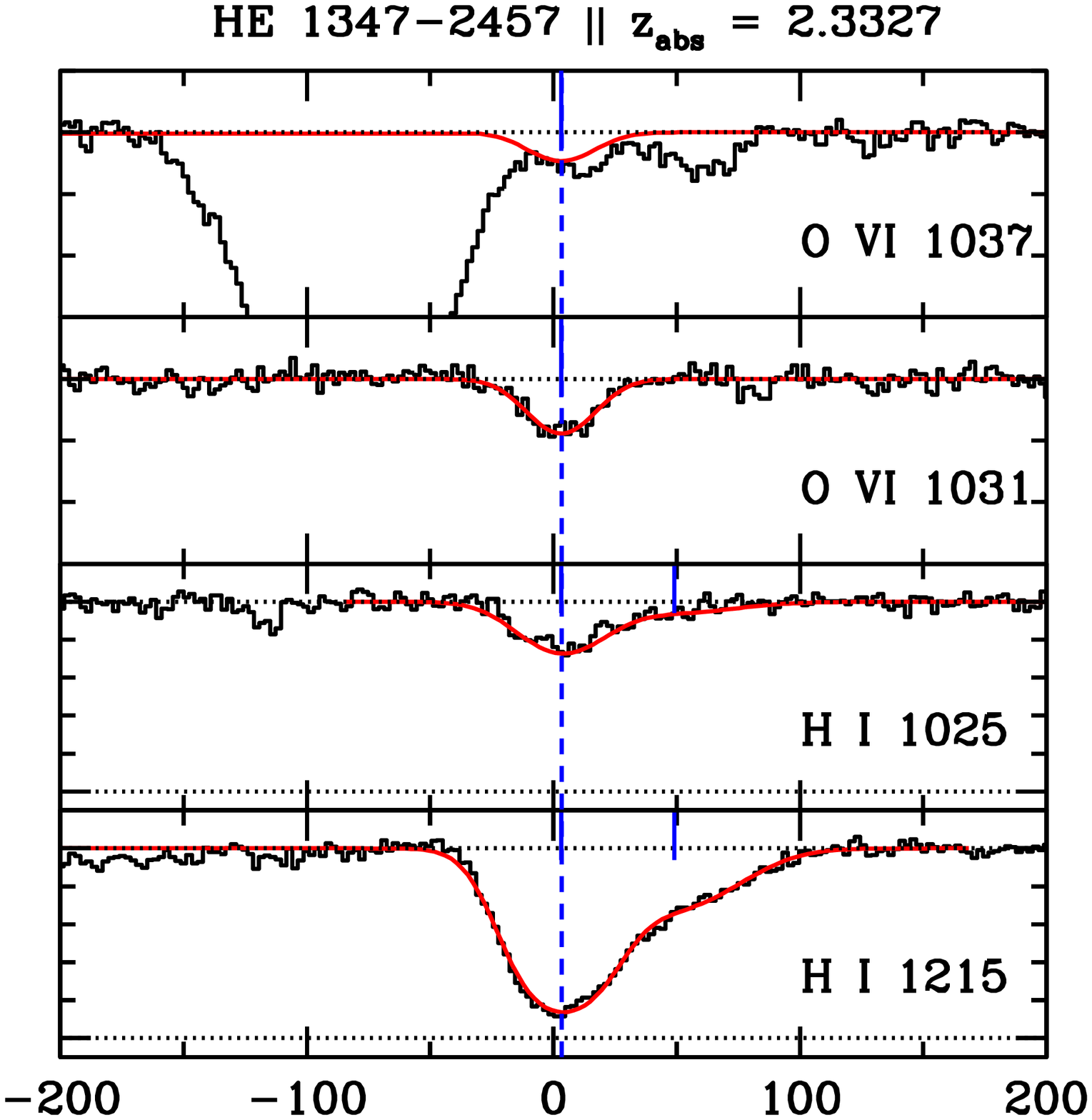} 
\includegraphics[height= 5.4cm,width= 4.1cm,bb=40 163 590 718,clip=]{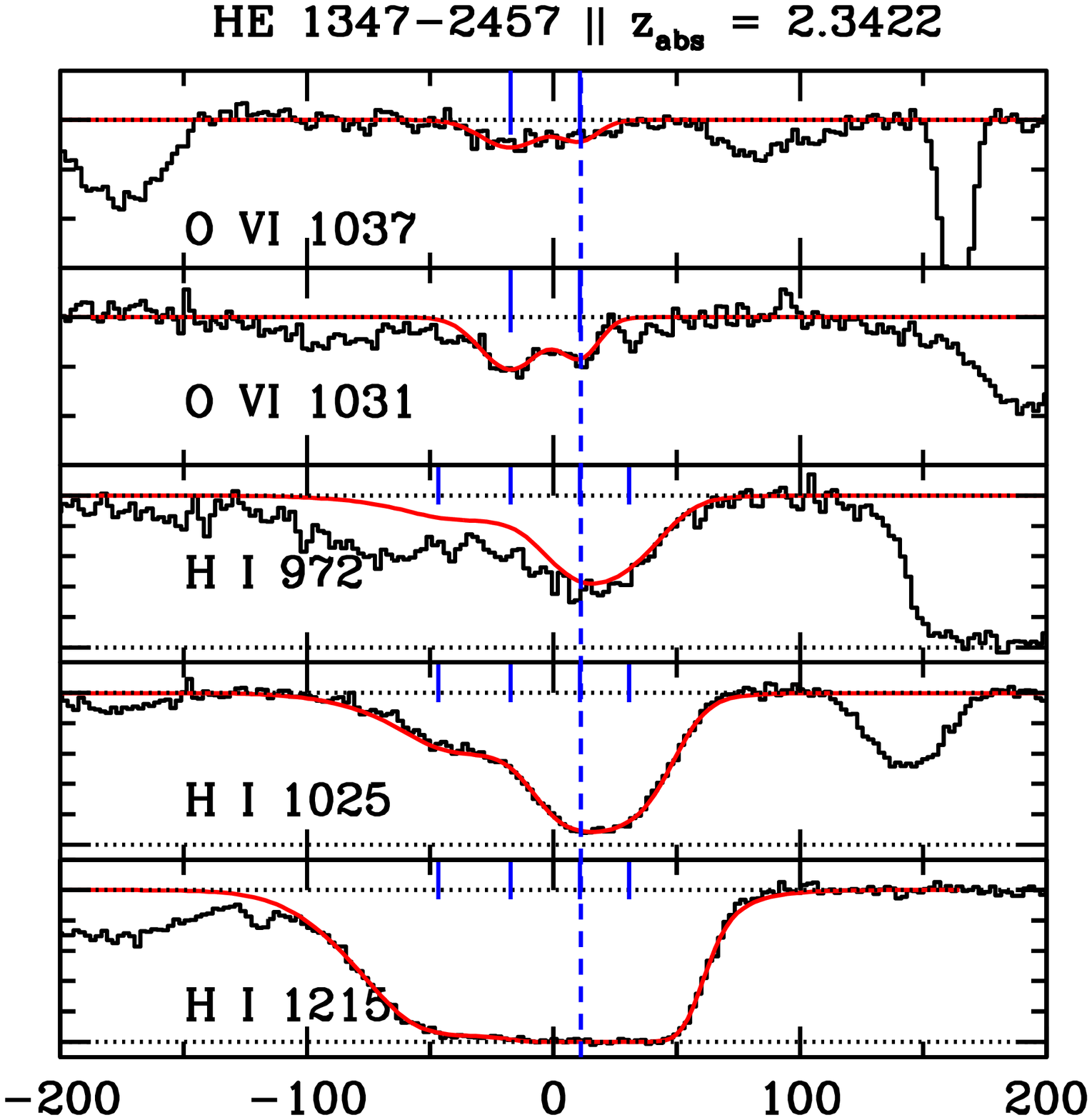} 
}} 
\centerline{\hbox{  
\includegraphics[height= 5.4cm,width= 4.1cm,bb=40 163 590 718,clip=]{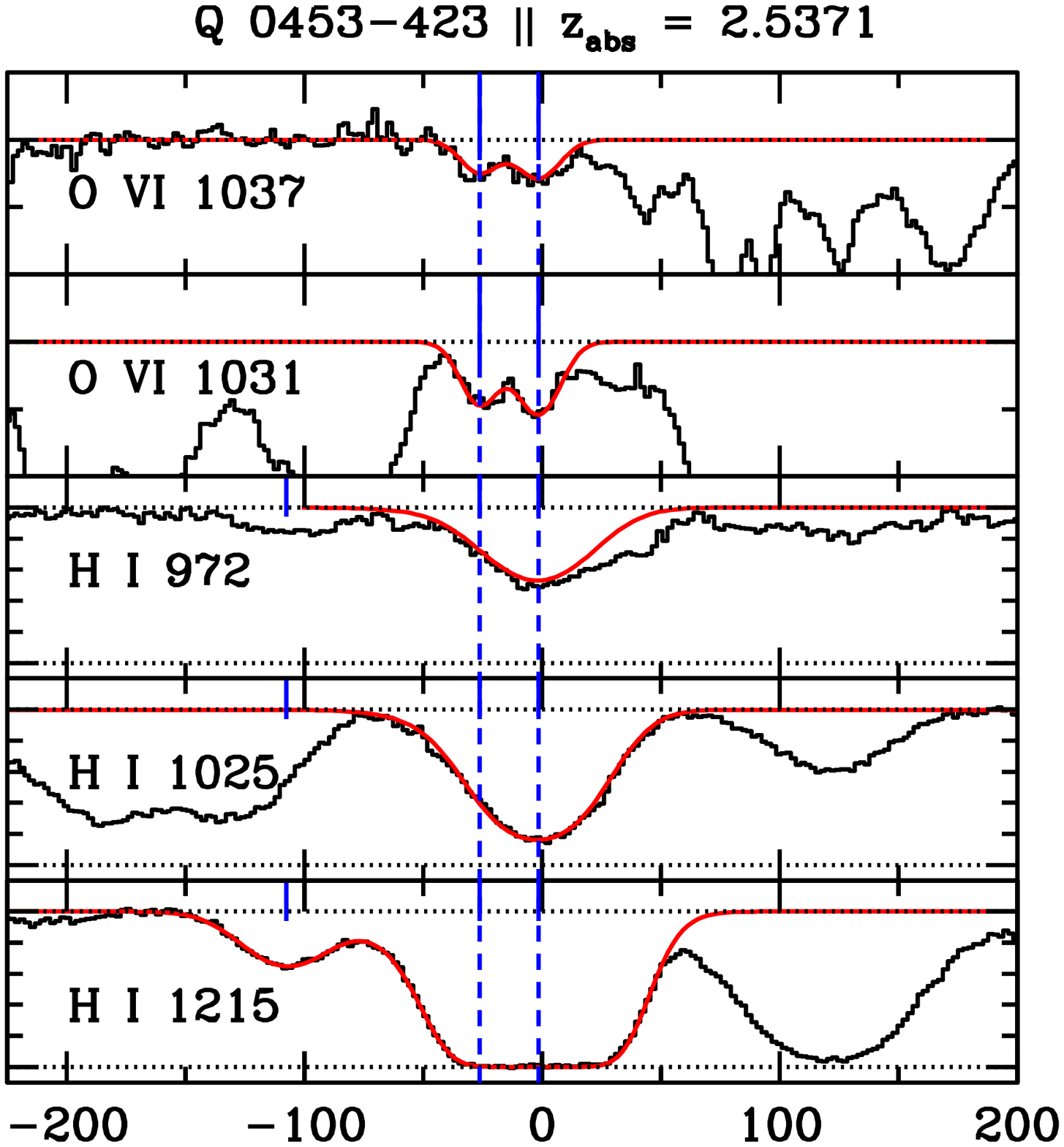} 
\includegraphics[height= 5.4cm,width= 4.1cm,bb=40 163 590 718,clip=]{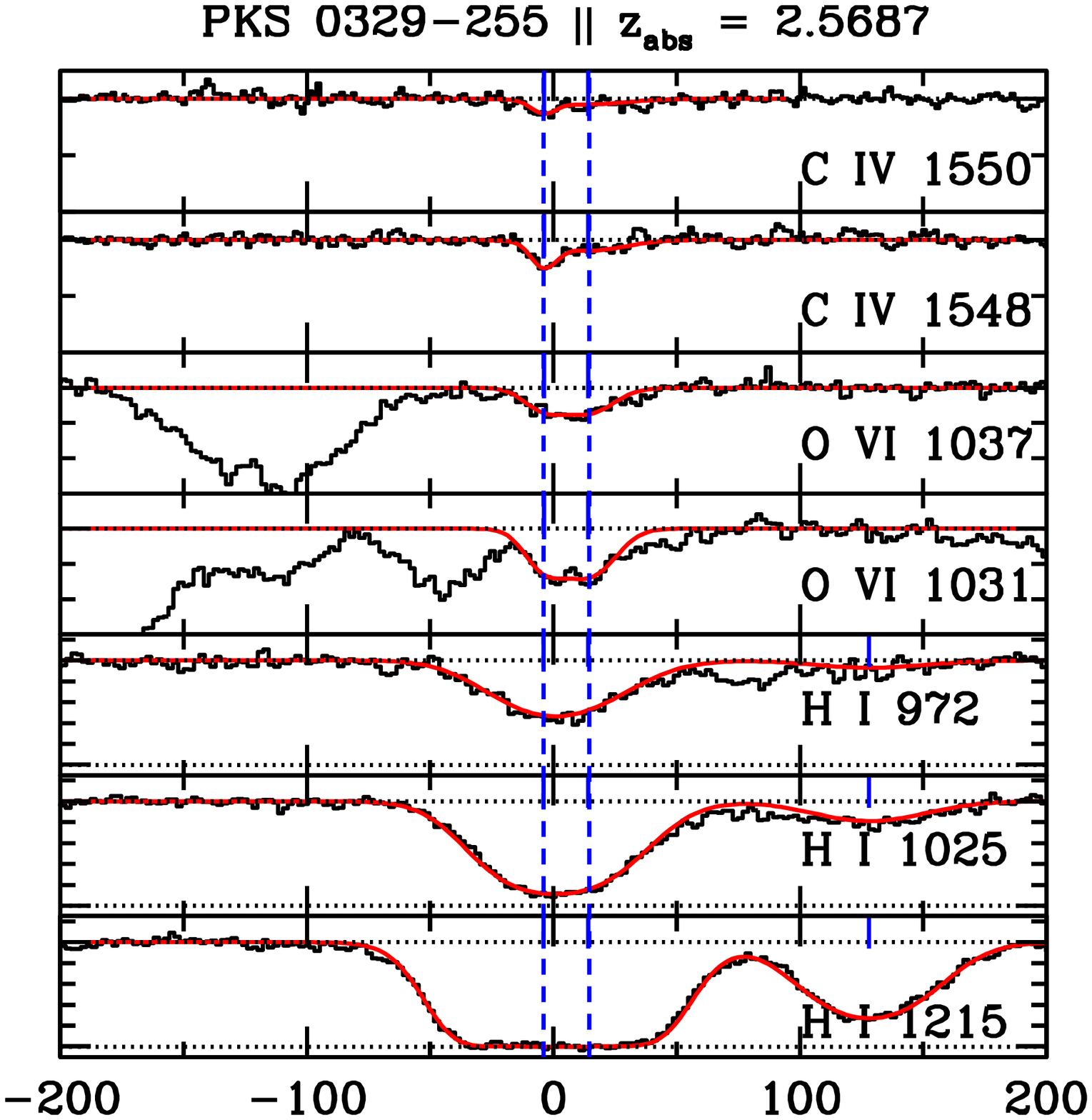} 
\includegraphics[height= 5.4cm,width= 4.1cm,bb=40 163 590 718,clip=]{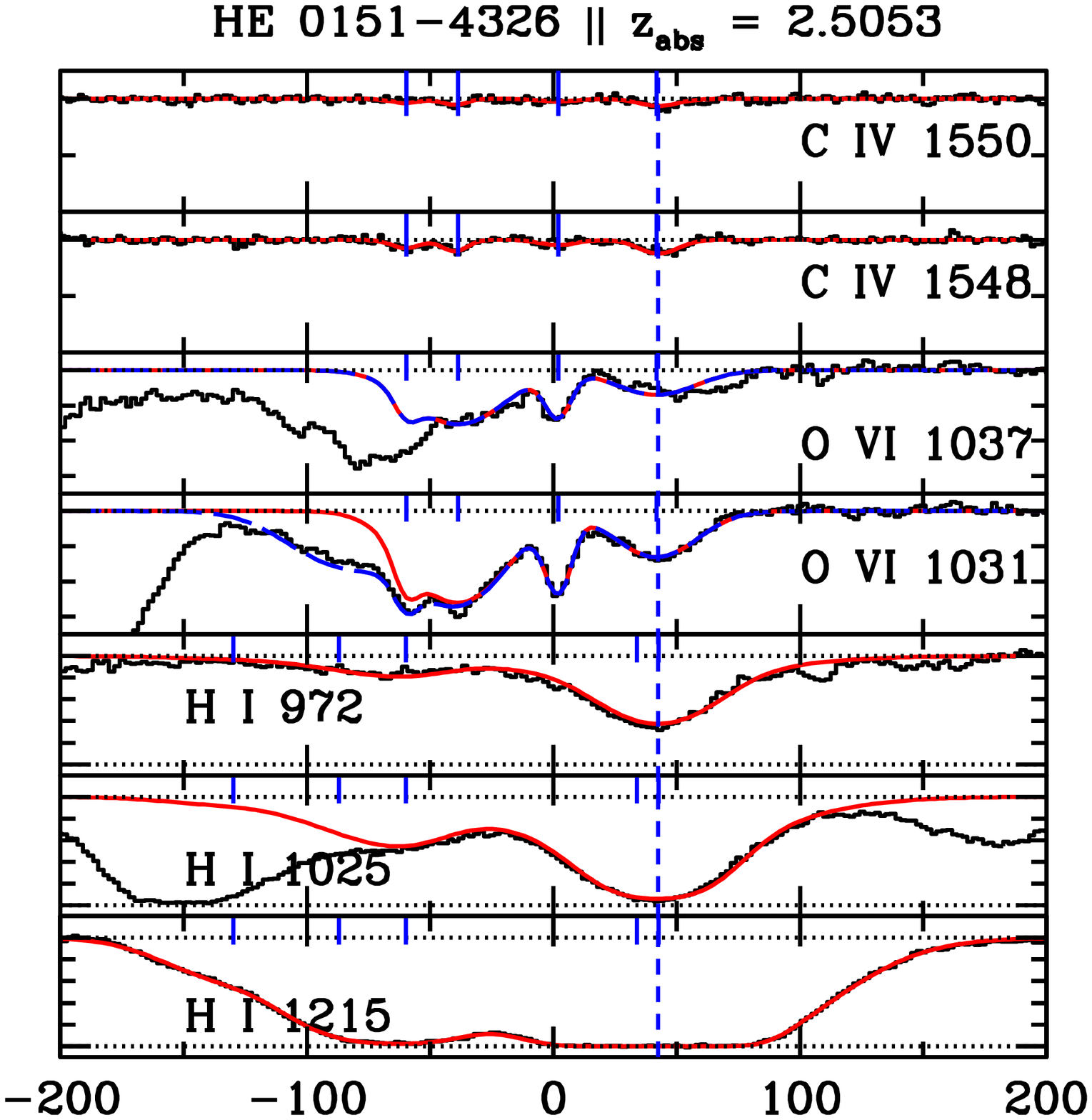} 
\includegraphics[height= 5.4cm,width= 4.1cm,bb=40 163 590 718,clip=]{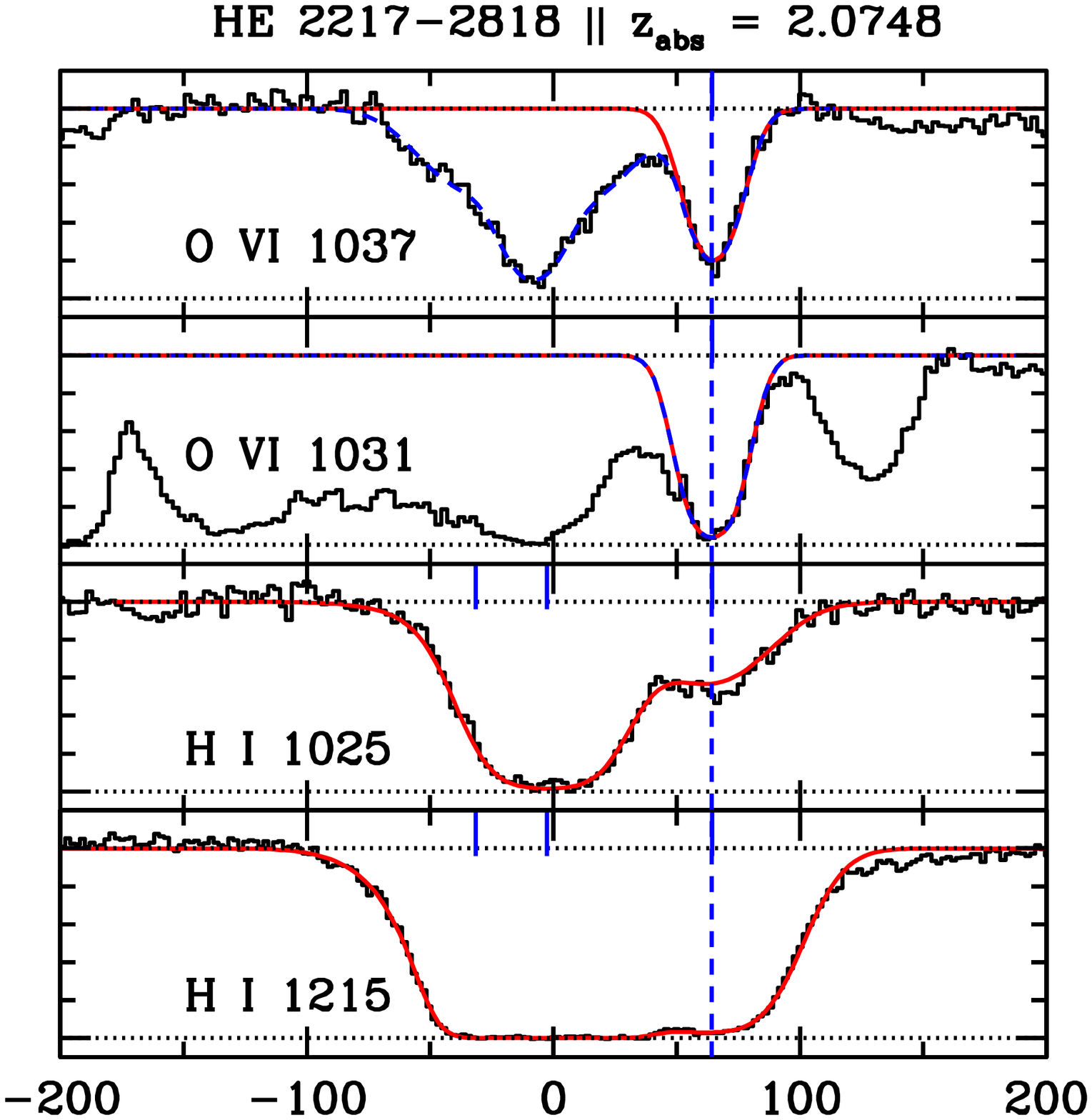} 
}} 
}} 
\vskip 0.15in 
\caption{Velocity plots of the 13 well aligned systems that are used to measure 
temperature and turbulent motion of the absorbing gas. 
The best fitted Voigt profiles are over-plotted on the observed data. 
The vertical tick marks show the positions of individual components. 
The vertical dashed lines mark the components used for this study.  
The occasional dashed curves over-plotted on the data show the contamination 
from other intervening absorption. The absorption redshift that defines the 
zero velocity and the name of the background QSO are indicated at 
the top of each panel. 
} 
\vskip -1.0in 
\begin{picture}(595,842)(0,0) 
\put(205,825){\large Relative Velocity [km s$^{-1}$]} 
\put(00,1100){\begin{sideways} \large Normalised Flux \end{sideways}} 
\end{picture}  
\label{vplot_aligned} 
\end{figure*}

\subsection{Thermal and non-thermal contributions to $b$-parameters} 
\label{bnt_T}

In the previous section we find that on an average $b$-parameters of 
\os\ measured at low-$z$ are higher than that measured at high-$z$. 
To explore this further we use a subsample of \os\ absorbers where 
\os\ and \hi\ absorption are well aligned. 
In general the line width of any species can be decomposed into 
thermal ($b_{\rm th}$) and non-thermal ($b_{\rm nt}$) parts, 
i.e. ~ $b^{2} = b_{\rm th}^{2}+b_{\rm nt}^{2}$. For species located in 
the same physical region, the non-thermal part is supposed 
to be identical whereas the thermal part scales inversely with 
the mass of the ion. 

Line saturation in the case of \lya\ and blending with \hi\ lines in the case 
of \os\ make the robust estimation of $b$-parameters for \hi\ and \os\ difficult 
in our sample. In addition we need to ensure that there is a good 
alignment between \os\ and \hi\ absorption.  
Hence we select systems using the following two criteria : 
(1) The component structure of \hi\ is well defined and one of the available 
Lyman series lines is unsaturated (i.e. ``Class-A" absorbers as defined in 
Table~\ref{allsystems}). 
(2) The \os\ profiles are well defined and the velocity offset between 
\hi\ and \os\ absorption centroids are consistent with zero within 3$\sigma$ 
uncertainty. We also avoid systems with low ion absorption lines 
as \hi\ seems to be predominantly associated with the low-ionization phase 
 when these species are present (see section~\ref{model}). 
Thus by using these selection criteria we minimize the probability that these 
absorbers have a multiphase structure.    

We find only 13 systems with 19 Voigt profile components (identified by 
vertical dashed lines in Fig.~\ref{vplot_aligned}) satisfying the conditions 
listed above. This is only $\sim$~15\% (i.e. 13 out of 84 systems) of our full 
sample. There are 6 systems showing only \os\ and \hi\ absorption 
(i.e. ``\os\ only" systems). In the remaining 7 systems both 
\os\ and \cf\ absorption are seen. In 6 of them \os\ and \cf\ components are 
remarkably aligned. In these cases we estimate the temperature and $b_{\rm nt}$ 
using both $b(\hi)$--$b(\os)$ and $b(\hi)$--$b(\cf)$ pairs.  
Only in the case of the \zabs = 2.0748 system towards HE~2217--2818, the 
corresponding \cf\ component is $\sim$~4.2 \kms\ away from the \os\ 
component and hence it is not used in our analysis. 

The results of the decomposition in thermal and non-thermal broadening 
are summarized in Table~\ref{tab_turb}. Columns \#1, \#2 and \#3 list, 
respectively, the QSO name, system redshift ($z_{\rm sys}$) and 
the velocity of the component ($v_{\rm rel}$) with respect to the 
systemic redshift. Columns \#4, \#5 and 
\#6 give $b$-values of \hi, \os\ and \cf\ components, respectively.  
The corresponding column densities are given in columns \#7, \#8 and 
\#9 respectively. The non-thermal contribution to the broadening 
and the estimated temperature are listed in columns \#10 and \#11, 
respectively. The values in parenthesis are calculated using 
$b(\hi)$--$b(\cf)$ pairs. Column \#12 lists the upper limits on 
the temperature as calculated from $b(\hi)$ assuming pure thermal broadening. 
It can be seen from the table that the temperatures estimated from 
the $b(\hi)$--$b(\cf)$ pairs are consistent with those derived from 
the $b(\hi)$--$b(\os)$ pairs, whereas $b(\cf)$ gives slightly lower 
$b_{\rm nt}$ compared to that obtained using $b(\os)$.  

\begin{figure} 
\centerline{
\vbox{
\centerline{\hbox{ 
\includegraphics[height= 8.4cm,width= 8.4cm,angle= 0]{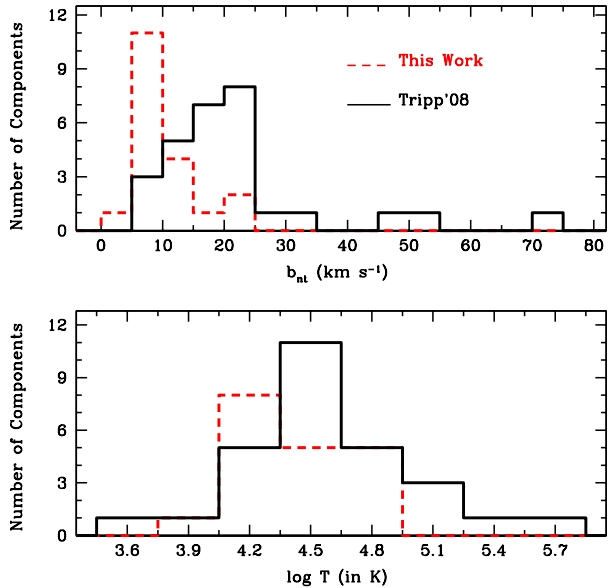} 
}}
}}
\caption{
Distributions of temperature (lower panel) and turbulent velocity 
(upper panel) in the well aligned sample are shown in dashed histogram. 
The solid histogram in both panels shows the corresponding distribution at 
low-$z$ as derived by \citet{Tripp08}.} 
\label{b_temp} 
\end{figure}

In Fig.~\ref{b_temp}  we show the distribution of $b_{\rm nt}$ 
(top) and temperature (bottom) as calculated from the 
$b(\hi)$--$b(\os)$ pairs. The solid histogram shows the results 
by \citet{Tripp08} for the low-$z$ well aligned \os\ absorbers 
for comparison. 
The values of non-thermal velocity in our sample are found to be in the range 
$ 3.6 \le b_{\rm nt} \le 21.2$ \kms\ with a median value of 8.2~\kms. 
The median value of $b_{\rm nt}$ for the \citet{Tripp08} sample 
is $\sim$~20.0 \kms. 
From the top panel of Fig.~\ref{b_temp} it is apparent that the $b_{\rm nt}$ 
distributions at high and low-$z$ are significantly different. This is 
confirmed by a two sided KS test with a probability that the two distributions  
differ much greater than 99.9\%. 

The median value of the temperature distribution in both high and low redshift 
samples is found to be $\sim3\times10^{4}$~K.  
The KS test shows that the temperature distribution 
of low-$z$ sample is not significantly different from high-$z$ sample 
($\sim$ 38\% probability that the difference is occuring by chance). 
It is interesting to note that while none of the components has temperature 
log~$T \ge 5.3$ (which would favor collisional ionization for \os), 
42\% of the components (i.e. 8 out of 19 components) show 
$4.6 \le$~log~$T \le 5.0$, which is warmer than the temperatures expected in 
photoionization equilibrium. These higher temperatures can be obtained 
in a rapidly cooling over-ionized gas that was shock heated through 
mechanical processes such as galactic winds. 
We thus compare the observed $N(\os)/N(\hi)$ in these well aligned components 
with the non-equilibrium collisional ionization models of \citet{Gnat07} 
assuming the temperature derived from $b(\hi)$ (given in column \#12 of 
Table~\ref{tab_turb}) and find that the observed ratios can not be reproduced 
by these models (even when we use maximum gas temperature) for gas phase 
metallicity less than solar. Thus it seems that the ionization state is probably 
maintained by the UV background radiation which is expected to 
dominate when $T \le 10^{5}$~K \citep[see Fig. 9 of ][]{Muzahid11}.

%

\begin{table}
\caption{Results of Spearman rank correlation analysis} 
\begin{tabular}{cccp{0.8cm}cc}
\hline  
  &   & \multicolumn{2}{c} {This Work}  & \multicolumn{2}{c} {\citet{Tripp08}} \\ 
\cline{3-4} \cline{5-6} 
Observable-1 & Observable-2 &  $\rho_{\rm s}$ & $\rho_{\rm s}/\sigma$ & 
$\rho_{\rm s}$ & $\rho_{\rm s}/\sigma$ \\  
\hline \hline  

$b_{\rm nt}(\os)$  & $N(\os)$ &$+$0.41 &$+$1.7&  $+$0.31&$+$1.6 \\ 
$b_{\rm nt}(\os)$  & $N(\hi)$ &$+$0.21 &$+$0.9&  $+$0.07&$+$0.4 \\  
$b_{\rm nt}(\os)$  & $b(\os)$ &$+$0.78 &$+$3.3&  $+$0.99&$+$5.2 \\ 
$b_{\rm nt}(\os)$  & $b(\hi)$ &$-$0.05 &$-$0.2&  $+$0.56&$+$2.9 \\ 
log~$T(\os)$       & $b(\os)$ &$+$0.30 &$+$1.3&  $+$0.21&$+$1.1 \\ 
log~$T(\os)$       & $b(\hi)$ &$+$0.92 &$+$3.9&  $+$0.83&$+$4.3 \\ 
\hline 
\hline 
\end{tabular}
\label{temp_corr_tab}
\end{table}

In Table~\ref{temp_corr_tab} we summarize the Spearman rank correlation 
analysis to search for possible correlations between $b_{\rm nt}$, log~$T$ 
and other observables of \os\ absorbers discussed here as well as in 
\citet{Tripp08}. In all the correlation analysis presented in this paper, 
the correlation coefficient and its significance are denoted by $\rho_s$ and 
$\rho_s/\sigma$, respectively.
As expected, $b_{\rm nt}$ and log~$T$ are strongly correlated to $b(\os)$ 
and $b(\hi)$, respectively, with a very high correlation coefficient 
($\gtrsim$~0.8) and at $>$~3$\sigma$ significance level in both samples. 
A weak correlation (at $<$~2$\sigma$ level) is seen between 
$b_{\rm nt}$ and $N$(\os), both at high and low redshift. 
Recent simulations of \citet{Cen11} suggest such a trend of higher 
non-thermal contribution at higher $N$(\os).  
On the other hand \citet{Oppenheimer09} have introduced density dependent 
turbulence in their simulations in order to reproduce the equivalent width
distribution and the $b-N$ correlation of low-$z$ \os\ absorbers. 
The lack of strong correlation between $b_{\rm nt}$ and $N(\os)$ or $N(\hi)$ 
in our sample suggesting that this density dependence of turbulence 
may be weak at high-$z$.

\subsection{Column density distributions} 
\label{cdd}

\begin{figure*} 
\centerline{
\vbox{
\centerline{\hbox{ 
\includegraphics[height= 9.4cm,width= 8.4cm,angle= 0]{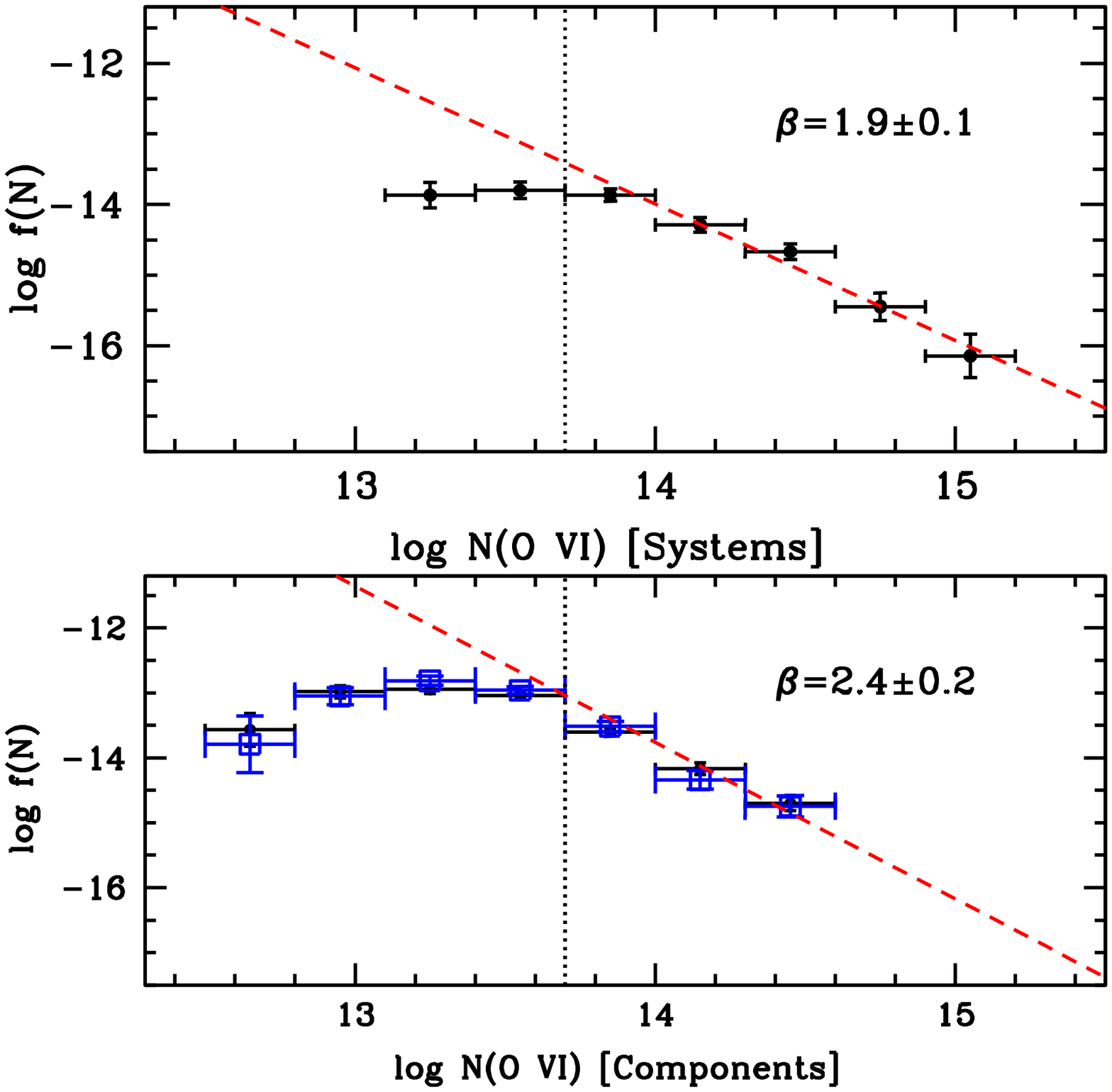}  
\includegraphics[height= 9.4cm,width= 8.4cm,angle= 0]{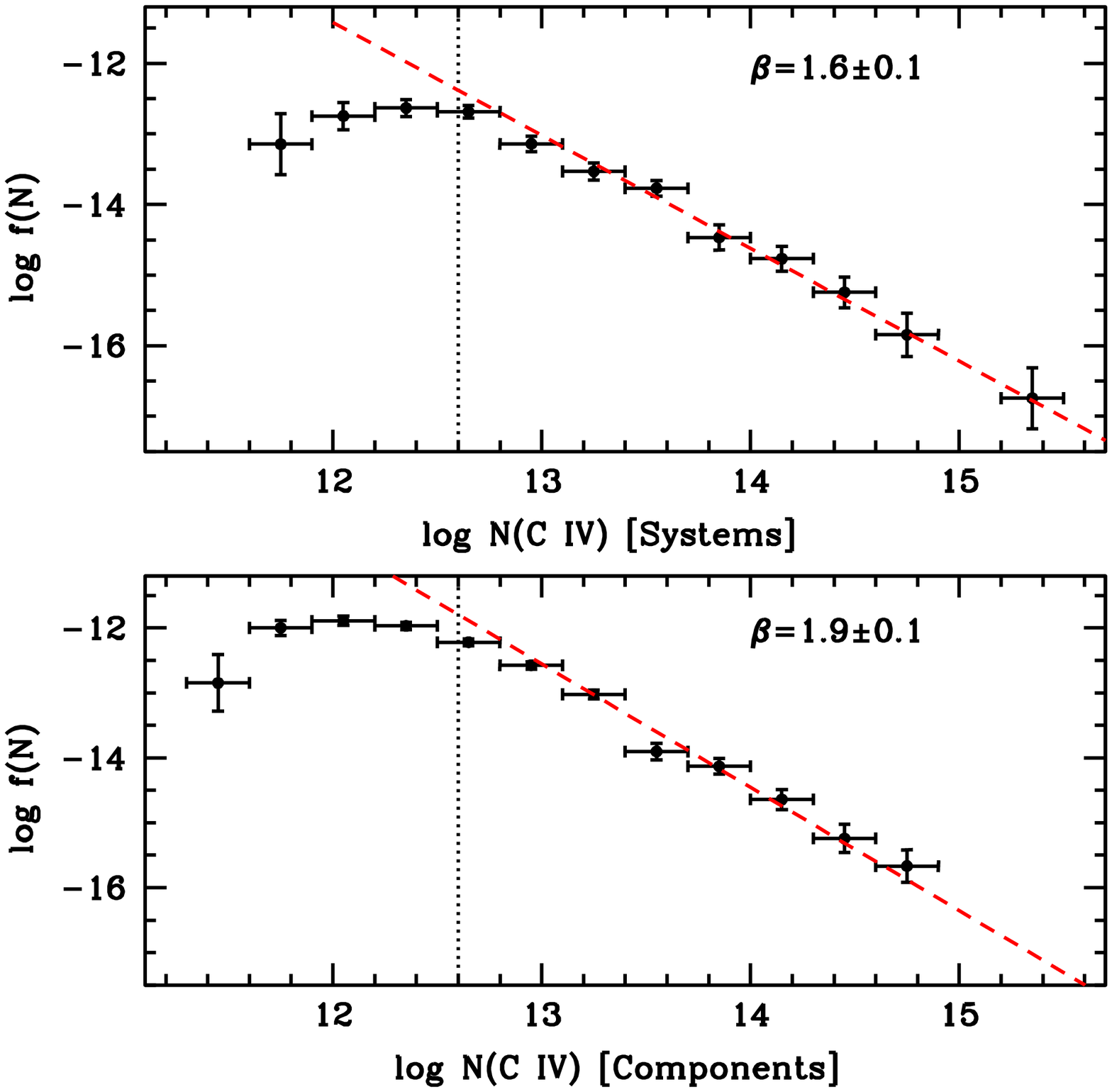}  
}}
}}
\caption{{\it Left:} Column density distribution functions of \os\ systems 
(upper panel) and \os\ components (lower panel). The $y$-axis is the 
number of systems/components per column density interval per unit 
redshift path length. The bin size along the column density axis is 
$10^{0.3} {\rm cm}^{-2}$ and the 1$\sigma$ error bars in the y-axis 
are calculated using Poisson statistics. 
The (blue) squares are taken from \citet{Bergeron05}. The dashed 
straight lines are power laws of the form $f(N) = BN^{-\beta}$. 
The power law indices, $\beta$, are obtained using the maximum likelihood method 
where no binning is involved. 
Vertical dotted lines show the column density above which 
our sample is complete. {\it Right:} Same as left but for \cf .
} 
\label{CDDF} 
\end{figure*} 

In this section, we study the column density distribution functions (CDDF) 
of \cf\ and \os\ systems in our sample. We use the usual parametrization of 
the column density distribution function, i.e., 
\begin{equation}
f(N)dN = B N^{-\beta} dN,
\end{equation}
where, $f(N)$ is the number of systems/components per unit column 
density interval per unit redshift path length defined as,  
\begin{equation} 
dX = (1+z)^{2}[\Omega_{\Lambda}+\Omega_{\rm m}(1+z)^{3}]^{-1/2} dz . 
\end{equation}

We use the maximum likelihood method to estimate the power law index, $\beta$. 
The column density distribution functions of \os\ systems (top) 
and components (bottom) are shown in the left panel of  Fig.~\ref{CDDF}.  
The (blue) squares in the bottom panel correspond to the \os\ component 
CDDF derived from the fits by \citet{Bergeron05} for only 12 sight lines. 
It is in good agreement with our results based on the full sample. 
As mentioned above, we cover a redshift path ($\Delta z$) of 7.62 
(and $\Delta X$ = 24.85) with the full sample where the S/N 
is~$> 10$ per pixel.  
This figure shows that our survey is not severely affected by 
incompleteness for log~$N(\os) > 13.7$ as indicated by the vertical 
dotted line.   
%
\begin{table}
\caption{Summary of power law index measurements for \cf\ CDDF}
\begin{tabular}{lcccc}
\hline 
Redshift & $>$~log $N(\cf)$ & \multicolumn{2}{c}{$\beta_{\cf}$} & References \\ 
\cline{3-4}
         &                 & Systems & Components & \\ 
\hline 
\hline 
1.9--3.1  & 12.6 &  1.6$\pm$0.1  &  1.9$\pm$0.1  & 1 \\ 
1.9--3.1  & 13.0 &  1.7$\pm$0.1  &  2.0$\pm$0.1  & 1 \\   
          & 13.4 & 1.64$\pm$0.10 &       ....    & 2 \\ 
2.9--3.5  & 13.0 &  1.8$\pm$0.1  &       ....    & 3 \\ 
          & 13.0 & 1.71$\pm$0.07 &       ....    & 4 \\  
	  & 12.3 & 1.44$\pm$0.05 &       ....    & 5 \\ 
	  &      &      1.6      &       1.8     & 6 \\ 
~~~$<$~1.0 & 13.2 &$1.50^{+0.17}_{-0.19}$ &  ...  & 7 \\ 	  
\hline 
\hline 
\end{tabular} 
\footnotesize 
References:-- (1) This work; (2) \citet{Petitjean94}; 
(3) \citet{Songaila01}; (4) \citet{Dodorico10}; 
(5) \citet{Ellison00}; (6) \citet{Boksenberg03}; 
(7) \citet{Cooksey10}
\label{tab_betaCIV}
\end{table}
%

A maximum likelihood fit to our data gives a power law index of 
$\beta_{\os} = 1.9\pm0.1$ for \os\ systems and $\beta_{\os} = 2.4\pm0.2$ 
for \os\ components for log~$N(\os)>13.7$ . 
We would like to mention that in the subsample $dd$ where both 
lines in the doublet are unblended, we find  $\beta_{\os} = 2.2\pm0.3$ for 
systems and $\beta_{\os} = 2.7\pm0.4$ for components . These values are somewhat 
steeper than, albeit consistent with, the full sample within the measurement 
uncertainties. 
For a sample of low redshift 
($z<0.15$) \os\ absorbers with Ly$\alpha$  rest frame equivalent width  
($W_{\rm Ly\alpha}$) $\ge 80~$m$\AA$, \citet{Danforth05} found 
$\beta_{\os} = 2.2\pm0.1$ for the components.
In a latter paper studying an extended sample of \os\ absorbers,
\citet{Danforth08} found $\beta_{\os}$ to be 1.98$\pm$0.11. 
In addition, we have performed a maximum likelihood analysis on the low-$z$ sample 
of \citet{Tripp08} and found $\beta_{\os} = 2.3\pm0.2$ for components  
and 2.0$\pm$0.2 for systems for log~$N(\os)>13.7$.
Therefore the $\beta_{\os}$ measurements at low-$z$ are consistent with our 
estimations.   

The right panels of Fig.~\ref{CDDF} show the \cf\ column density distribution 
function for components (lower panel) and systems (upper panel). 
Incompleteness limit is clearly log~$N(\cf)$($\rm cm^{-2}$)~$\ge 12.6$ as shown 
in the figure by a vertical dotted line. 
Using maximum likelihood analysis, we find, $\beta_{\cf}$, to be 1.6$\pm$0.1  
for systems and 1.9$\pm$0.1 for components for log~$N(\cf)>12.6$. 
Our results are consistent with the earlier findings  
\citep[]{Petitjean94,Songaila01,Dodorico10} within 1$\sigma$ uncertainty 
when we use similar column density cutoff 
(i.e. log~$N(\cf)>13.0$; see Table~\ref{tab_betaCIV}). 
%

\subsection{\os\ and \cf\ cosmological densities} 
\label{omega}

In this section we compute the contribution of \os\ and \cf\ absorbers to 
the baryon density using our column density estimates. The cosmic density of 
the \os\  absorbers can be expressed as : 
\begin{equation} 
\Omega_{\os} = \left(\frac{H_{0}\ m_{\rm O}}{c~\rho_{\rm cr}}\right) 
\left(\frac{\sum_{\rm i}{N_{\rm i}(\os)}}{\Delta X}\right),  
\label{eqn_omega1}
\end{equation} 
with an associated fractional variance 
\citep[as proposed by][]{Storrie-Lombardi96b}: 
\begin{equation} 
\left(\frac{\delta \Omega_{\os}}{\Omega_{\os}}\right)^{2}= 
\frac{\sum_{\rm i}[{N_{\rm i}(\os)}]^{2}}{[\sum_{\rm i}N_{\rm i}(\os)]^{2}}~,
\end{equation} 
where $H_{0}$ is the Hubble constant, $m_{\rm O}$ is the atomic mass of oxygen,  
$\Delta X$ and $\rho_{\rm cr}$ are the total redshift path and the current 
critical density, respectively.  
We find $\Omega_{\os} = (1.0\pm0.2) \times 10^{-7}$ for log~$N(\os)>13.7$. 
If we include the LLS in our calculation then 
$\Omega_{\os} = (1.3\pm0.3) \times 10^{-7}$ for the same column density cutoff. 
Using the column densities of 12 intervening systems listed in Table~2 of 
\citet{Simcoe02} we calculate $\Omega_{\os}$ for their sample. For 
their uncorrected redshift path (i.e. $\Delta X = 6.9$), $\Omega_{\os}$ 
turns out to be $(1.2\pm0.5)\times10^{-7}$ for log$N(\os)>13.7$ which is in 
good agreement with what we find here. For the 12 lines of sight our 
estimated value of $\Omega_{\os}$ is in excellent agreement with 
\citet{Bergeron05}. 

It is also possible to calculate the fractional contribution to the 
cosmological density of the baryons associated with the \os\ phase, 
provided the ionization fraction of \os\ ($f_{\os}$) and the 
metallicity ($Z$) of the gas are known, viz.,      
\begin{equation} 
\Omega_{\rm IGM}^{\os} = \left(\frac{H_{0}\ \mu m_{\rm H}}{c \rho_{\rm cr}} \right) 
\left(\frac{1}{f_{\os}Z({\rm O/H})_{\odot}}\right) 
\left(\frac{\sum_{\rm i}{N_{\rm i}(\os)}}{{\Delta X}}\right) .     
\label{eqn_omega2}
\end{equation} 
Where $\mu = 1.3$ is the mean atomic weight and $m_{\rm H}$ is the mass of 
the hydrogen atom. For log~$N(\os)>13.7$ the value of $\Omega_{\rm IGM}^{\os}$ 
is : 
$\Omega_{\rm IGM}^{\os} = 0.0011[(h/0.71)(Z/0.1Z_{\odot})(f_{\os}/0.2)]^{-1}$. 
For $f_{\os} = 0.2$ and $Z= 0.1Z_{\odot}$, we get 
$\Omega_{\rm IGM}^{\os}/\Omega_{\rm b}$ = 0.028 i.e. \os\ 
absorbers contribute 2.8\% to the total baryon density 
at redshift $z \sim 2.3$.    
For a more typical IGM  metallicity of $Z = 0.01Z_{\odot}$ 
this contribution can go up to $\sim$ 30\%.  
At low redshift ($z < 0.4$) \citet{Danforth08} derived 
$\Omega_{\rm IGM}^{\os}/\Omega_{\rm b} = 0.073\pm0.008$ down to 
log~$N(\os) = 13.4$ and $0.086\pm0.008$ down to log~$N(\os) = 13.0$ (for 
$f_{\os} = 0.2$ and $Z= 0.1Z_{\odot}$).

Using Eqn.~(\ref{eqn_omega1}) for log~$N(\cf)>12.6$, we find 
$\Omega_{\cf} = (2.4\pm0.6)\times 10^{-8}$ without including 
Lyman limit systems. Inclusion of LLS increases the value up to
$(5.3\pm2.1)\times10^{-8}$. 
This value compares well with the previous estimations  
\citep[]{Songaila01,Simcoe04,Scannapieco06,Dodorico10} 
for a similar redshift range as can be seen from Table~\ref{Omega_CIV}. 
For completeness, the recent $\Omega_{\cf}$ measurements for $z>$~5  
are also summarized in this table.  
%

\begin{table}
\caption{Summary of $\Omega_{\cf}$ measurements}
\begin{tabular}{lcccc}
\hline 
Redshift  &  & $\Omega_{\cf} \times(10^{-8})$ &  & References \\ 
\hline 
\hline 
$<$~1.0    & ~~~~~~~~~~~~~~~~~ & 6$\pm$1        &~~~~~~~~~~~~~~~~~ &  1 \\ 
1.9--3.1   & ~~~~~~~~~~~~~~~~~ & 5.3$\pm$2.2    &~~~~~~~~~~~~~~~~~ &  2 \\ 
2.0--2.5   & ~~~~~~~~~~~~~~~~~ & 6.83           &~~~~~~~~~~~~~~~~~ &  3 \\   
2.0--2.5   & ~~~~~~~~~~~~~~~~~ & 5.0$\pm$1.0    &~~~~~~~~~~~~~~~~~ &  4 \\   
$\sim$~2.2 & ~~~~~~~~~~~~~~~~~ & 7.5$\pm$2.2    &~~~~~~~~~~~~~~~~~ &  5 \\ 
$\sim$~2.5 & ~~~~~~~~~~~~~~~~~ & 3.8$\pm$0.7    &~~~~~~~~~~~~~~~~~ &  6 \\ 
2.5--3.0   & ~~~~~~~~~~~~~~~~~ & 4.79           &~~~~~~~~~~~~~~~~~ &  3 \\   
2.5--3.0   & ~~~~~~~~~~~~~~~~~ & 3.2$\pm$0.7    &~~~~~~~~~~~~~~~~~ &  4 \\  
5.3--6.0   & ~~~~~~~~~~~~~~~~~ & $\lesssim$~1.3 &~~~~~~~~~~~~~~~~~ &  7 \\ 
$\sim$~5.8 & ~~~~~~~~~~~~~~~~~ & 0.44$\pm$0.26  &~~~~~~~~~~~~~~~~~ &  8 \\   
\hline 
\hline 
\end{tabular} 
\footnotesize
References:-- (1) \citet{Cooksey10}; (2) This work; 
(3) \citet{Songaila01}; (4) \citet{Dodorico10}; 
(5) \citet{Scannapieco06}; (6) \citet{Boksenberg03}; 
(7) \citet{Becker09}; (8) \citet{Ryan-Weber09}  
\label{Omega_CIV}
\end{table}
\begin{figure} 
\centerline{
\vbox{
\centerline{\hbox{ 
\includegraphics[height= 8.4cm,angle= 0]{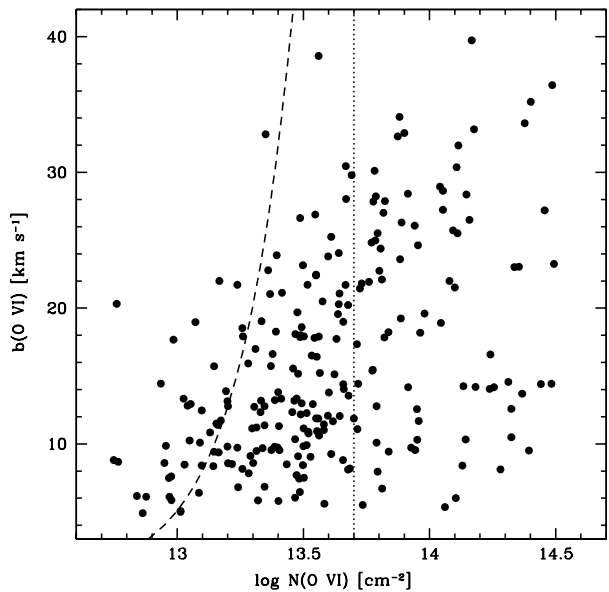}  
}}
}}
\caption{$N(\os)$ against $b(\os)$ measured in individual 
         components. The short dashed curve and the dotted line show 
	 the 5$\sigma$ detection threshold and the $N(\os)$ limit for the  
	 sample completeness respectively.}  
\label{b_N} 
\end{figure} 

\subsection{$b-N$ correlation} 

A correlation between column density and $b$-parameter of \os\ components 
has been predicted by theoretical studies of hot ionized gas 
\citep[see][]{Edger86,Heckman02} as a natural consequence of radiatively 
cooling hot gas passing through a coronal regime. \citet{Heckman02} have 
shown that such a correlation exists in a wide variety of astrophysical 
environments (such as Galactic disk, halo, High Velocity Cloud (HVC), 
Large Magellanic Cloud (LMC), Small Magellanic Cloud (SMC), starburst 
galaxies, IGM etc.). These authors have also shown that the relationship 
between the log~$N(\os)$ and log~$b$ is linear for the broad lines 
(i.e. $b(\os)>$~40 \kms) but rolls over and steepens for the narrower lines. 
\citet{Danforth06}, \citet{Tripp08} on the other hand, report no convincing 
evidence of such correlation for the low redshift 
intergalactic \os\ absorbers. \citet{Lehner06} revisited the \citet{Heckman02} 
model and found that their sample of \os\ absorbers is consistent with the 
model but the observed $N(\mbox{Ne\,{\sc viii}})$ is much less than 
the model prediction. 
%
\begin{table}
\caption{ Results of correlation analysis between $N(\os)$ and $b(\os)$}
\begin{tabular}{lccccrr}
\hline 
Sample & $>$~log~$N(\os)$ & Components & & $\rho_{s}$ & $\rho_{s}/\sigma$ \\ 
\hline 
\hline 
Total    &  13.7  &  82 &~~~~~~ &0.05 & 0.4 \\  
$dd$     &  13.7  &  19 &~~~~~~ &$-$0.36 & $-$1.5 \\ 
$bd$     &  13.7  &  22 &~~~~~~ &0.17 & 0.8 \\ 
$bb$     &  13.7  &  41 &~~~~~~ &0.24 & 1.5 \\ 
\hline 
\hline 
\end{tabular}
\label{corr_tab}
\end{table}

In Fig.~\ref{b_N} we plot the \os\ $b$-values against 
column densities in individual components for our full sample. 
The dashed curve shows the 5$\sigma$ detection threshold of our 
data assuming S/N~$\sim$~10~in the forest. The limiting equivalent width  
for a given $b$-value has been calculated using the prescription by 
\citet{Hellsten98}. This limiting equivalent width is then converted 
to a column density assuming the optically thin case. 
The vertical dotted line at log~$N(\os)=13.7$ shows our sample completeness 
(see section \ref{cdd}). For log~$N(\os)>$~13.7 the $b-N$ space is uniformly 
populated by the data points indicating the 
lack of any significant correlation. 
The results of Spearman rank correlation analysis performed between 
$N(\os)$ and $b(\os)$ in our full sample and various subsamples are 
given in Table~\ref{corr_tab}. No statistically significant 
correlation is found in any of these cases when appropriate column density 
limit (i.e. log~$N(\os)>$~13.7) is considered. 
%

\begin{figure*} 
\centerline{
\centerline{
\hbox{ 
\includegraphics[width= 8.4cm,angle= 0]{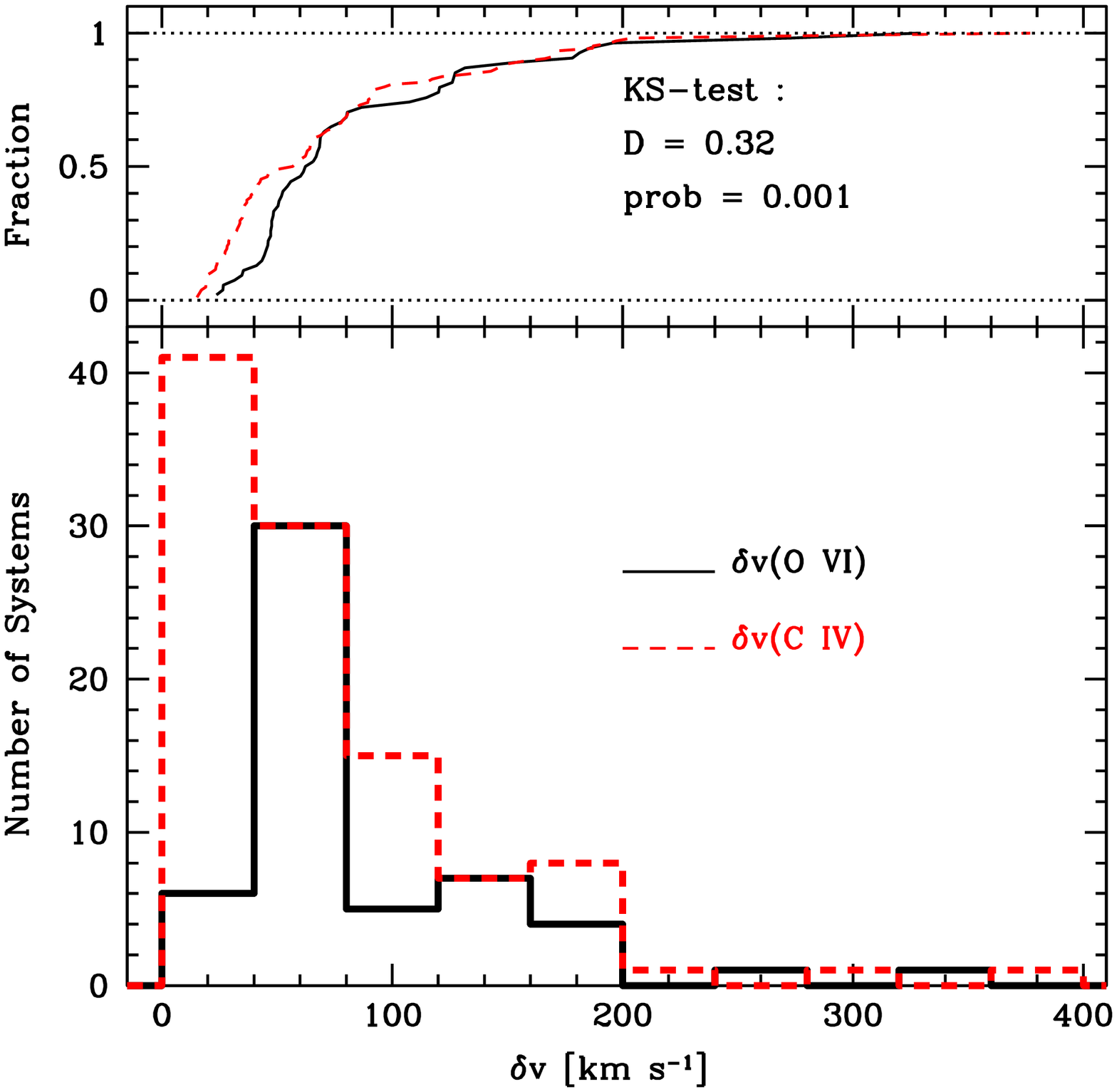}
\includegraphics[width= 8.4cm,angle= 0]{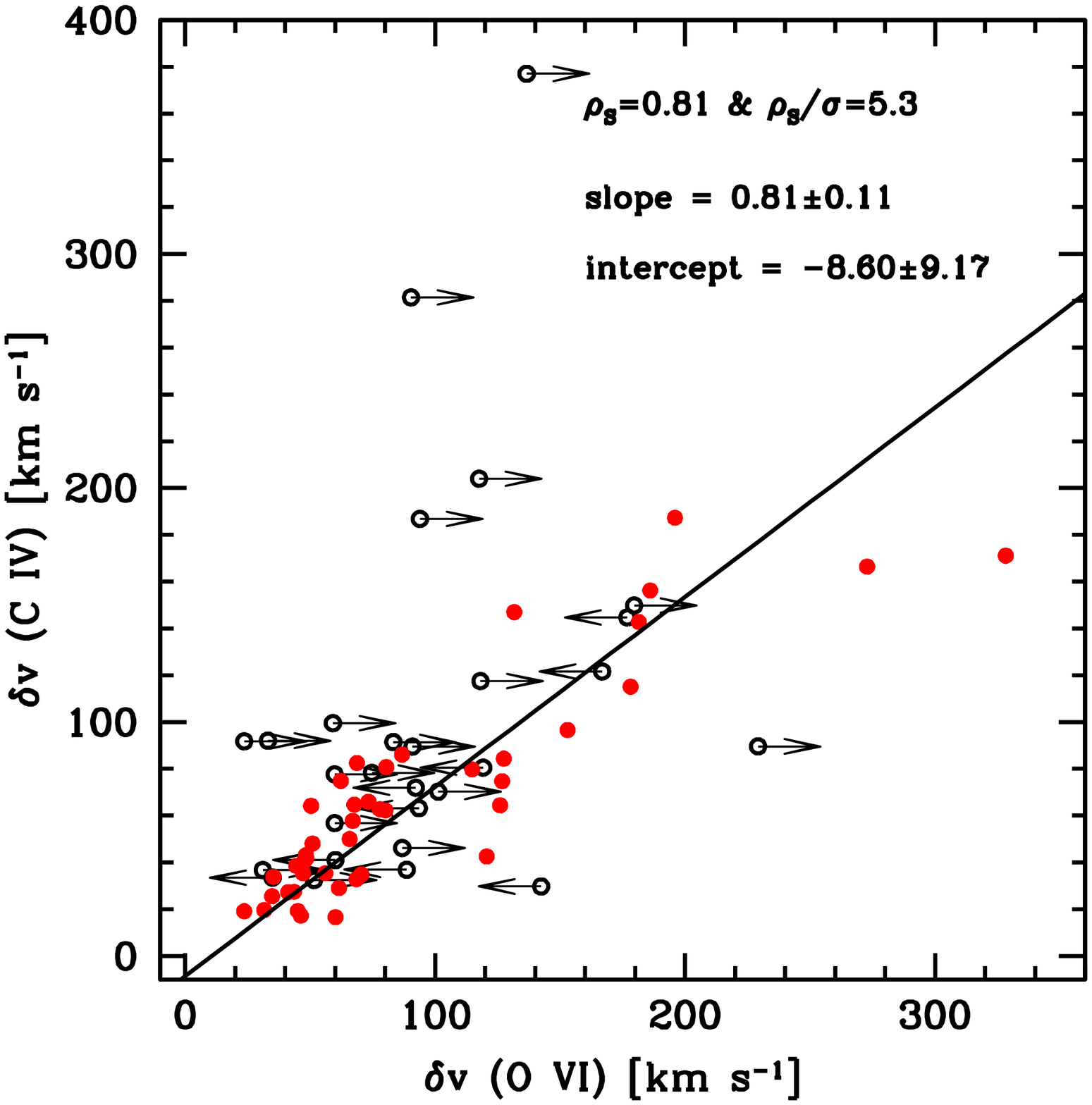}  
}
}}
\caption{{\it Left :}~$\delta v$ distributions of \os\ (solid histogram) 
and \cf\ (dashed histogram) systems. Corresponding cumulative 
distributions are plotted in the upper panel. 
{\it Right :}~$\delta v(\cf)$ versus 
$\delta v(\os)$. The (red) solid circles are for the systems with 
$\delta_{\rm type}=0$ whereas the arrows indicate systems with 
$\delta_{\rm type}= \pm1$.  
The straight line shows the least square fit to the 
data for systems with $\delta_{\rm type}=0$.    
}
\label{delv_dist}
\end{figure*} 

\section {Analysis based on Apparent optical depth} 
\label{kinemat}

%
In this section, we study the line kinematics of \os\ and \cf\ absorption 
using the line spread ($\delta v$) and the velocity shift ($\Delta v$) 
between the \cf\ and \os\ optical depth weighted redshifts
defined in section~\ref{vsa}.
For most of the discussions presented below, unless otherwise stated,  
we restrict ourselves to the $\delta_{\rm type}$ = 0 subsample defined 
in Table~\ref{allsystems}.

\subsection {Line spread ($\delta v$) distribution}  
\label{lsd}

In the left panel of Fig.~\ref{delv_dist}, we show the $\delta v$ distribution of 
\os\ systems (solid histogram) and \cf\ systems (dashed histogram, including all 
\cf\ systems). The median value of $\delta v$(\os) is 66 \kms~\ and the maximum 
observed value is 340 \kms. The median value of $\delta v$(\cf) is 58 \kms. 
The $\delta v$ distributions of \os\ and \cf, appear fairly 
similar except in the first velocity bin. If the latter is considered, the  
KS-test gives a probability of 0.1\% that the two distributions are drawn from 
the same population (see top panel in Fig.~\ref{delv_dist}). 
However if it is excluded, then the KS-test shows only a 
$\sim$ 52\% probability that the two distributions differ. 
The lack of \os\ systems with $\delta v(\os) <$~40 \kms\ could be related to 
the difficulty in detecting narrow systems in the \lya\ forest. 

In the right panel of Fig.~\ref{delv_dist}, we plot the \cf\ line spread  
against that of \os. The (red) filled circles are for systems with 
$\delta_{\rm type} = 0$. A strong correlation between $\delta v(\cf)$ and 
$\delta v(\os)$ is apparent 
from the figure. The Spearman rank correlation coefficient for the robust 
measurements is $\sim 0.81$ with $\sim 5.3\sigma$ significance. 
The slope ($0.81\pm0.11$) and the intercept ($-8.60\pm9.17$) of the best 
fitted straight line to the (red) solid circles indicates that the spread of 
\cf\ absorption is systematically smaller than that of \os\ for a given 
$\delta v(\os)$. 

\begin{figure*} 
\centerline{
\centerline{ 
\vbox{
\hbox{  
\includegraphics[width= 6.0cm,height=4cm,angle= 0]{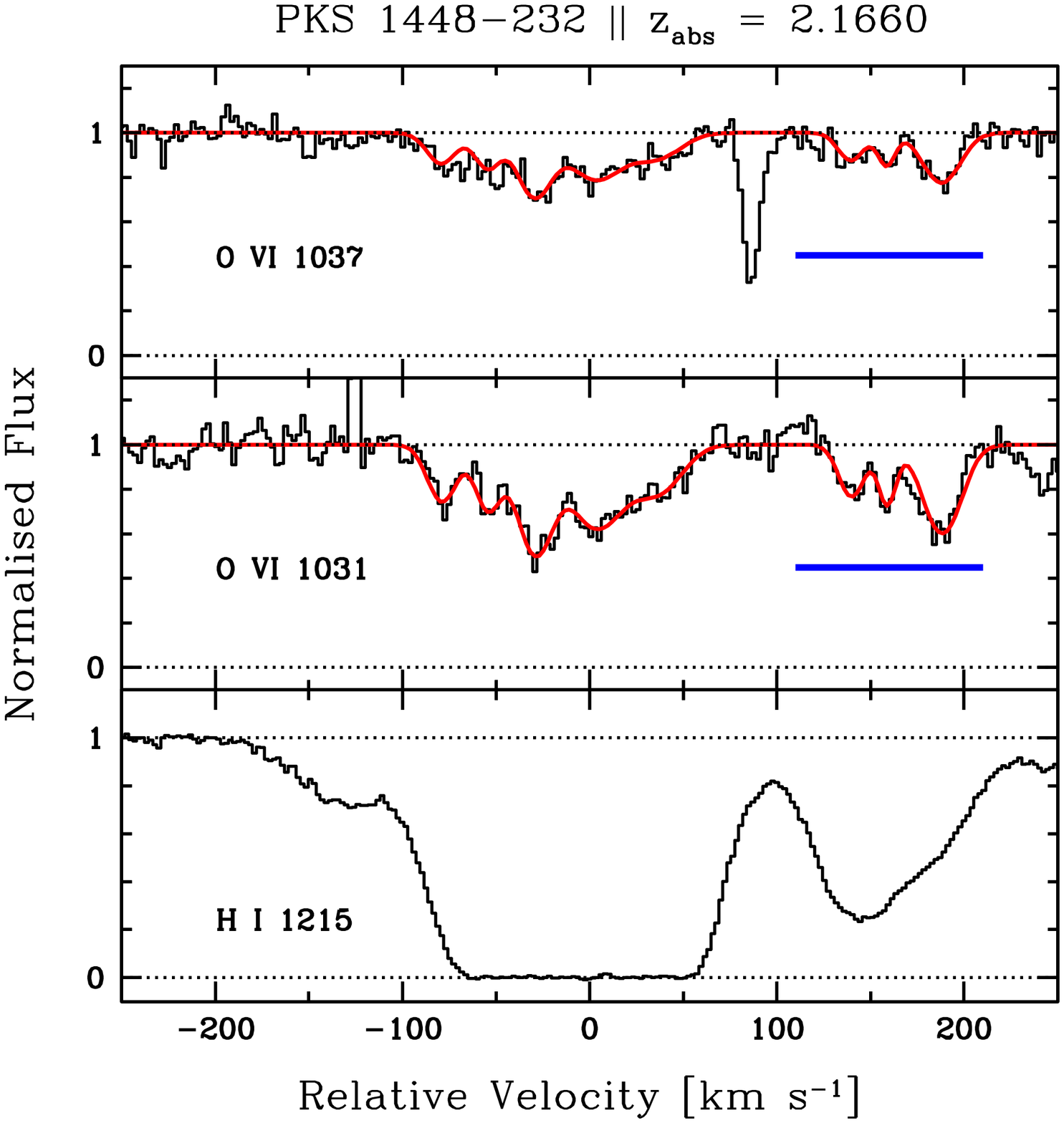}
\includegraphics[width= 6.0cm,height=4cm,angle= 0]{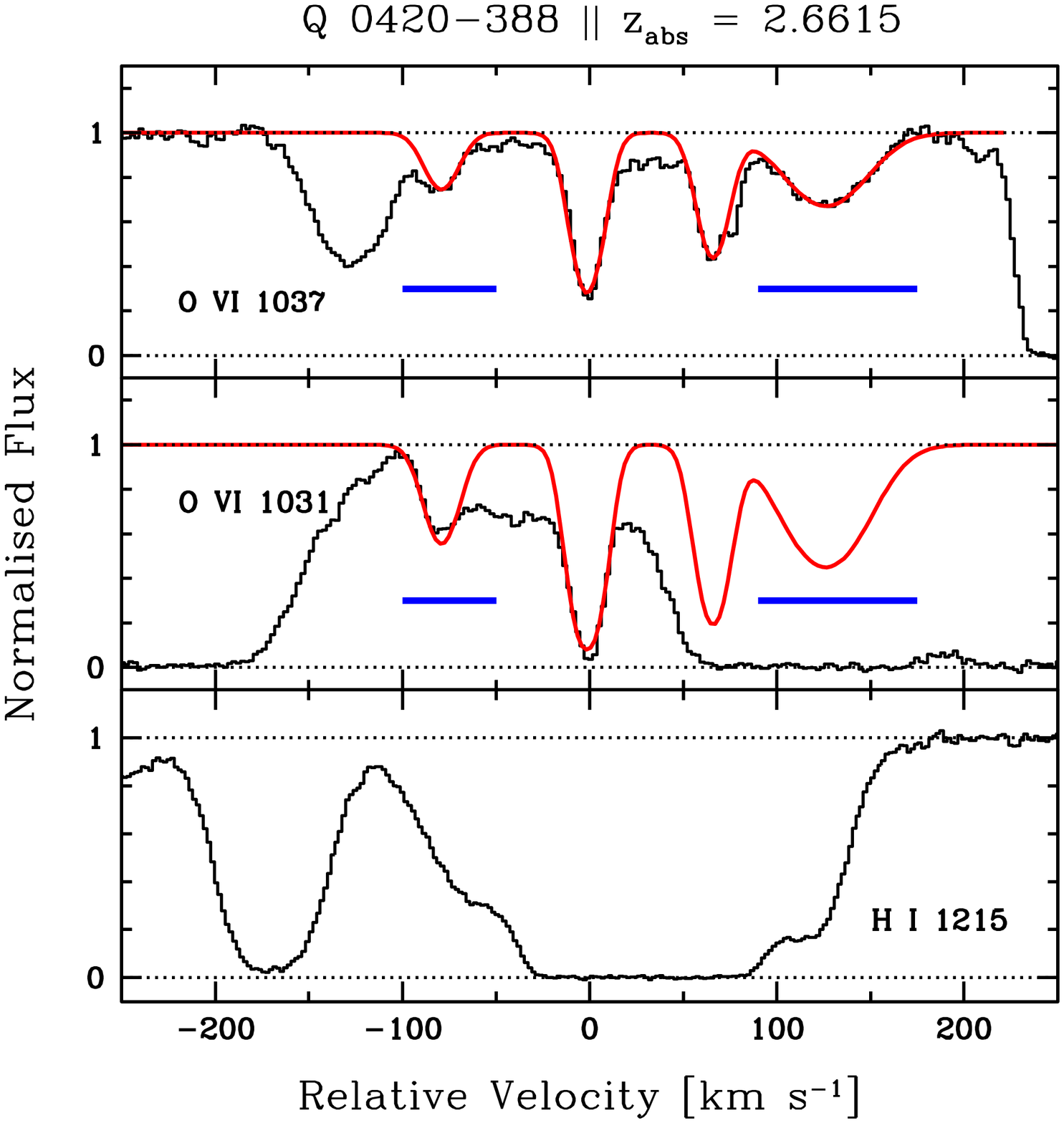}  
\includegraphics[width= 6.0cm,height=4cm,angle= 0]{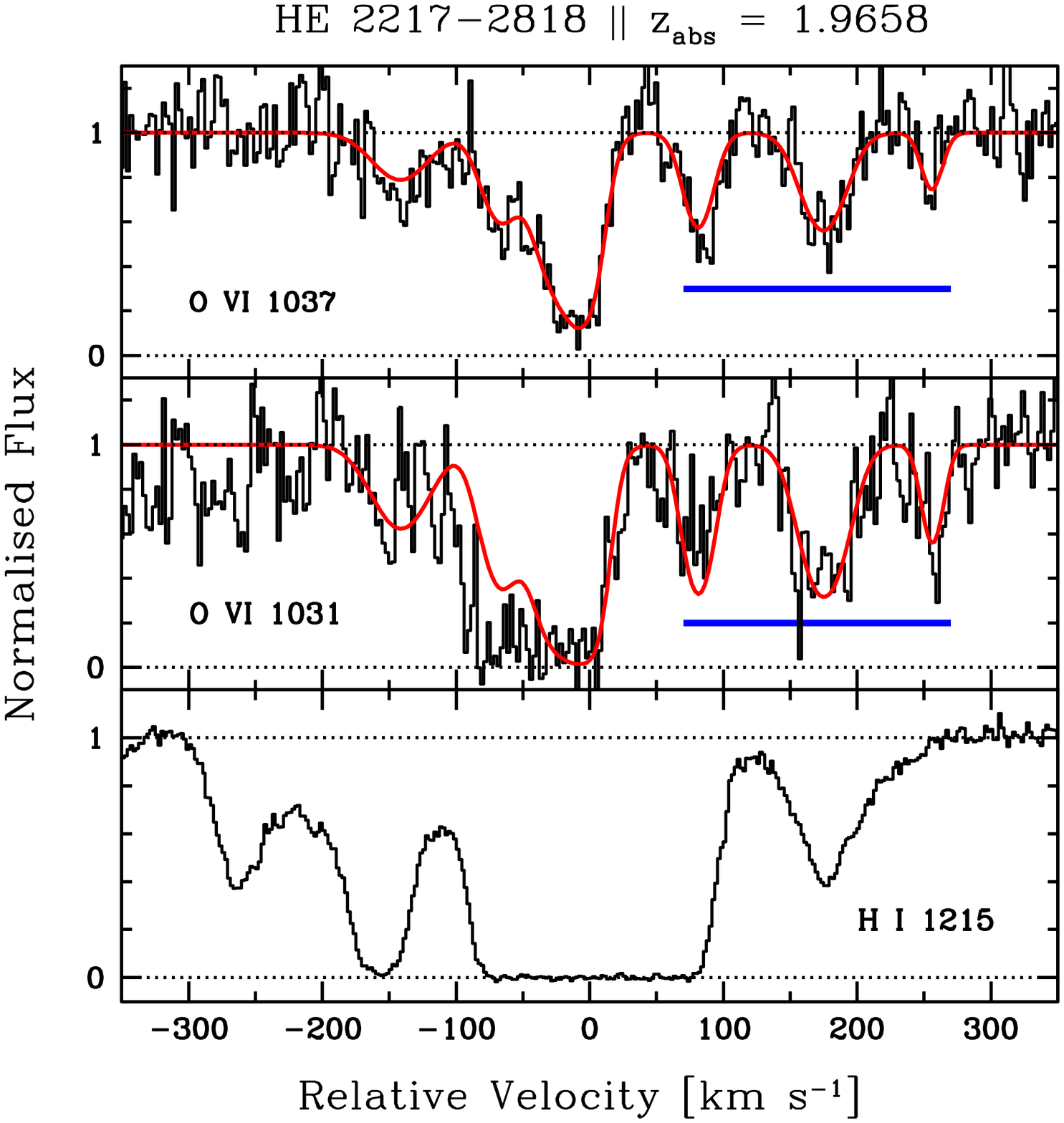}  
}
\hbox{ 
\includegraphics[width= 6.0cm,height=4cm,angle= 0]{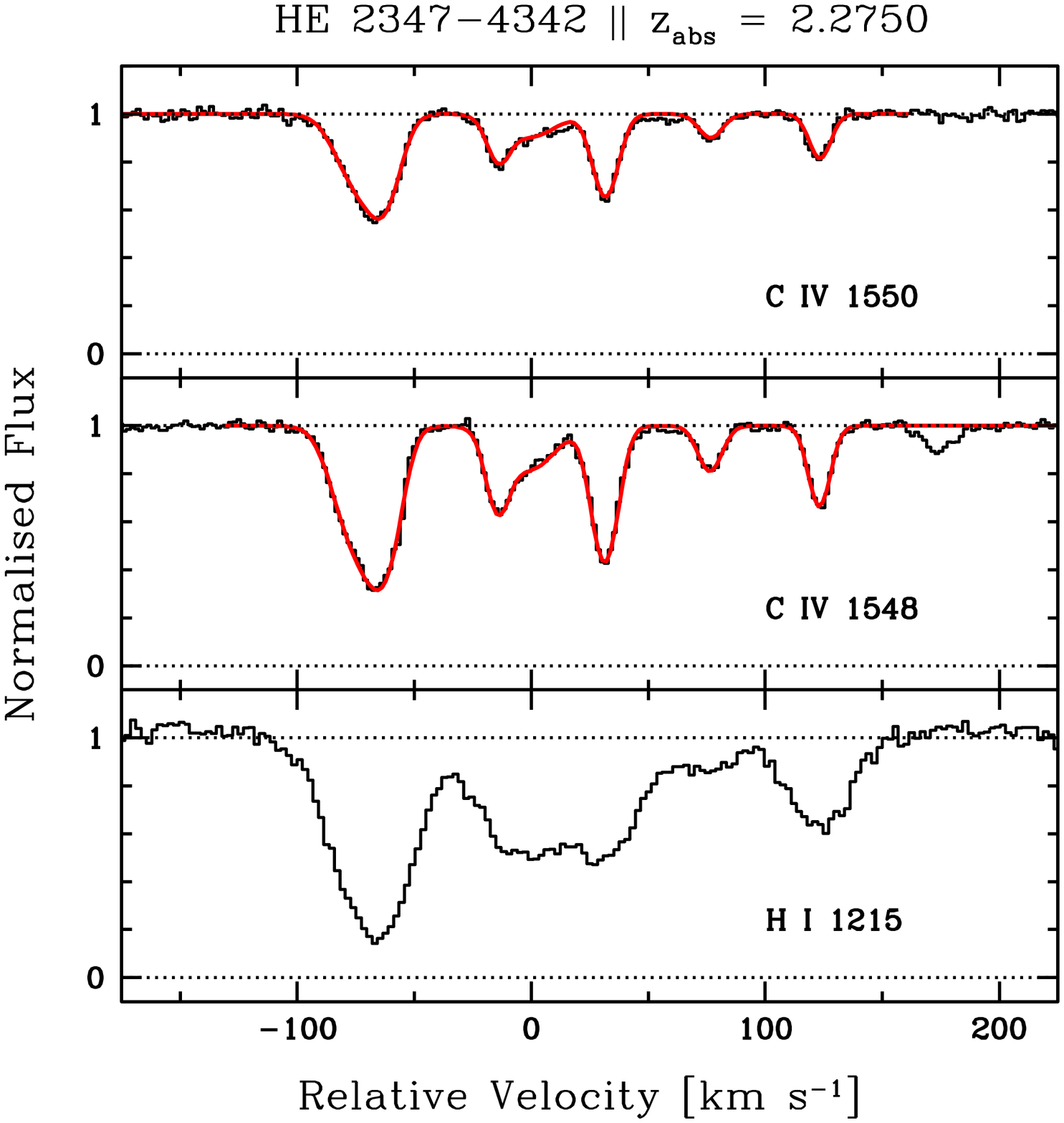}
\includegraphics[width= 6.0cm,height=4cm,angle= 0]{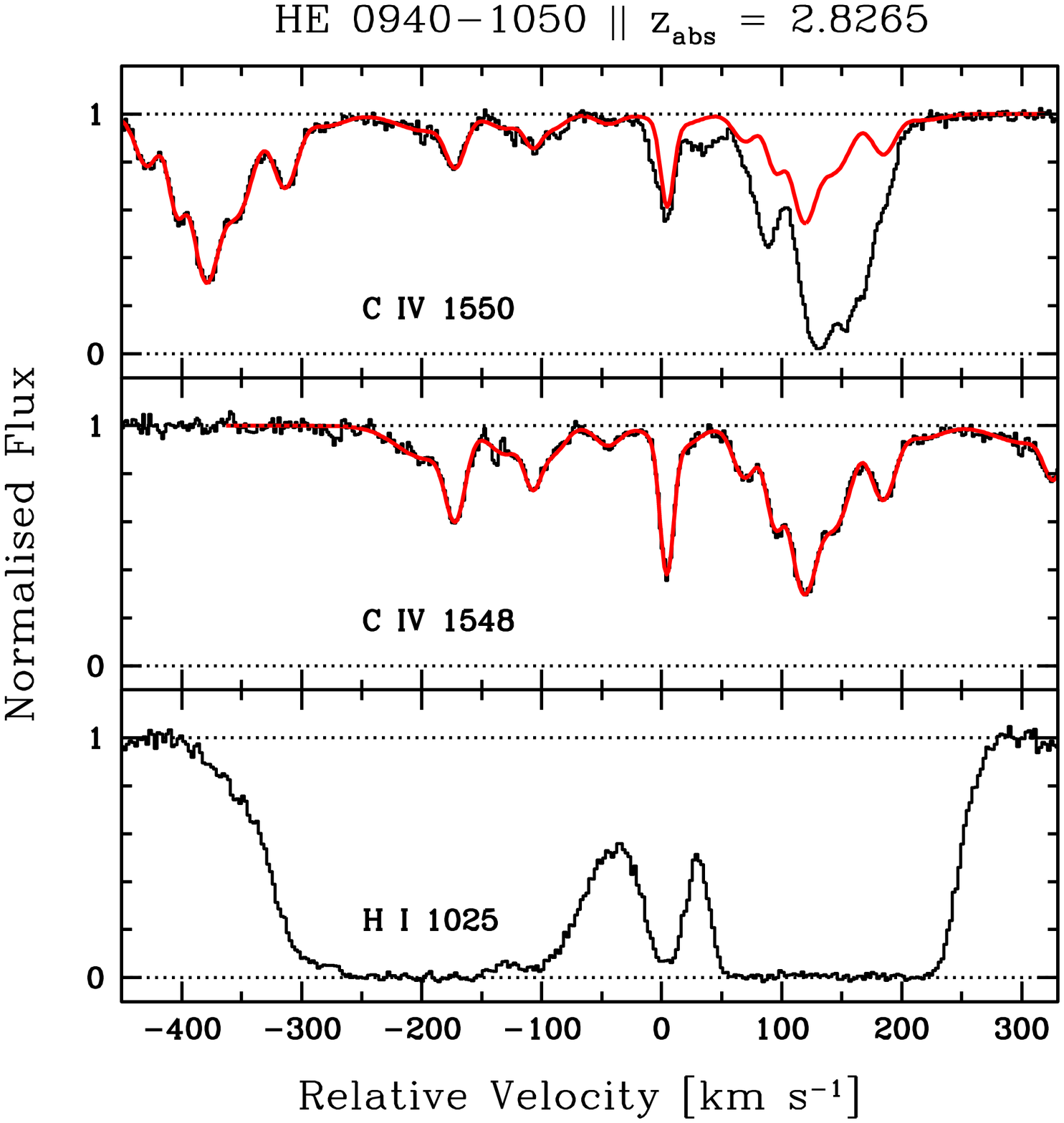}  
\includegraphics[width= 6.0cm,height=4cm,angle= 0]{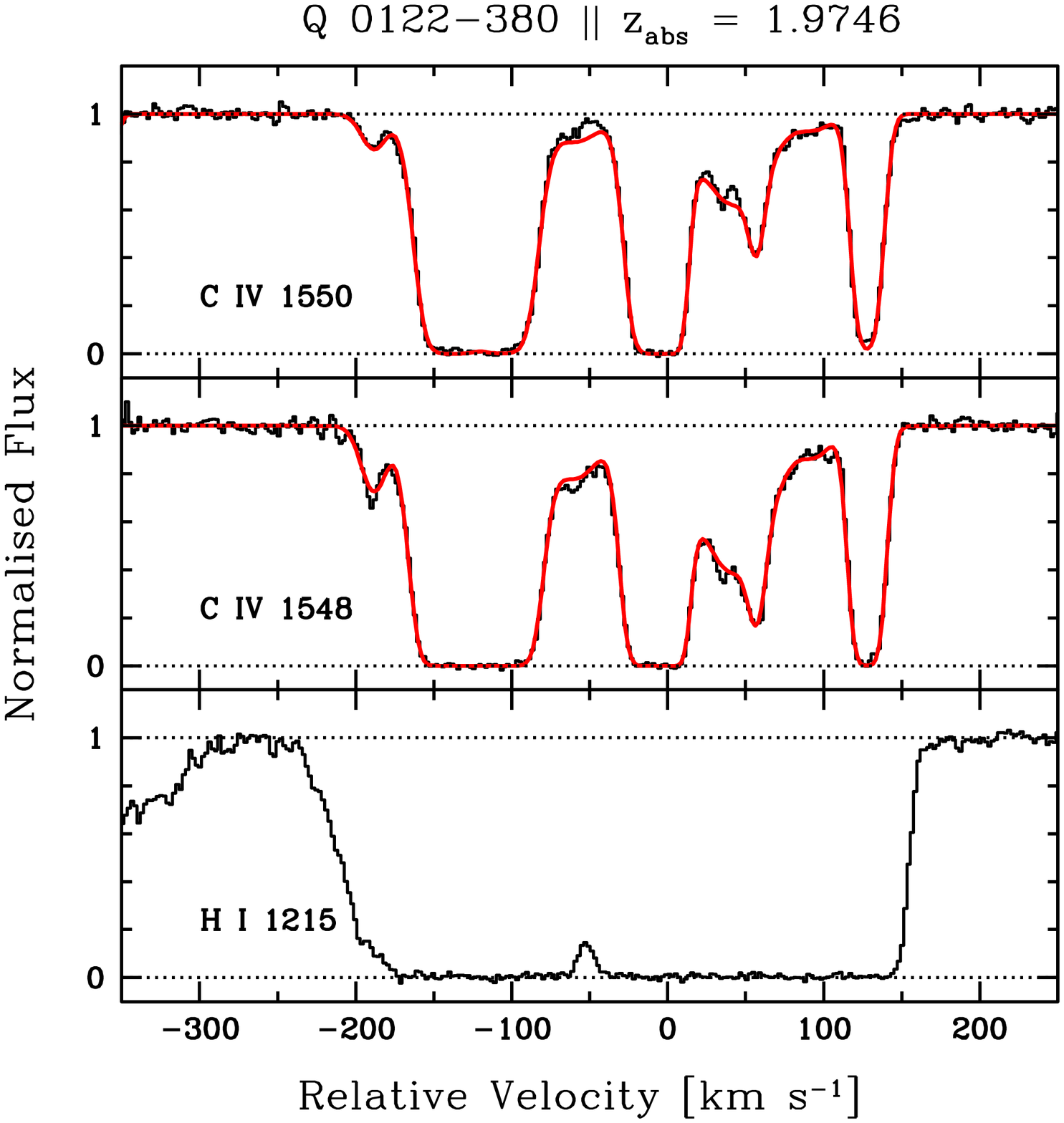}  
}}  
}}
\caption{Absorption profiles of systems with $\delta v(\os) > 200$~\kms\ 
(top panel) and $\delta v(\cf) > 200$~\kms\ (bottom panel) together with 
best fitted Voigt profile. The high velocity components identified with 
solid horizontal bars are showing high ionization.  
The absorption redshift that defines the zero velocity and 
the name of the background QSO are also indicated.
} 
\label{large_delv}
\end{figure*}

There are three \os\ (top panel of Fig.~\ref{large_delv}) and 
three \cf\ systems (bottom panel of Fig.~\ref{large_delv}) 
with $\delta v > 200$~\kms. 
These large velocity spreads could be related to either (a) large 
scale winds as seen in Lyman break galaxies 
\citep{Adelberger03,Adelberger05}, (b) redshift clustering of absorbing 
gas \citep[][]{Scannapieco06} or (c) mere chance coincidence 
of randomly distributed absorbers \citep[][]{Rauch97a}. 
The core of the \lya\ absorption in all three systems where 
$\delta v(\os) > 200$~\kms\ is highly saturated whereas the high velocity 
components show weak \lya\ absorption and relatively strong metal absorption. 
This could mean that the gas in these high velocity components 
is highly ionized and possibly of high metallicity.
On the other hand, if the three systems where $\delta v(\cf) > 200$~\kms\ 
follow the correlation seen in Fig.~\ref{delv_dist}, we expect them 
to have $\delta v(\os) > 200$~\kms. 
However, $\delta v(\os)$ could not be measured due to either low S/N 
(in two cases) or blending with strong Ly$\alpha$ lines. 
The system at \zabs = 2.2750 towards HE~2347$-$4342 is showing unsaturated 
well connected wide spread Ly$\alpha$ absorption with signature of high 
ionization. 
For the other two systems strong \cf\ is seen with heavily saturated \lya. 
The system at \zabs = 2.8265 towards HE~0940$-$1050 which shows largest 
velocity spread ($\delta v(\cf) \sim$ 375 \kms) for \cf\ in our 
sample is possibly associated with a Lyman break galaxy (LBG) 
at an impact parameter of 150~$\rm h^{-1}$ kpc \citep{Crighton11}. 
Hence it is extremely important to have a detailed spectroscopic survey 
of galaxies around the redshifts of such large $\delta v$ systems 
to understand the possible origin of \os\ absorbers. 
%

\subsection {Correlation of $\delta v$~with other parameters} 

\begin{figure*} 
\centerline{
\vbox{
\centerline{\hbox
{
\includegraphics[width= 8.4cm,angle= 0]{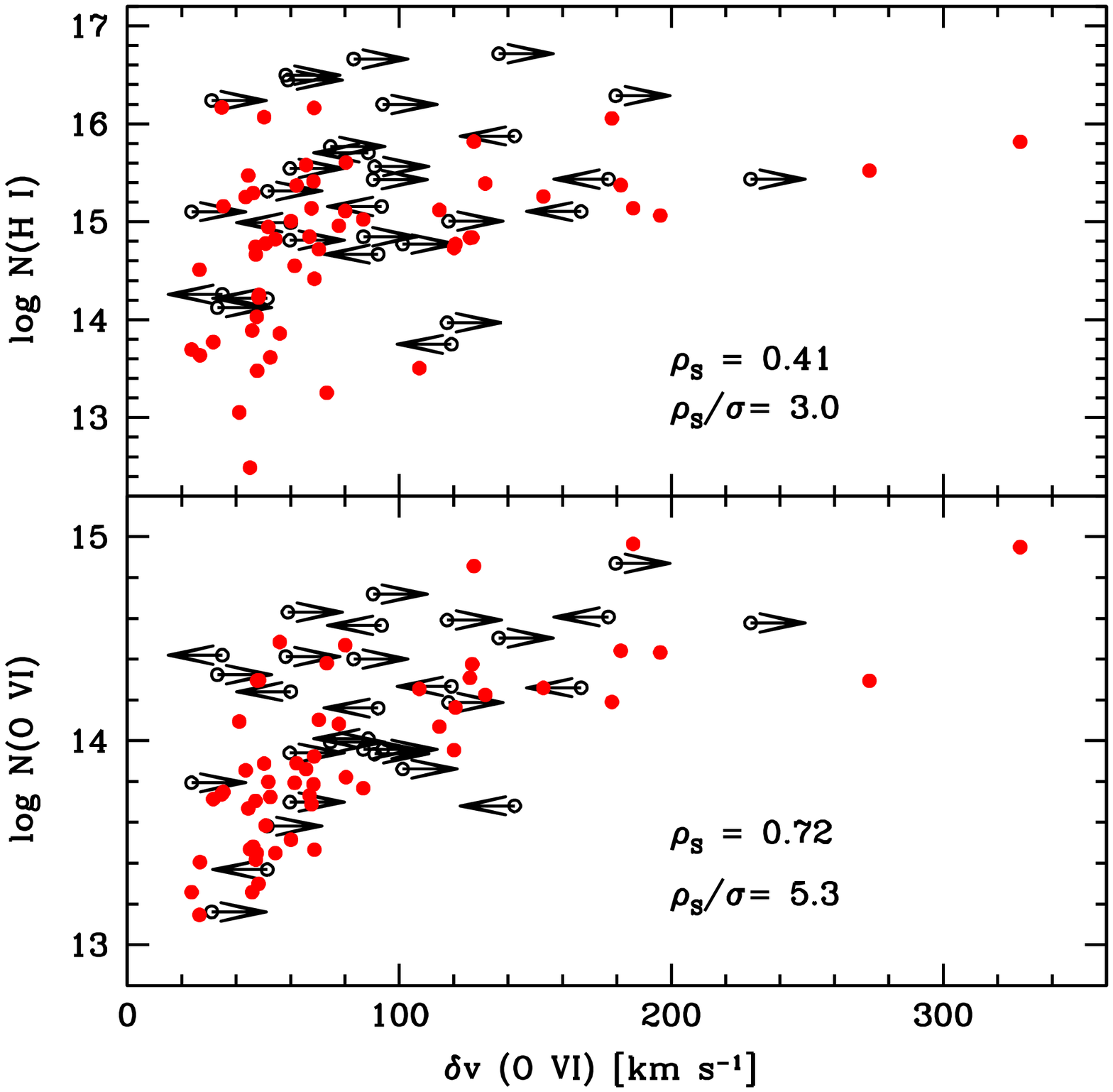}  
\includegraphics[width= 8.4cm,angle= 0]{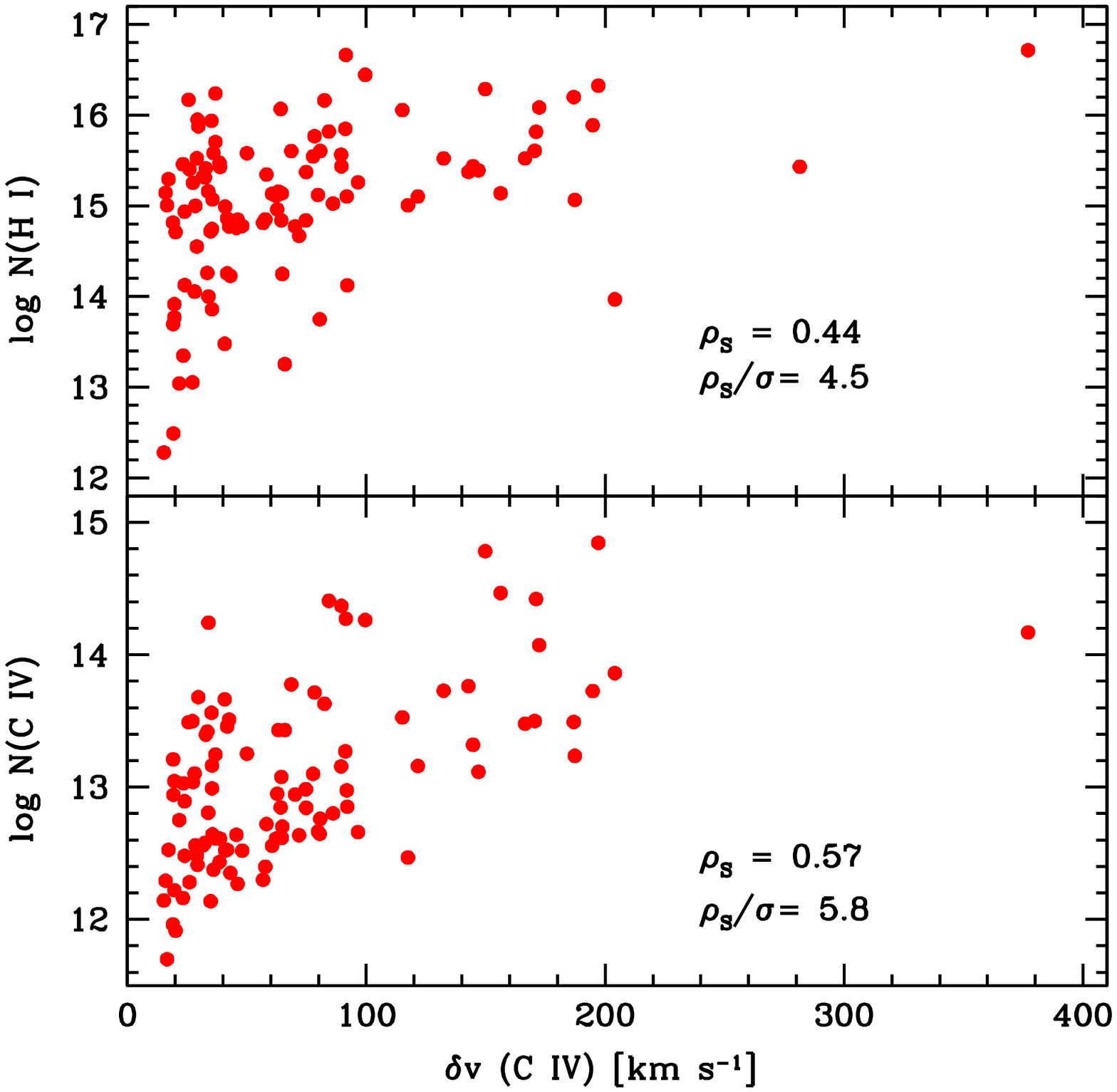}
}}
}}
\caption{Relationship between $\delta v$ and other measurable quantities.
	The results of Spearman test performed for the systems using only 
	the detections of ~$\delta v$ (filled circles) are mentioned 
	in each panel.}
\label{delv_OVI}
\end{figure*}

%
In this section we explore any possible correlation between $\delta v$ and 
other observable parameters (see Fig.~\ref{delv_OVI}). 
The strongest correlation is seen between $\delta v$(\os) and $N$(\os).  
The Spearman rank correlation coefficient is 0.72 and the correlation
is confirmed at 5.3$\sigma$ level. 
While systems with low $\delta v$ are seen over a wide range of 
$N$(\os), the systems with $\delta v\ge100$ \kms\ are seen when
log~$N$(\os) $\gtrsim 14$. Thus the above mentioned correlation 
is due to the lower envelop one can see in the figure.
As $\delta v$ is mainly related to the number of components, this 
lower envelop can not be attributed to a detection bias. This is confirmed 
by the presence of a significant correlation (i.e. $\rho_s = 0.64$ with a   
4.0$\sigma$ significance) even when we restrict our analysis to systems 
with log~$N$(\os)~$>$~13.7.  

Interestingly, a similar correlation is also seen between $N$(\cf) and 
$\delta v$(\cf) (i.e. $\rho_s = 0.57$ with a 5.8$\sigma$ significance). 
The correlation is significant even when we restrict our analysis to 
systems with log~$N$(\cf)~$>$~12.6 (i.e. $\rho_s = 0.42$ with a 
3.7$\sigma$ significance).
These results are consistent with what is seen in Fig.~9 of \citet{Songaila06},  
who found \cf\ systems with peak optical depth greater than 0.4 to show larger 
velocity extent compared to systems with lower peak optical 
depth. \citet{Songaila06} suggested that some of the wider \cf\ systems found 
among those with high peak optical depth could be associated with galaxy 
outflows.

We also find a 3$\sigma$ correlation between $\delta v(\os)$ and $N(\hi)$    
and a 4.5$\sigma$ correlation between $\delta v$(\cf) and $N(\hi)$. 
It is clear from Fig.~\ref{delv_OVI} that for both \os\ and \cf, 
$\delta v\ge 100$ \kms\ is seen mainly in systems with log~$N$(\hi) $\gtrsim$~15. 
Such high column densities are generally associated with high over density 
regions \citep[see e.g.,][]{Schaye01}. 
In the following section, we show that the systems with log~$N(\hi)>$~15.0 
have associated low ion absorption. Low ions usually trace higher density 
regions if the gas is photoionized by the meta-galactic UV radiation 
\citep[see e.g.,][]{Rauch97a}. All these suggest that the \os\ or \cf\ 
absorbing gas with high velocity spread is probably related to overdense 
regions. 
We do not find any strong correlation between $\delta v$(\os) or $\delta v$(\cf) 
with either redshift, $N$(\os)/$N$(\cf) or $N$(\os)/$N$(\hi). 
%

\begin{figure*} 
\centerline{
\vbox{
\centerline{\hbox{ 
\includegraphics[width= 6.0cm,angle= 0]{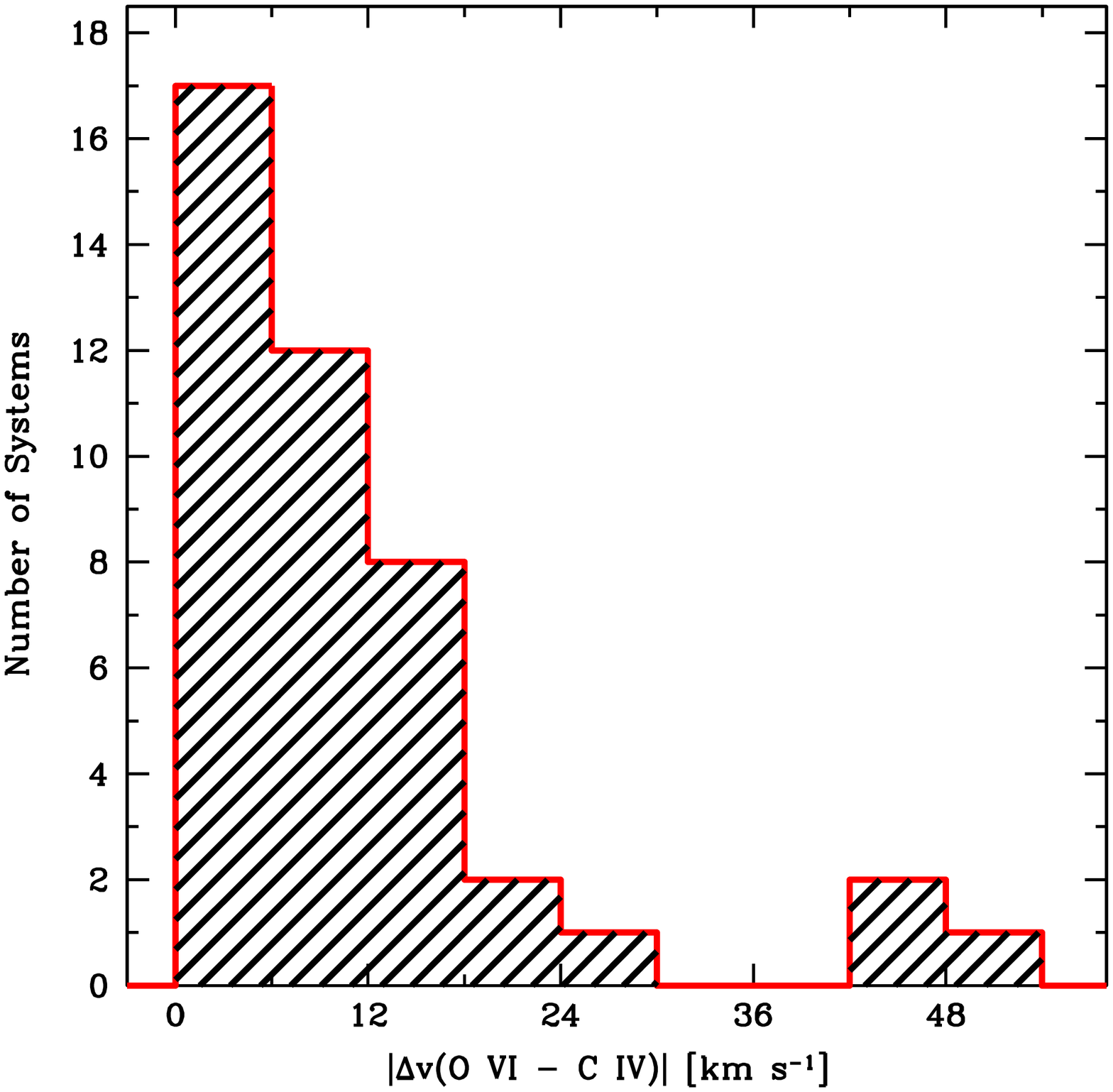}  
\includegraphics[width= 6.0cm,angle= 0]{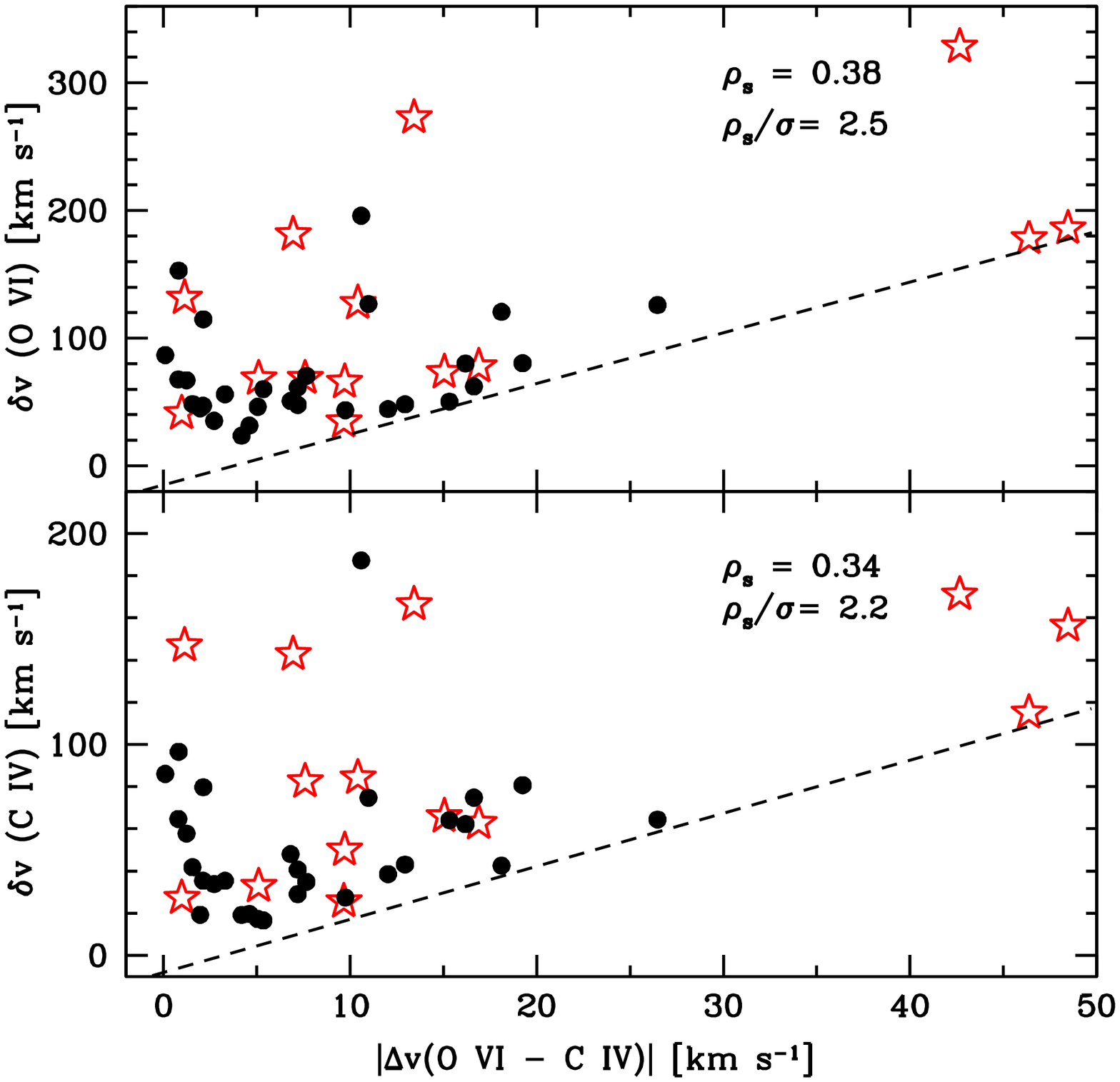}  
\includegraphics[width= 6.0cm,angle= 0]{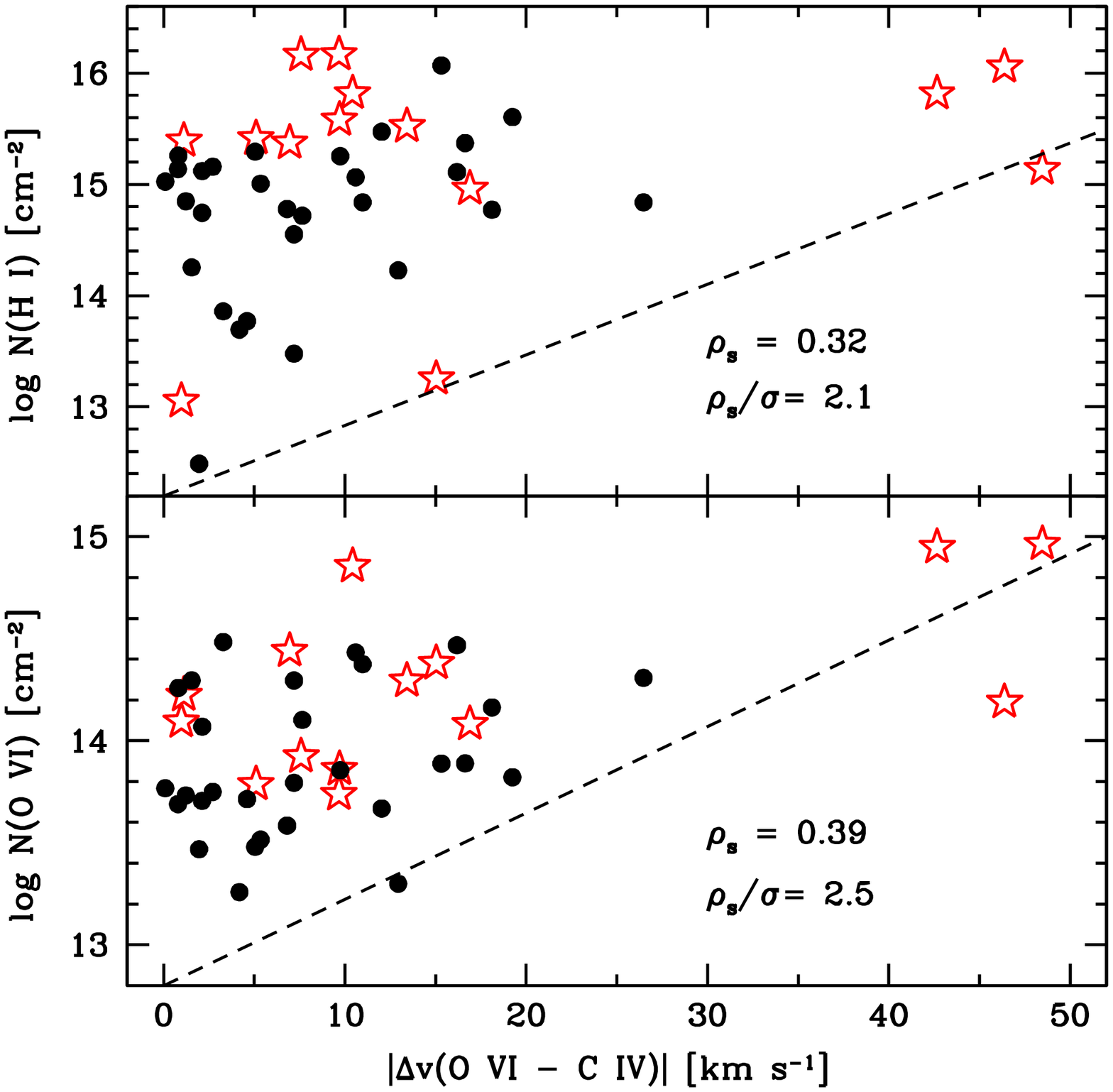}
}}
}}
\caption{{\it Left} : The distribution of the velocity shift between the 
optical depth weighted redshifts of \cf\ and \os.   
{\it Middle} : The velocity shift, $|\Delta v(\os - \cf)|$ versus 
$\delta v(\cf)$ (bottom) and $\delta v(\os)$ (top). 
{\it Right} : The velocity shift, $|\Delta v(\os - \cf)|$ versus 
$N(\os)$ (bottom) and $N(\hi)$ (top). The dashed lines in the middle and right 
panels are for illustrative purpose to emphasize the possible presence 
of a lower envelop. The (red) stars in both the panels represent systems 
with low ions. The results of Spearman rank correlation analysis for all the 
data points are also summarized in each panel.  
}
\label{Delv_N}
\end{figure*} 

\subsection{Velocity shift ($\Delta v$) distribution} 

In this section, we discuss the velocity offset between the \os\ and \cf\ 
absorption originating from the same system. 
Here, we do not attempt to estimate the offset between individual 
components as our main aim is to find the optical depth weighted phase separation 
between \cf\ and \os\ and how this is related to other measurable quantities. 
In most of the cases, the \os\ absorption is wider than the \cf\ one (see section 
\ref{lsd}) and has a larger number of components.  
We use only systems where both \cf\ and \os\ profiles are well defined 
at least in one of the doublets and calculate the optical depth 
weighted redshifts (i.e. $\bar{z}_{\os}$ and $\bar{z}_{\cf}$). The velocity 
shift between these two redshifts is then, 
\begin{equation} 
|\Delta v(\os - \cf)| = [|(\bar{z}_{\os}-\bar{z}_{\cf})|/(1+\bar{z}_{\cf})]\times c, 
\end{equation} 
where $c$ is the speed of light in \kms. The values of $|\Delta v(\os - \cf)|$
measurements are summarized in Table~\ref{allsystems}.
In the left most panel of Fig.~\ref{Delv_N}, we plot the distribution 
of $|\Delta v(\os - \cf)|$. The values of the velocity shift are found to 
be in the range  $0 \le |\Delta v(\os - \cf)| \le 48$ \kms\ with a median 
value of 8~\kms. We find that only $\sim$~9\% of the systems show 
$|\Delta v(\os - \cf)| > 20$ \kms\ whereas $\sim$~33\% of the systems have 
$|\Delta v(\os - \cf)| \le 5$ \kms. 
It is to be remembered that the systems with large variation in 
$N(\os)/N(\cf)$ ratio from one component to another tend to 
have large values of $|\Delta v(\os - \cf)|$. Hence $|\Delta v(\os - \cf)|$ 
can be seen as a measure of the ionization inhomogeneity among components. 
Our analysis suggests that \os\ systems with large ionization 
inhomogeneities across the profile are rare. In addition, the median 
value of the velocity shift distribution for systems with low ions 
($\sim10.4$~\kms) is higher than that for systems without low ions 
($\sim6.8$~\kms). Hence, the ionization inhomogeneity seems to be more 
important in the systems where low ions are detected. 
But due to the small number of data points the 
KS test do not show any significant difference between the shift 
distributions for systems with and without low ions. 
We find a mild (at $\sim2.5\sigma$ significance level) correlation between 
either $\delta v(\os)$ or 
$N$(\os) and $|\Delta v(\os - \cf)|$ (see Fig.~\ref{Delv_N}). 
These correlations are dominated by the fact 
that systems with  $|\Delta v(\os - \cf)| \ge 20$ \kms~are predominantly coming 
from systems with $\delta v(\os)\ge$~100~\kms\ and 
log~$N$(\os)(cm$^{-2}$)~$\ge$~14. 
A similar trend is also seen for $\delta v (\cf)$ and $N$(\hi) 
possibly due to the presence of 
a lower envelop. However, we find none of the other parameters 
(i.e., $z$, $N(\os)/N(\hi)$, $N(\cf)/N(\hi)$ and $N(\os)/N(\cf)$) showing 
any significant correlation with  $|\Delta v(\os - \cf)|$.  
%

\section{Analysis based on total column densities} 
\label{model}
Here we study the distributions of measured column densities of 
different species, their ratios and dependencies between 
them. Upper limits are not considered for the analysis.

\subsection{Redshift evolution of column density ratios} 
\begin{figure*} 
\centerline{
\vbox{
\centerline{\hbox{ 
\includegraphics[height=6.5cm,width= 6.0cm,angle= 0]{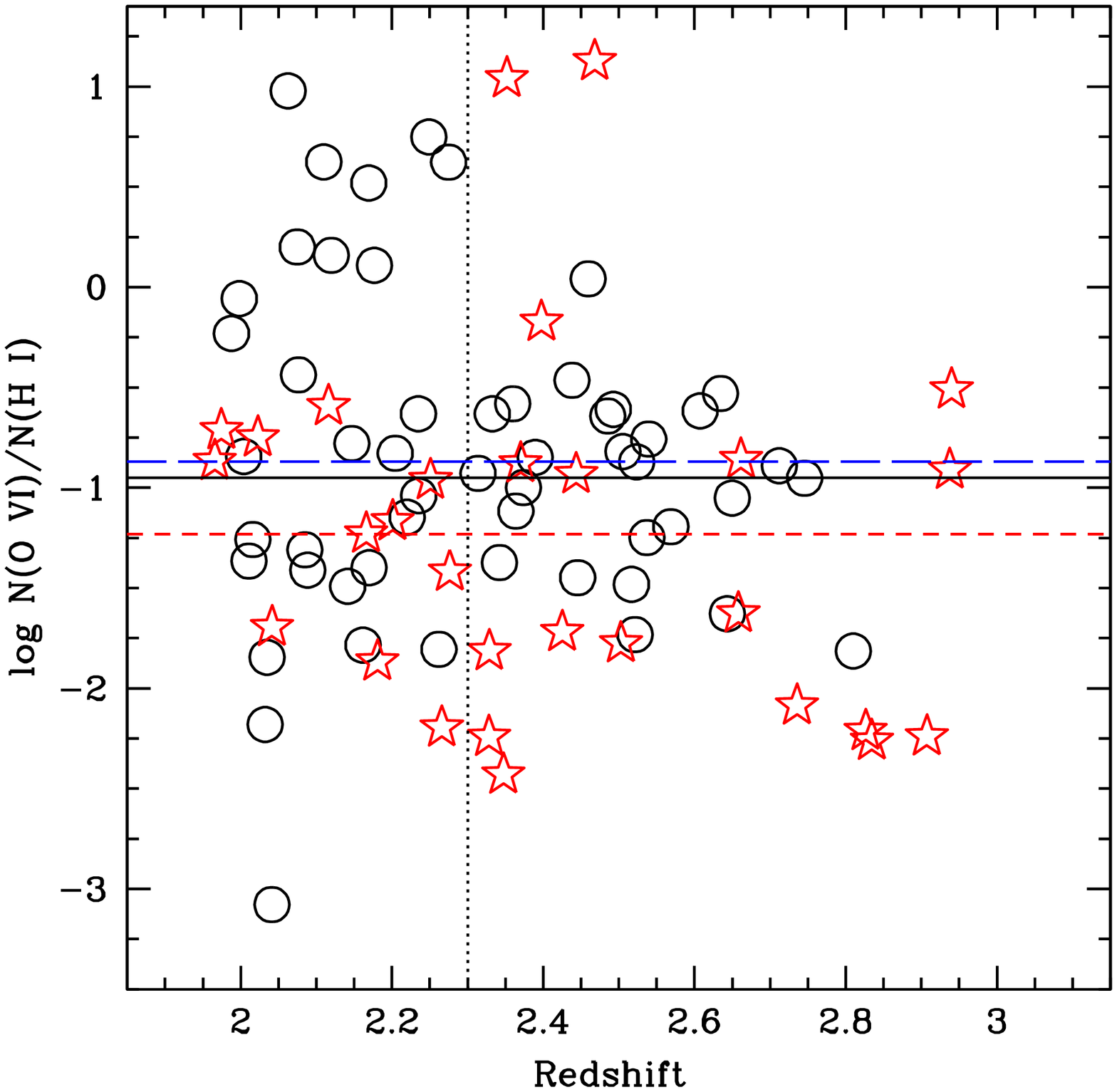}  
\includegraphics[height=6.5cm,width= 6.0cm,angle= 0]{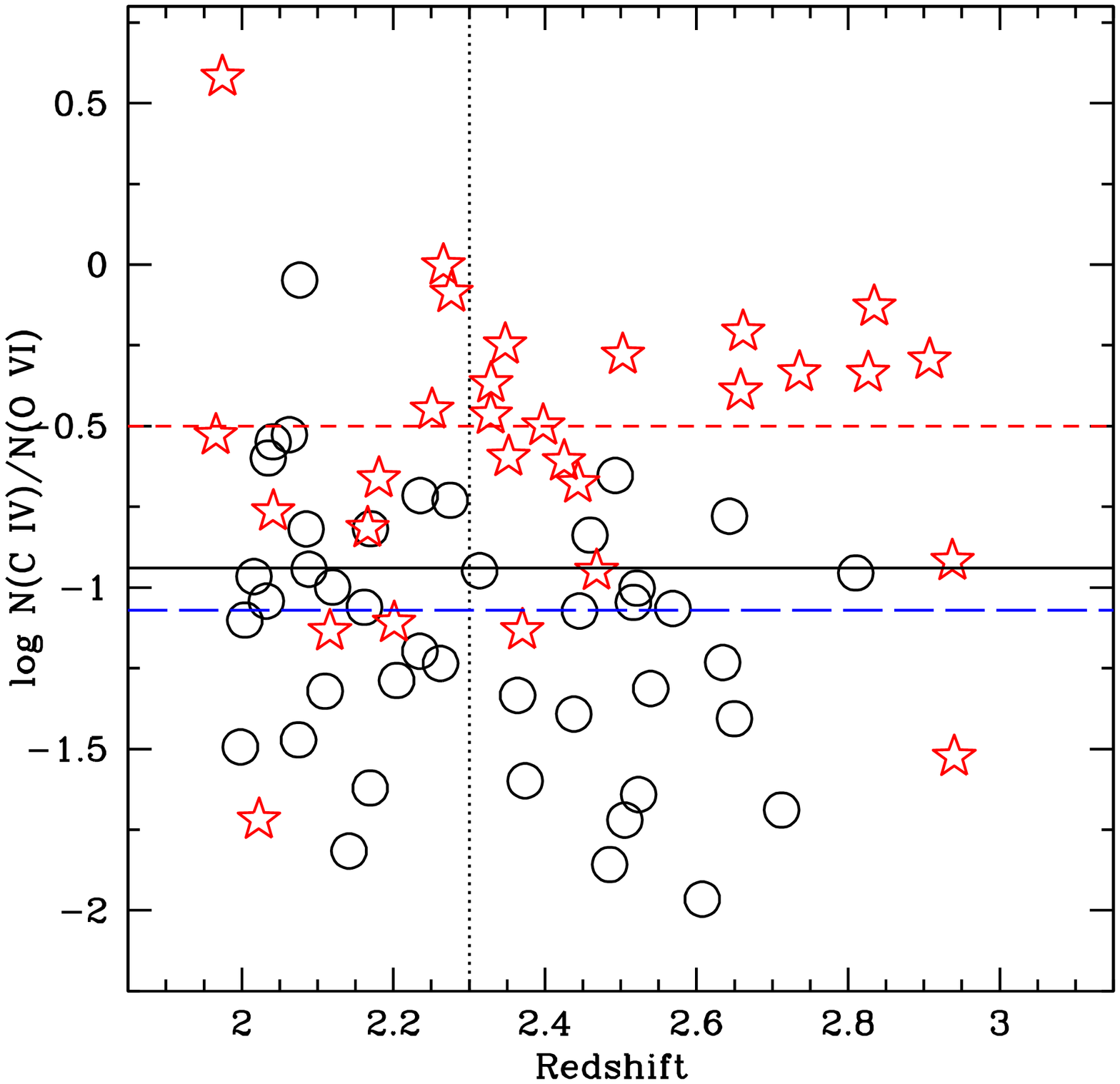}  
\includegraphics[height=6.5cm,width= 6.0cm,angle= 0]{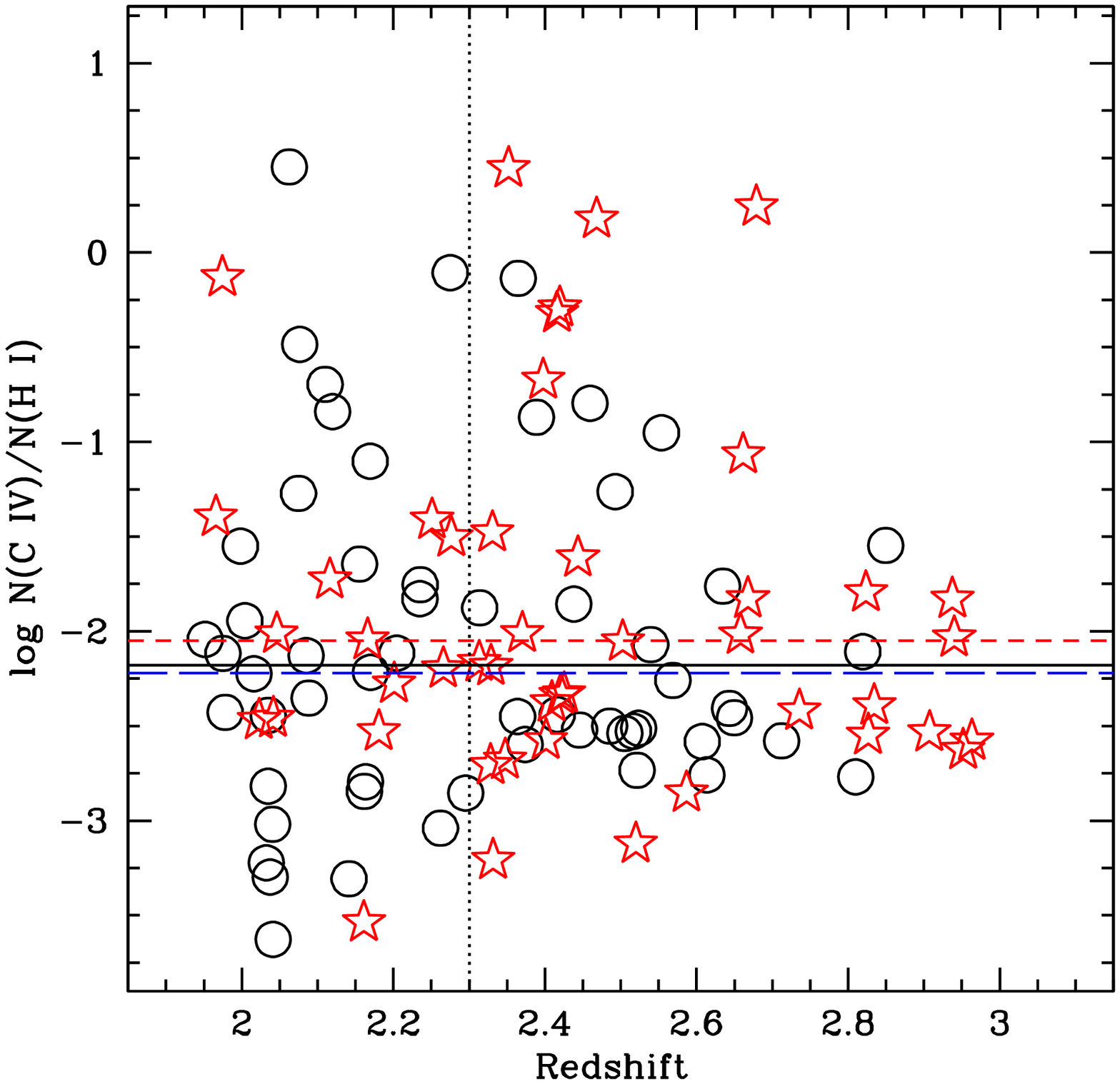}  
}}
}}
\caption{ Redshift evolution of \os\ to \hi\ ({\it Left}) 
	      \cf\ to \os\ ({\it Middle}) and \cf\ to \hi\ ({\it Right}) 
	      column density ratio.        
	      In all the panels the (red) stars are the systems with low ions 
	      and (black) open circles are the systems without any 
	      low ions. The vertical dotted line shows the median 
	      redshift of our sample. The median values of column density ratio 
	      for the full sample (solid horizontal line), systems with low ions 
	      (short dashed line) and systems without low ions (long dashed line) 
	      are also indicated in each panel.  
}
\label{red_evlv} 
\end{figure*} 
%

In Fig.~\ref{red_evlv} we plot the redshift evolution of various column 
density ratios. In all the panels (red) stars represent systems with  
detectable low ions and (black) open circles are for systems without low ions. 

The Spearman rank correlation analysis performed between various column 
density ratios and redshift do not reveal any significant correlation. 
We also find that the presence of low ions does not influence this result. 
One of the interesting features of the left most panel in Fig.~\ref{red_evlv}  
is the scarcity of data points with log~$N(\os)/N(\hi)>$~$-$0.5 for 
$z >$~2.5. \citet{Bergeron05} identified such systems as 
a separate population of \os\ absorbers with high metallicity 
\citep[see also][]{Schaye07}. Such high values of $N(\os)/N(\hi)$ ratio are 
also seen in proximate \os\ absorbers \citep[][]{Fox08,Tripp08}. Therefore,  
these absorbers may either trace high metallicity gas or regions ionized 
by AGN like sources. Hence, confirming the redshift evolution of these 
absorbers will be very important. It is to be noted that similar trend 
(i.e., lack of data points with high $N(\cf)$ to $N(\hi)$ ratio for $z>$ 2.5) is 
apparent from the right most panel of Fig.~\ref{red_evlv}.  
The median values of various column density ratios are also shown in 
Fig.~\ref{red_evlv} with horizontal lines. The difference of median 
values for systems with and without low ions is maximum for 
$N(\cf)/N(\os)$ ratio (factor~$\sim$~4). A two sided KS test shows 
that their distributions are different with very high significance 
(D = 0.55 \& Prob. = 2.6$\times10^{-5}$). 
%

\subsection{Relationship between column densities of various species 
and viability of photoionization model}  

In this section, we investigate the total column densities of various 
species and any correlation between them. We compare these with 
simple photoionization model predictions using ``CLOUDY v(07.02)'' 
\citep[]{Ferland98}. We would like to point out here that such a single phase 
photoionization model may oversimplify the problem in view of the  
possible multiphase structure of the absorbing gas, but such models are often 
used as a guideline to draw broad conclusions on the nature of 
the \os\ absorbers    
\citep[e.g. see,][]{Simcoe02,Carswell02,Bergeron02,Bergeron05,Schaye07,Muzahid11}. 

The model assumes the absorbing gas to be an optically 
thin plane parallel slab illuminated by the meta-galactic UV background 
radiation contributed by QSOs and Lyman break galaxies at the median redshift 
(i.e. $z$ = 2.3) of our survey. We use ``HM05" background radiation available 
in CLOUDY v(07.02) based on the UV spectrum calculated using the method 
described in \citet{Haardt96}.  
For simplicity, we assumed solar relative abundances and run our model 
with two different values of metallicities (i.e. $Z$ = 0.1 and 0.01 
$Z_{\odot}$). In this model, we incorporate the relation between the 
particle density and the observed \hi\ column density as given 
in \citet{Schaye01}, i.e., 
\begin{eqnarray} 
N(\hi)&\sim& 2.7\times10^{13} {\rm cm^{-2}} (1+\delta)^{3/2} 
T_{4}^{-0.26}~\Gamma_{12}^{-1} \nonumber \\ 
&&\times \left(\frac{1+z}{4}\right)^{9/2}
\left(\frac{\Omega_{\rm b}h^{2}}{0.02}\right)^{3/2} 
\left(\frac{f_{\rm g}}{0.16}\right)^{1/2} \times \kappa, 
\label{nh-NH}
\end{eqnarray} 
where, $\delta$~denotes over-density, $f_g$ is the gas mass fraction  
normalised to it's universal value (i.e. $\Omega_b/\Omega_m$). 
All other symbols have their usual meaning. We have introduced a fudge 
factor called $\kappa$ and run this model for three different values of 
$\kappa$ (0.5, 1 and 2). For the original equation of \citet{Schaye01}, $\kappa=1$. 
Hence, for a given $N(\hi)$, $\kappa~ > 1$ implies lower density (higher ionization 
parameter) compared to what is predicted by \citet{Schaye01}. The temperature 
is calculated from the standard $T-\delta$ relation given by \citet[]{Hui97} i.e. 
$T = T_{0}~\delta^{\gamma -1}$ with $T_0 = 2\times10^4$~K and $\gamma = 1.1$. 
This temperature is very close to the photoionization temperature for the density 
range of interest here.

\begin{figure*} 
\centerline{\vbox{
\centerline{\hbox{ 
\includegraphics[height=5.2cm,width= 4.5cm,angle= 0]{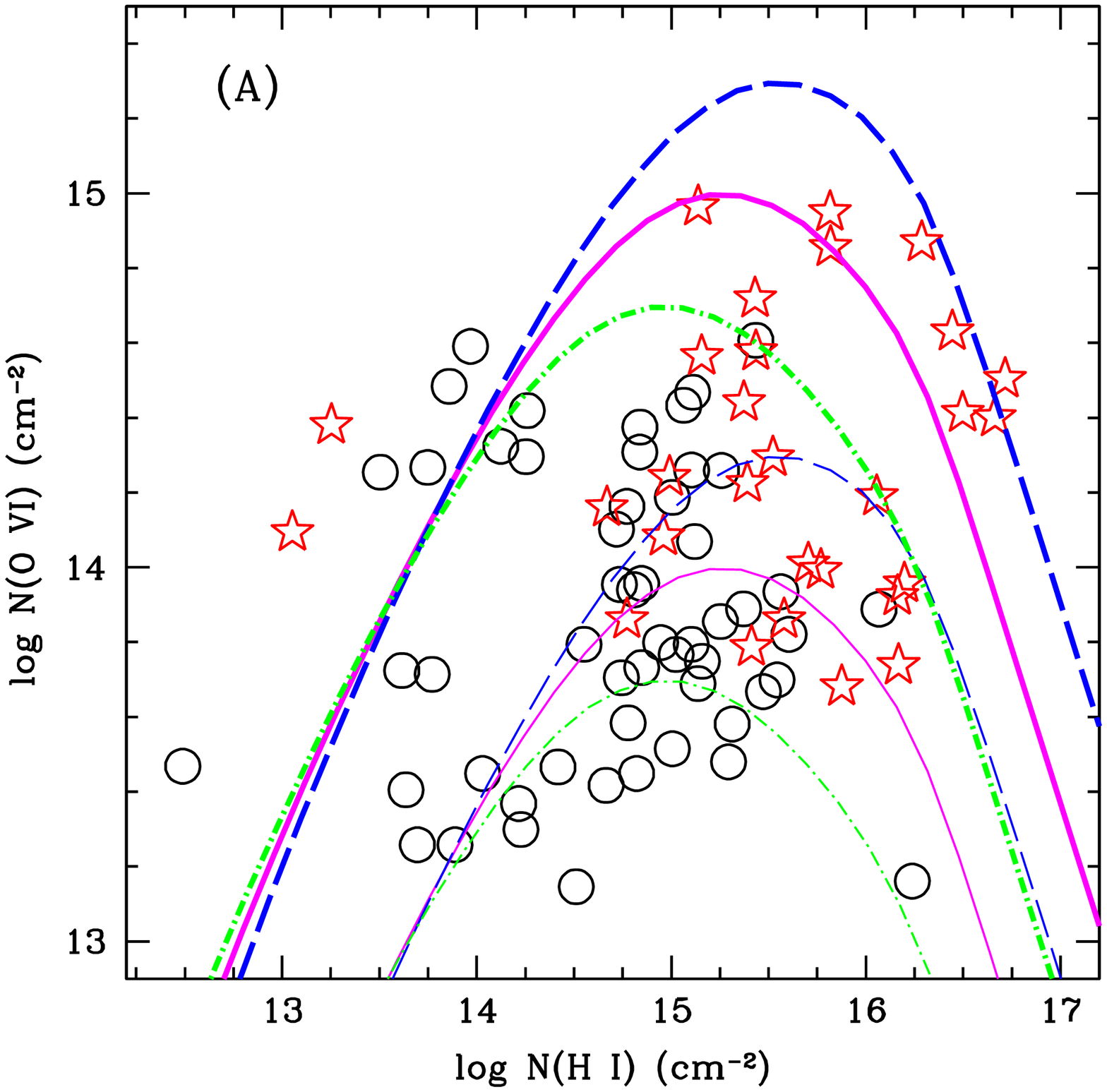}  
\includegraphics[height=5.2cm,width= 4.5cm,angle= 0]{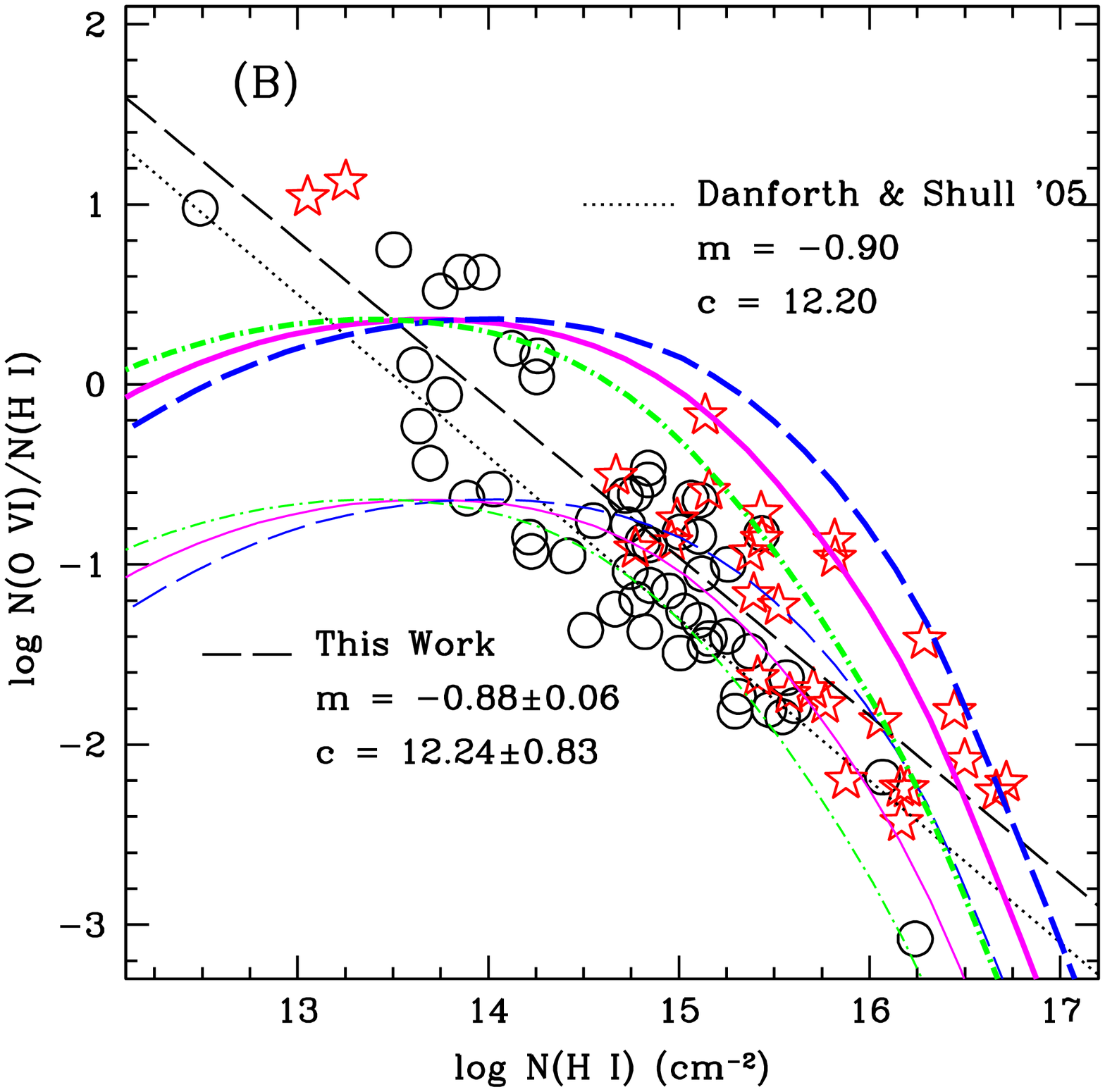} 
\includegraphics[height=5.2cm,width= 4.5cm,angle= 0]{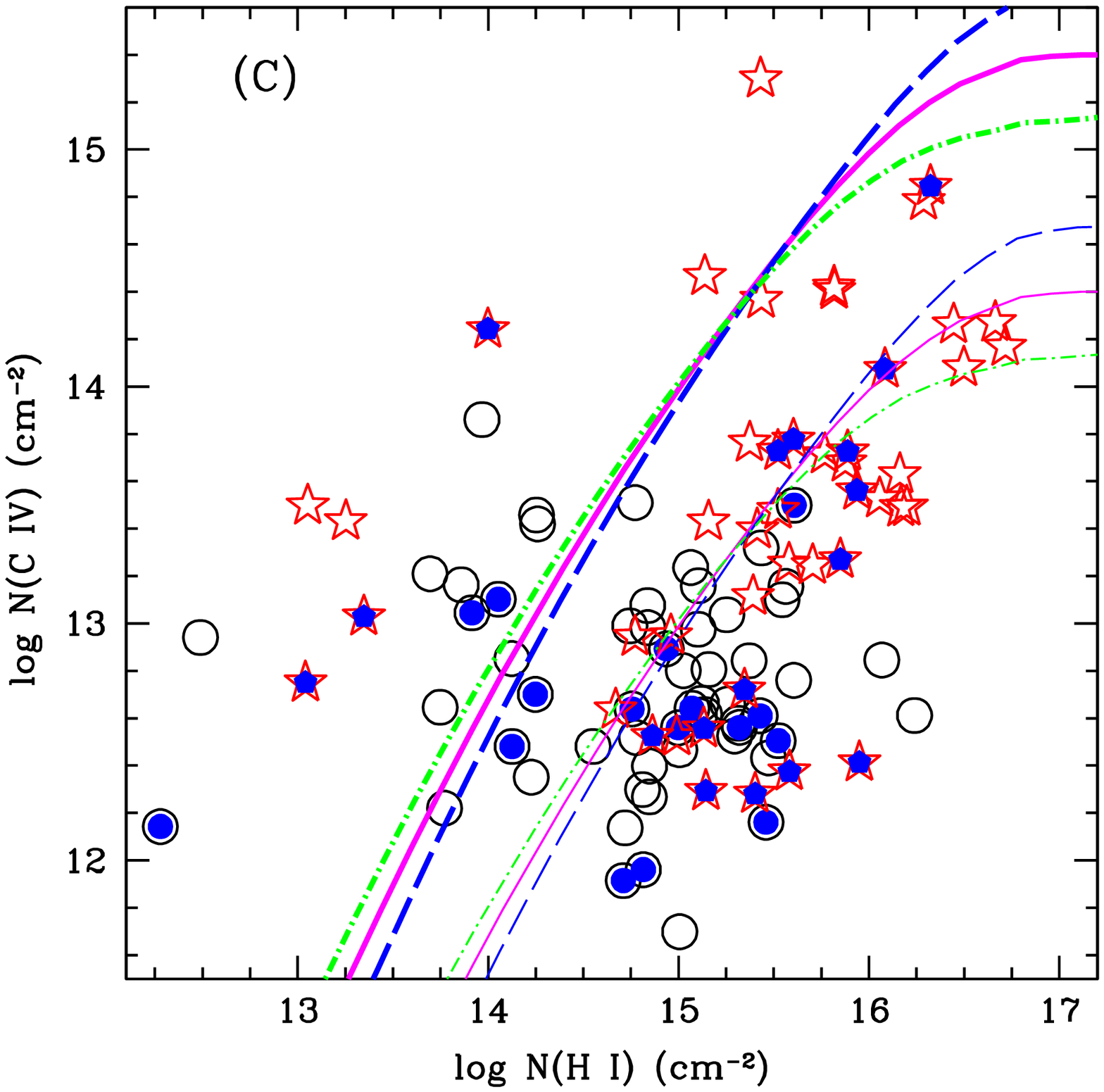}
\includegraphics[height=5.2cm,width= 4.5cm,angle= 0]{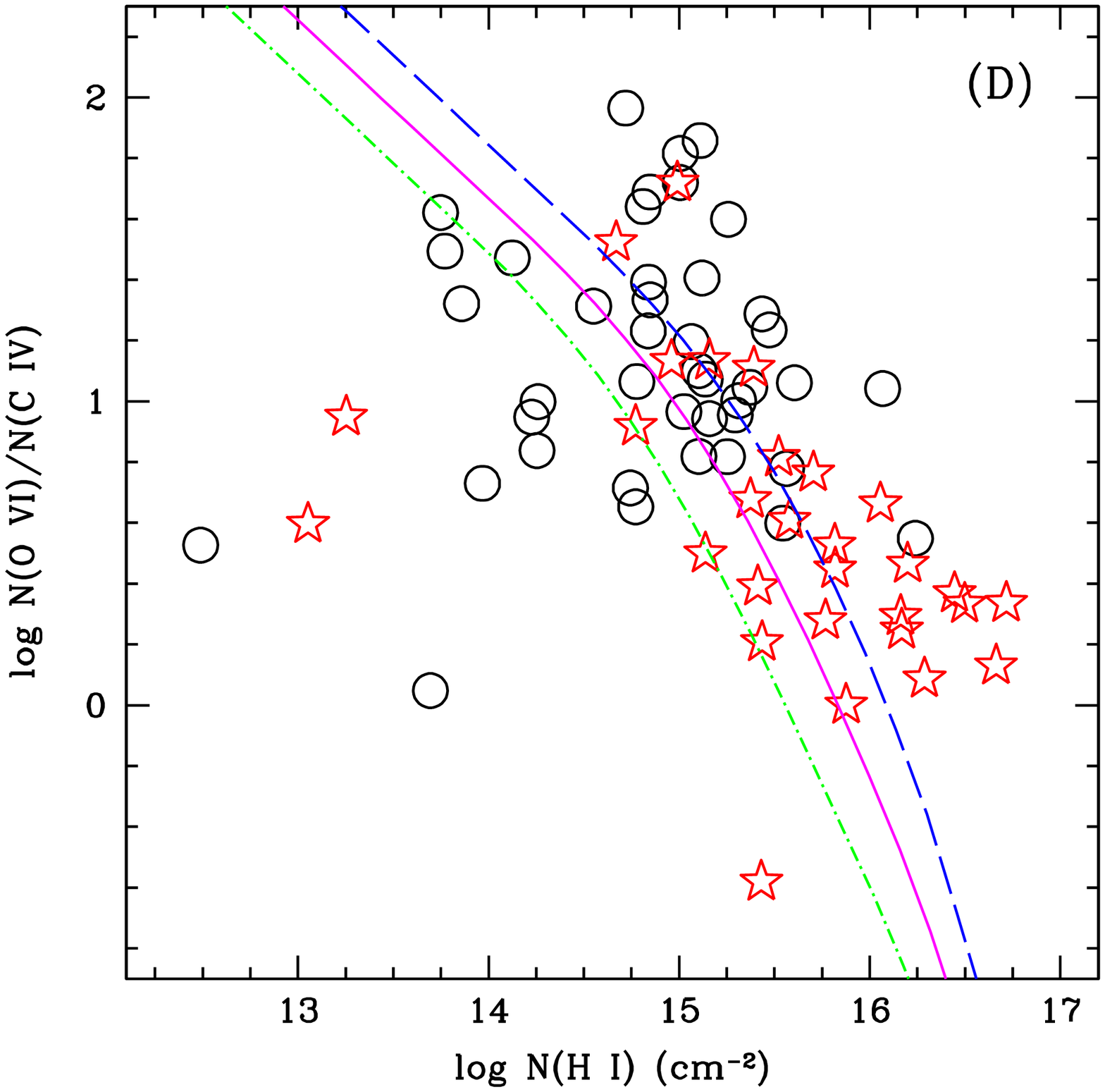} 
}} 
}} 
\caption{{\bf(A)} \os\ column density against \hi\ column density. 
In all panels the (red) stars and the (black) circles represent 
systems with and without low ions respectively.  
The dot-dashed, solid, and long-dashed curves indicate the results of 
the photoionization model for $\kappa$~=~0.5, 1.0, and 2.0 respectively. 
The thin and thick curves correspond to metallicities 
$0.01Z_{\odot}$ and $0.1Z_{\odot}$.
{\bf (B)} \os\ to \hi\ column density ratios in \os\ systems as a function 
of $N$(\hi). The long-dashed line shows the power law fit to our data. 
The dotted line shows the fit obtained by \citet{Danforth05} for their 
low-$z$ \os\ sample.  
{\bf (C)} \cf\ column density against \hi\ column density. The shaded ones 
are the systems where only upper limits on $N$(\os) can be estimated. 
{\bf (D)}~ $N(\os)/N(\cf)$ column density ratio against $N(\hi)$. Since 
the $N(\os)$ to $N(\cf)$ column density ratio is independent of metallicity 
we show the curves for only $Z = 0.01Z_{\odot}$ . 
}
\label{model1}
\end{figure*} 

\subsubsection{$N(\os)$ vs $N(\hi)$} 

In the panel-(A) of Fig.~\ref{model1} we plot $N(\os)$ against $N(\hi)$ for 
individual \os\ systems. Note that $N(\os)$ is varying   
only $\sim$~2 dex over a $\sim$~5 dex variation in $N(\hi)$. 
The Spearman rank correlation analysis shows only a mild correlation 
(at $\sim 2.5\sigma$ level) between $N$(\hi) and $N$(\os) when we consider 
the full sample. 
The subsamples of systems with and without low ions do not show any 
trend between $N$(\hi) and $N$(\os) individually.   

It is apparent from the figure, that apart from two 
systems, low ions are detected in systems with higher \hi\ column density 
(i.e. log~$N(\hi) > 14.6$). The median values of log~$N(\hi)$ are 
15.58 and 14.82 for systems with and without low ions respectively. 
The KS test shows that the distributions of $N(\hi)$ in systems with and 
without low ions are different with very high significance 
(D = 0.52 and Prob. = 2.1$\times10^{-7}$). It is interesting to note 
that 36 out of 46 systems where we detect low ions show log~$N(\hi)>15.0$. 
This could mean that $N(\hi)$ is predominantly 
coming from the low ionization phase of the absorbing gas whenever low 
ions are detected. This is why we do not use systems with 
low ions (even when metals are aligned with \hi) when we measure thermal 
and non-thermal contribution to the $b$-parameter in section~\ref{bnt_T}. 
Although there is no statistically significant trend between 
$N(\os)$ and $N(\hi)$, the \os\ column density seems to be 
systematically higher when low ions are present. 
The median value of log~$N(\os)$ for systems with low ions is found 
to be 0.4 dex higher compared to that of the systems without low ions. 
A two sided KS test shows that the $N(\os)$ distribution between systems 
with and without low ions are indeed different 
(D = 0.44 and Prob. = $7.3\times10^{-4}$). 

It is evident from the model curves that the systems with log~$N(\hi)>14.5$  
can be roughly reproduced by our simple model for metallicity ranging from 
0.01 to 0.1~$Z_{\odot}$. The systems in the top-right corner of the figure 
can also be reproduced with higher $\kappa$ values (i.e. $\kappa \gtrsim 5$) 
and with low metallicity (i.e.~$\sim$~0.01$Z_{\odot}$). 
It is to be remembered that associating entire \hi\ column density 
measured in a system 
to the \os\ bearing phase will essentially underestimate the metallicity. 
If, in case of systems with low ions, most of the \hi\ comes from the 
lower ionization phase then the (red) stars will move towards the left in 
this plot requiring higher metallicities.  
Most interestingly,  
the systems with log~$N(\hi)<$~14.5 and log~$N(\os)>14.0$ can not be 
reproduced by our simple model with metallicity $Z = 0.01 Z_{\odot}$ 
(thin curves) irrespective of the $\kappa$ values used. These systems 
require metallicity $\gtrsim 0.1 Z_{\odot}$ provided the ionization is 
dominated by the meta-galactic UV background. 

In the low redshift studies, the $N$(\hi)/$N$(\os) vs $N$(\hi) plot is 
used as an indicator of multiphase medium. For collisional ionization, 
the \os\ ionization fraction peaks at 
$\sim$~10$^{5.5\pm0.3}$~K \citep[see][]{Sutherland93, Gnat07} 
and hence \os\ is the most promising species in the UV-optical regime 
to probe the hard-to-detect hot gas phase in the intergalactic medium, 
namely, the WHIM. On the other hand most of the \lya\ lines are believed 
to trace relatively cool ($T$$\sim$ few $\times10^{4}$~K) 
photoionized gas, namely, the warm ionized medium. To investigate the 
relative importance of the warm photoionized and hot collisionally 
ionized gas, it is customary to plot the ``multiphase ratio'' i.e. 
$N(\hi)/N(\os)$ against $N(\hi)$ \citep[see][]{Shull03,Danforth05}. 
Panel-(B) of Fig.~\ref{model1} is a very similar plot and it reveals a strong 
anti-correlation between $N(\os)/N(\hi)$ and $N(\hi)$. This is not very 
surprising because it is the manifestation of $N(\os)$ being not strongly 
correlated with $N(\hi)$. Note that the best fit straight line to our data 
(dashed line) gives a slope and intercept very similar to what 
is obtained at low redshift (dotted line) by \citet{Danforth05}. 
We find that the systems with low ions show slightly higher 
$N(\os)/N(\hi)$ value for a given $N(\hi)$. However, the slope of the 
best fitted straight lines for systems with and without low ions are 
very similar. 
%

\subsubsection{$N(\cf)$ vs $N(\hi)$} 

Unlike $N$(\os), $N$(\cf) seems to be correlated with $N(\hi)$
(see panel-(C) of Fig.~\ref{model1}). 
A Spearman rank correlation analysis suggests a correlation with rank coefficient 
$\rho_{\rm s}$~=~0.41 and a 4.2$\sigma$ significance level. 
We find that this correlation is mainly dominated by the systems with 
low ions. In fact no trend is seen between $N$(\cf) and $N$(\hi) for systems 
without low ions.  
The shaded points in this panel represent the systems where we have upper 
limits on \os\ column density as listed in Table~\ref{upp_lim}, and 
they are predominantly lying in the  
lower half plane. The median value of log~$N(\cf)$ (i.e. 12.64) in these 
systems is 0.5 dex less than that for the rest of the systems. 
A two sided KS test indeed shows that the distribution of these points 
are different from the rest of the data points with $\sim$~98\% probability. 
The non-detection of \os\ in most of these systems is consistent 
with the average $N(\os)$/$N(\cf)$ ratio in our sample. 

The trend of increasing $N(\cf)$ with $N(\hi)$ is well reproduced by our 
simple model. Unlike in the case of $N$(\os), the difference in the predicted 
curves at a given $N$(\hi) and metallicity is very small for the range of 
$\kappa$ used in our models (in particular for $14\le$~log~$N$(\hi)~$\le 16$). 
Again here, most of the systems with log~$N(\hi)>$~14.5 can be well 
accommodated with the model curves for metallicity ranging from 
0.001 to 0.1 $Z_{\odot}$. On the other hand systems with log~$N(\hi)<$~14.5 
and log~$N(\cf)\gtrsim13$ require metallicity $Z \gtrsim 0.1 Z_{\odot}$ 
irrespective of $\kappa$.  
%

\subsubsection{$N$(\os)/$N$(\cf) vs $N$(\hi)} 

In panel-(D) of Fig.~\ref{model1}, the $N(\os)$/$N(\cf)$ ratio 
is plotted against the \hi\ column density. A strong 
anti-correlation ($\sim~4.3\sigma$ level) is seen between 
$N(\os)/N(\cf)$ and $N(\hi)$. Since $N(\os)$ shows a mild correlation 
whereas $N(\cf)$ shows strong correlation with $N(\hi)$, such 
anti-correlation is expected. 
It is to be noted that the anti-correlation seen in the full sample is 
again dominated by the systems with low ions. No trend is suggested by 
Spearman rank correlation analysis for the systems without low ions. 
As the \os\ to \cf\ column density ratio mainly depends on the ionization 
parameter in this single phase model both the thick and thin curves 
fall on top of each other. Note that the observed anti-correlation 
is well reproduced by our model apart from a few systems with very 
low $N(\hi)$.

\section{Summary \& Conclusions}  
\label{discuss}

We presented  a detailed analysis of 84 \os\  and 105 \cf\ systems at 
$z \sim 2.3$ detected in high resolution ($R \sim$~45,000) spectra 
of 18 bright QSOs observed with VLT/UVES. Here we summarize the main results 
of our survey.   

\vskip 0.20in
\par\noindent 
$\bullet$ {\bf Multiphase nature of the gas : }
Consistent with the previous studies, we show that \os\ components have 
systematically wider Doppler parameters ($b$) compared to those  
of \cf\ components. We also show that the line spread ($\delta v$) 
of \cf\ and \os\ are strongly correlated. Therefore we conclude that the 
metal absorbing regions in the IGM consist of multiphase gas correlated 
over large velocities. 

We do not find any trend between $N(\os)$ and $N(\hi)$, over five orders of 
magnitude spread in $N(\hi)$ but there is a $3\sigma$ level correlation between 
line spread of \os\ and $N(\hi)$ which possibly suggests that the \hi\ and \os\ 
occur in different phases of a correlated structure. 
Indeed, \citet{Fox11} recently argued that the 
constancy of $N(\os)$ over a range of $N(\hi)$ can be reconciled if 
\os\ absorption originate from conductive, turbulent or shocked boundary 
layers between warm and hot plasma. Such models are considered to explain 
the multiphase structure seen in different components of our galaxy 
\citep[][]{Savage94,Savage03,Spitzer96,Sembach03,Collins04,Collins05,
Lehner11}. 
Existing simulations of metals in the IGM roughly predict such a multiphase 
correlated structure where \os\ absorption originates from low density and 
extended region whereas \cf\ originates from regions of slightly higher density 
\citep[e.g.,][]{Rauch97a,Kawata07,Fangano07,Oppenheimer09,Cen11}. 

\vskip 0.1in
\par\noindent 
$\bullet$ {\bf Thermal state of the gas and non-thermal velocities : } 
The observations presented here are inconsistent with most of the gas 
associated with the \os\ absorption (at $z \sim 2.3$) being in 
collisional ionization equilibrium (i.e. with $T \sim (2-3)\times10^{5}$~K). 
We draw this conclusion mainly from the $b$(\os) distribution.
Using a subsample of well aligned \os\ components we find the median gas 
temperature is $\sim3\times10^{4}$~K with none of the components having 
$T > 2\times10^{5}$~K. However, 42\% of these well aligned 
components show $4.6 \le$ log~$T$ $\le 5.0$, which is warmer than the 
temperature expected in pure photoionization equilibrium.   
In case of rapidly cooling over-ionized gas such temperatures are 
expected provided the gas abundance is close to the solar value 
\citep[see][]{Gnat07}. 
The estimated non-thermal contribution to the $b$-parameter using 
the $b(\os)$--$b(\hi)$ pairs are in the range 
$3.6\le b_{nt}$(\kms)~$\le 21.2$ with a median value of 8.2 \kms. 
This is consistent with the previous measurements of intergalactic 
turbulence by \citet{Rauch96,Rauch01}. 

\vskip 0.1in
\par\noindent 
$\bullet$ {\bf Column density distribution : }
The \os\ column density distribution is well fitted by a power-law 
with indices $\beta = 1.9\pm0.1$ and $2.4\pm0.2$ for systems and 
components respectively for log~$N$(\os)~$>$~13.7. We find the 
distribution is flatter in the case of \cf\ with respective $\beta$ 
values of $1.6\pm0.1$ and $1.9\pm0.1$ down to log~$N(\cf) = 12.6$. 
For both \cf\ and \os\ the $\beta$ values measured for the components 
are higher compared to that measured for systems. 
This is a natural consequence of the fact that the systems with 
higher column density having systematically higher number of 
components. 
While we could not make precise comparisons of these results with the 
predictions of existing simulations, it is interesting to note that some 
of the simulations (with or without feedback) produce similar trends  
\citep[see for example Fig. 10 of][]{Rauch97a}. \citet{Cen11} have found a 
steeper slope for the $N(\os)$ distribution compared to that 
of \cf. They attributed this to the existence of transient structures 
stemming from the shock heated regions in the neighborhood of galaxies. 

\vskip 0.1in
\par\noindent 
$\bullet$ {\bf Gas kinematics : }
A strong correlation (5.3$\sigma$ level) is seen between the velocity 
spreads ($\delta v$) of \os\ and \cf. However, $\delta v(\cf)$ is 
systematically lower than $\delta v(\os)$. 
We find that both $\delta v(\os)$ and $\delta v(\cf)$ are strongly correlated 
($>5\sigma$ level) with their respective column densities and slightly less 
(3 and 4.5$\sigma$ respectively) correlated with $N$(\hi). 
We note that \os\ and/or \cf\ systems with large velocity spread also show 
associated low ion absorption lines. As the low ions (as well as strong \hi\ 
absorption) are expected to originate from high density regions, we conclude 
that systems with large velocity spread are probably associated with regions of 
high density.  
We measured the velocity offset, $|\Delta v (\os -\cf)|$, between optical 
depth weighted redshifts of \cf\ and \os\ absorption, which is found to be 
in the range $0 \le |\Delta v (\os -\cf)| \le 48$~\kms\ with a median value 
of 8 \kms. The systems with low ions seem to have higher velocity shift 
which possibly indicates higher ionization inhomogeneity in these absorbers.   
We do not find any strong correlation between 
$|\Delta v (\os -\cf)|$ and other observable parameters.

\vskip 0.1in
\par\noindent 
$\bullet$ {\bf Total column densities and their ratios : }   
The total column densities of different species (i.e. \hi, \cf, and \os) 
seem to be affected by the presence of low ions. The median values of 
$N(\hi)$, $N(\os)$ and $N(\cf)$ are found to be higher when low ions 
are present. A two sided KS test suggests that the column density 
distributions for systems with and without low ions are significantly 
different. 
Almost $\sim$~80\% of the systems with low ions show log~$N(\hi)>$~15.0 
indicating that considerable \hi\ absorption may be originating from the 
low ionization phase. We find a strong correlation 
($\sim 4.3\sigma$) between $N$(\cf) and $N$(\hi) which is dominated by 
the systems with low ions.  
We do not find any clear evidence for the column density ratios 
(i.e. $N(\os)/N(\hi)$, $N(\cf)/N(\hi)$ and $N(\os)/N(\cf)$) to evolve with 
redshift over the range 1.9~$\le z \le$~3.1. We find only a tentative 
evidence for number of systems with high log~$N(\os)/N(\hi)$ 
(e.g., $\ge -$0.5 dex) to decrease with increasing redshift. 
A similar trend is also present for $N(\cf)/N(\hi)$ ratio.

\vskip 0.1in
\par\noindent 
$\bullet$ {\bf Comparison between low and high-$z$~ \os\ absorbers : } 
The $b$-parameter distribution of \os\ components of our sample is 
significantly different from that of the low-$z$ sample of \citet{Tripp08}. 
The median value of $b(\os)$ at low-$z$ is twice as high as at
high-$z$ \citep[see also][]{Fox11}. 
In addition, the median $b_{\rm nt}$ value for the well aligned components in 
our sample is a factor $\sim2$ less than what is found by \citet{Tripp08}. 
Interestingly the median value of the temperature measured in these well aligned 
components is found to be the same (i.e. $T\sim3\times10^4$K) in both 
high and low-$z$ sample. All these are consistent with the non-thermal 
contribution to $b(\os)$ being higher at low-$z$.  
In the models of \citet{Evoli11} where the turbulence induced by supernova 
driven winds is considered, the volume-weighted $b$-parameter remains roughly 
constant between $z = 2$ to $z = 0$ [see their Fig. 6]. Such a case is not 
supported by our observations. We speculate that the excess of turbulence 
at low-$z$ may originate from shocks due to structure formation.   

At low redshift, a few thermally broadened Ly$\alpha$ absorbers (BLAs) are 
seen with signature of high temperature ($T\sim10^{6}$~K) 
gas \citep[]{Sembach04,Richter04,Richter06,Danforth10,Savage11}. 
We do not detect such systems in our high-$z$ sample. This may be related to 
difficulties in detecting broad and shallow absorption features in 
the dense Ly$\alpha$ forest. 

At $z \sim 2.3$ we find the cosmic density of \os\ absorbers,  
$\Omega_{\os}$ = (1.0$\pm$0.2)$\times 10^{-7}$ for log~$N(\os) > 13.7$. 
This should be treated as a lower limit as (a) we have not applied the 
correction factor to redshift path length due to Ly$\alpha$ line blanketing 
in the forest, (b) the CDDF of \os\ is steep and $\Omega_{\os}$ 
may increase when contributions of numerous low column density systems 
are included. 
We calculate the lower limits on the baryonic content of the \os\ 
absorbers assuming (i) ionization fraction of \os,  
$f_{\os} $ = 0.2, (ii) the average metallicity of the \os\ bearing gas 
is 10\% of the solar value as assumed in the low-$z$ studies of \os\ absorbers 
\citep[see,][]{Tripp00,Savage02,
Sembach04,Danforth05,Lehner06,Danforth08}. Most of these papers have 
shown that low redshift \os\ absorbers harbor roughly 5--10\% of the 
baryons in the nearby universe. 
We find this contribution to be 2.8\% at $z \sim 2.3$. 
The correction to the redshift path length due to line blanketing 
\citep[as suggested by][]{Simcoe02} 
will increase it up to $\sim$~7\%. Therefore within allowed uncertainties,  
$\Omega_{\rm IGM}^{\os}$ at $z \sim 0$ is consistent 
with what we found at $z \sim 2.3$. If \os\ predominantly traces a hot  
phase of the IGM (i.e. $T > 10^{5}$~K) at every epoch then most  
numerical simulations \citep[][]{Cen99,Dave01,Fang01,Chen03,Cen11} 
suggest a strong increase in $\Omega_{\rm IGM}^{\os}$ with decreasing 
redshift. These simulations also suggest that the fraction of baryons 
at $T < 10^{5}$~K decreases with decreasing $z$. 
Therefore the near constancy of $\Omega_{\rm IGM}^{\os}$ with redshift probably 
means that the \os\ absorbers at high and low redshift may not originate from 
regions with similar physical conditions. 

\section{acknowledgment} 
We thank the anonymous referee for useful comments that 
significantly improve this paper. 
SM thanks CSIR for providing support for this work. RS and PPJ 
acknowledge support from the Indo-French Centre for the Promotion 
of Advanced Research under the programme No.4304--2. We thank 
Pushpa Khare, Anand Narayanan, Kandaswamy Subramanian for useful 
discussions.  

\def\aj{AJ}%
\def\actaa{Acta Astron.}%
\def\araa{ARA\&A}%
\def\apj{ApJ}%
\def\apjl{ApJ}%
\def\apjs{ApJS}%
\def\ao{Appl.~Opt.}%
\def\apss{Ap\&SS}%
\def\aap{A\&A}%
\def\aapr{A\&A~Rev.}%
\def\aaps{A\&AS}%
\def\azh{AZh}%
\def\baas{BAAS}%
\def\bac{Bull. astr. Inst. Czechosl.}%
\def\caa{Chinese Astron. Astrophys.}%
\def\cjaa{Chinese J. Astron. Astrophys.}%
\def\icarus{Icarus}%
\def\jcap{J. Cosmology Astropart. Phys.}%
\def\jrasc{JRASC}%
\def\mnras{MNRAS}%
\def\memras{MmRAS}%
\def\na{New A}%
\def\nar{New A Rev.}%
\def\pasa{PASA}%
\def\pra{Phys.~Rev.~A}%
\def\prb{Phys.~Rev.~B}%
\def\prc{Phys.~Rev.~C}%
\def\prd{Phys.~Rev.~D}%
\def\pre{Phys.~Rev.~E}%
\def\prl{Phys.~Rev.~Lett.}%
\def\pasp{PASP}%
\def\pasj{PASJ}%
\def\qjras{QJRAS}%
\def\rmxaa{Rev. Mexicana Astron. Astrofis.}%
\def\skytel{S\&T}%
\def\solphys{Sol.~Phys.}%
\def\sovast{Soviet~Ast.}%
\def\ssr{Space~Sci.~Rev.}%
\def\zap{ZAp}%
\def\nat{Nature}%
\def\iaucirc{IAU~Circ.}%
\def\aplett{Astrophys.~Lett.}%
\def\apspr{Astrophys.~Space~Phys.~Res.}%
\def\bain{Bull.~Astron.~Inst.~Netherlands}%
\def\fcp{Fund.~Cosmic~Phys.}%
\def\gca{Geochim.~Cosmochim.~Acta}%
\def\grl{Geophys.~Res.~Lett.}%
\def\jcp{J.~Chem.~Phys.}%
\def\jgr{J.~Geophys.~Res.}%
\def\jqsrt{J.~Quant.~Spec.~Radiat.~Transf.}%
\def\memsai{Mem.~Soc.~Astron.~Italiana}%
\def\nphysa{Nucl.~Phys.~A}%
\def\physrep{Phys.~Rep.}%
\def\physscr{Phys.~Scr}%
\def\planss{Planet.~Space~Sci.}%
\def\procspie{Proc.~SPIE}%
\let\astap=\aap
\let\apjlett=\apjl
\let\apjsupp=\apjs
\let\applopt=\ao
\bibliographystyle{mn}
\bibliography{mybib}

\end{document}